\documentclass[a4paper,11pt]{article}
\pdfoutput=1
\usepackage{jcappub}
\usepackage[T1]{fontenc}
\usepackage{xcolor}
\usepackage{todonotes}

\begin{document}

\title{Soundness of Dark Energy properties}

\author[a]{Eleonora Di Valentino,}
\author[b]{Stefano Gariazzo,}
\author[b]{Olga Mena,}
\author[c]{and Sunny Vagnozzi}

\affiliation[a]{Jodrell Bank Center for Astrophysics, School of Physics and Astronomy, University of Manchester, Oxford Road, Manchester, M13 9PL, United Kingdom}
\affiliation[b]{Instituto de F\'isica Corpuscular (IFIC), CSIC-Universitat de Val\`encia, Apartado de Correos 22085,  E-46071, Spain}
\affiliation[c]{Kavli Institute for Cosmology (KICC) and Institute of Astronomy, University of Cambridge, Madingley Road, Cambridge CB3 0HA, United Kingdom}

\emailAdd{eleonora.divalentino@manchester.ac.uk}
\emailAdd{gariazzo@ific.uv.es}
\emailAdd{omena@ific.uv.es}
\emailAdd{sunny.vagnozzi@ast.cam.ac.uk}

\abstract{Type Ia Supernovae (SNeIa) used as standardizable candles have been instrumental in the discovery of cosmic acceleration, usually attributed to some form of dark energy (DE). Recent studies have raised the issue of whether intrinsic SNeIa luminosities might evolve with redshift. While the evidence for cosmic acceleration is robust to this possible systematic, the question remains of how much the latter can affect the inferred properties of the DE component responsible for cosmic acceleration. This is the question we address in this work. We use SNeIa distance moduli measurements from the Pantheon and JLA samples. We consider models where the DE equation of state is a free parameter, either constant or time-varying, as well as models where DE and dark matter interact, and finally a model-agnostic parametrization of effects due to modified gravity (MG). When SNeIa data are combined with Cosmic Microwave Background (CMB) temperature and polarization anisotropy measurements, we find strong degeneracies between parameters governing the SNeIa systematics, the DE parameters, and the Hubble constant $H_0$. These degeneracies significantly broaden the DE parameter uncertainties, in some cases leading to ${\cal O}(\sigma)$ shifts in the central values. However, including low-redshift Baryon Acoustic Oscillation and Cosmic Chronometer measurements, as well as CMB lensing measurements, considerably improves the previous constraints, and the only remaining effect of the examined systematic is a $\lesssim 40\%$ broadening of the uncertainties on the DE parameters. The constraints we derive on the MG parameters are instead basically unaffected by the systematic in question. We therefore confirm the overall soundness of dark energy properties.}

\maketitle

\section{Introduction}
\label{sec:intro}

Unprecedented precision measurements of Cosmic Microwave Background (CMB) temperature and polarization anisotropies from the \textit{Planck} satellite, together with Type Ia Supernovae (SNeIa) luminosity distance measurements, and measurements of the clustering of the large-scale structure (LSS) of our Universe from a number of galaxy redshift surveys, have provided a remarkably accurate description of the Universe (see e.g.~\cite{Alam:2016hwk,Scolnic:2017caz,Aghanim:2018eyx}). The combination of these measurements points towards the so-called concordance $\Lambda$CDM model, describing a spatially flat Universe where most of the energy content takes the form of two dark components: dark matter (DM) and dark energy (DE), see for instance~\cite{Sahni:2004ai}. DM represents about 25\% of the total energy budget of the Universe, while visible matter (baryons) accounts for 5\%.

The remaining $\sim$70\% of the total energy is associated to a mysterious component, which does not behave like matter and is commonly referred to as dark energy. The DE component is responsible for the current accelerated expansion of the Universe, and thus constitutes a fundamental key towards understanding the fate of the latter. Yet, its nature and gravitational properties remain largely unknown. The simplest explanation ascribes the effects of DE to the presence of a vacuum energy density, \textit{i.e.}\ to a so-called cosmological constant. While remarkably consistent with observational data, this simple scenario also appears to be extremely fine-tuned. An example of this is the fact that the amounts of DM and DE happen to be of the same order of magnitude today, an intriguing coincidence if one recalls how differently these two components evolve along the expansion history of the Universe. Such fact is the so-called \textit{coincidence problem}. Relaxing the hypothesis that DE is due to a cosmological constant, or in other words letting the DE contribution to the Universe budget evolve with time, could alleviate the required fine-tuning. A wide variety of DE models alternative to the cosmological constant, featuring either new fundamental particles and fields or modifications to the gravitational sector, have been proposed in the literature: for a selection of both seminal and more recent approaches towards explaining DE and cosmic acceleration, see e.g.~\cite{Wetterich:1987fm,Ratra:1987rm,Caldwell:1997ii,Peebles:1998qn,Kamenshchik:2001cp,Bento:2002ps,Freese:2002sq,Li:2004rb,Barbieri:2005gj,Cicoli:2012tz,Rinaldi:2014yta,Hlozek:2014lca,Rinaldi:2015iza,Nunes:2016aup,Sola:2016ecz,Capozziello:2017buj,Benisty:2018qed,Visinelli:2018utg,Benisty:2018oyy,DAgostino:2019wko,Heckman:2019dsj} and references therein.

Among the plethora of available precision cosmological datasets, SNeIa hold a special place, having provided the first evidence for cosmic acceleration in the late 1990s~\cite{Riess:1998cb,Perlmutter:1998np}. The most important use of SNeIa in cosmology is as distance indicators, in the form of standard(izable) candles~\cite{Tripp:1997wt}. In practice, this rests upon two important ideas. The first is that SNeIa luminosities can be empirically standardized, by making use of empirical correlations between light-curve properties and intrinsic luminosities, as well as quality cuts. The second is that the standardization is independent of the SNeIa redshifts or environments: in other words, that the standardized intrinsic luminosities should not display important trends with host galaxy redshift and/or morphology. Another way of expressing this second caveat is to state that two different SNeIa in different hosts, with the same colour, light-curve stretch, and host stellar mass, should on average have the same intrinsic luminosity, independently of their redshift.

Over the past decade, a large number of studies have established the existence of correlations between host galaxy properties (including their redshifts) and intrinsic SNeIa luminosities, see for example~\cite{Gallagher:2008zi,Kelly:2009iy,Uddin:2017rmc}. In this regard, plenty of discussion has been recently devoted in the literature to address the issue of whether or not intrinsic SNeIa luminosities evolve with redshift~\cite{Kim:2019npy,Kang:2019azh,Rose:2020shp}. Related works have examined whether intrinsic SNeIa luminosities correlate with the host star formation rate~\cite{Rigault:2014kaa,Rigault:2018ffm,Rigault:2013gux,Childress:2014vka,Jones:2018vbn} or metallicity~\cite{Timmes:2003xx,Travaglio:2005yt,Moreno-Raya:2015jqq,Moreno-Raya:2016rlw}. Overall, most of these studies appear to agree that SNeIa found in low-mass, star-forming, low-metallicity hosts are on average fainter than those found in high-mass, passive, high-metallicity ones. What these studies do not always agree upon is the magnitude of such correlations~\cite{Campbell:2016zzh,2018A&A...615A..68R,Rose:2020shp}. It is also worth mentioning the recent findings of~\cite{Brout:2020msh}, which suggests that the intrinsic scatter of SNeIa distance moduli residuals after standardization (and in particular their correlation with host galaxy properties) is entirely, or almost entirely, due to dust-induced extinction. The same work also proposes a physically motivated two-component SNeIa color model which accounts for the effects of dust.

Various analyses motivated by these findings and adopting different redshift-dependent parametrizations of the standardized SNeIa distance moduli, allowing for SNeIa to have different intrinsic luminosities as a function of redshift, have been carried out in the literature~\cite{Nordin:2008aa,Ferramacho:2008ap,Linden:2009vh,Tutusaus:2017ibk,Tutusaus:2018ulu,LHuillier:2018rsv,Martinelli:2019krf,Sapone:2020wwz}. In particular, the recent works of~\cite{Kang:2019azh} appear to suggest a large redshift-dependent luminosity evolution which, if correct, could question the evidence for cosmic acceleration from SNeIa. The correctness of these results were later disputed in~\cite{Rose:2020shp}. More generally, recent literature has witnessed the ignition of a strong debate around whether or not SNeIa data (alone or in combination with low-redshift probes), once possible systematics thereof are accounted for, can prove the accelerated expansion of the Universe (see e.g.~\cite{Nielsen:2015pga,Rubin:2016iqe,Dam:2017xqs,Colin:2018ghy,Desgrange:2019npu}). Here, we do not seek to fuel this debate further, but wish to point out that, despite the central role SNeIa played in establishing cosmic acceleration, evidence for the latter remains very strong even if SNeIa data are not taken into account. The evidence for cosmic acceleration can indeed be established independently using other data, such as CMB lensing or Baryon Acoustic Oscillation (BAO) measurements, see e.g.~\cite{Sherwin:2011gv,Nadathur:2020kvq}.~\footnote{Note though that the results of~\cite{Sherwin:2011gv} hold only for an assumed primordial power-law spectrum and pure cold dark matter (see e.g.~\cite{Hunt:2008wp,Hunt:2015iua} for alternative possibilities featuring a bump in the primordial power spectrum and/or neutrino hot dark matter, which could in principle be consistent with no acceleration).}

While we believe that the evidence for dark energy/cosmic acceleration is robust to SNeIa observational systematics, the aforementioned discussions make it clear that the time is ripe to study whether and to what extent these systematics impact the inferred properties of DE. In other words, how sound are the properties of DE we infer from observations including SNeIa. In particular, we shall here ask the following question: ``\textit{Granted that the evidence for dark energy is very robust, to what extent are the inferred properties of dark energy affected when neglecting a possible redshift-dependence of intrinsic SNeIa luminosities?}'' 

It is worth noting that the possibility of a redshift dependence of intrinsic SNeIa luminosities is not a far-fetched one. Without even needing to resort to astrophysical arguments, such an evolution can take place in a wide variety of beyond-$\Lambda$CDM scenarios, such as models featuring a time variation of the gravitational constant or more generally of the fundamental constants (see e.g.~\cite{Calabrese:2013lga,Wright:2017rsu}). This can also occur in many models of dark energy or modified gravity (e.g.~\cite{Brans:1961sx,Nunes:2016plz,Jimenez:2020bgw}), particularly if featuring additional (light) scalar degrees of freedom which couple to visible matter either directly or through the electromagnetic sector (for instance~\cite{Carroll:1998zi,Burrage:2016bwy,Vagnozzi:2019kvw,Jimenez:2020ysu}), and/or in string-inspired scenarios (see e.g.~\cite{Sandvik:2001rv,Damour:2002mi}). This highlights the important fact that, when using SNeIa data to constrain DE models beyond the cosmological constant, it might be important to include the effect of possible redshift-dependent intrinsic SNeIa luminosities simply for theoretical consistency (see for instance recent detailed discussions in~\cite{Wright:2017rsu,Zumalacarregui:2020cjh}). Conversely, the absence of detection of redshift-dependent intrinsic SNeIa luminosities could be used to constrain several DE models.

In this work, we shall explore the impact of relaxing the assumption of intrinsic SNeIa luminosities being independent of redshift, focusing on a number of different DE scenarios and considering two SNeIa samples: the \textit{Pantheon}~\cite{Scolnic:2017caz} and Joint Light-curve Analysis (\textit{JLA})~\cite{Betoule:2014frx} samples. Following earlier work in~\cite{Tutusaus:2017ibk,Tutusaus:2018ulu}, we will adopt a phenomenological but simple, model-agnostic, power-law parametrization for this effect. We will study the selected models considering both SNeIa combined with CMB temperature and polarization anisotropy data alone, or a more complete dataset combination further including CMB lensing and low-redshift measurements of the expansion history. Taking into account low-redshift observations such as Baryon Acoustic Oscillation distance measurements is extremely important when constraining models of cosmic acceleration~\cite{Aubourg:2014yra}. On the model side, we shall consider four classes of DE models: \textit{(a)} a DE component with constant equation of state (EoS) $w \neq -1$; \textit{(b)} a dynamical DE component with a time-dependent EoS described by the widely used Chevallier-Polarski-Linder (CPL) parametrization~\cite{Chevallier:2000qy,Linder:2002et}; \textit{(c)} three models of interacting DE, featuring interactions between DM and DE, and \textit{(d)} a model-agnostic parametrization of modified gravity scenarios. Given that the modified gravity results are affected by the anomalous lensing amplitude as inferred by the smoothing of the CMB temperature anisotropy power spectrum, we also check whether the inferred value of the phenomenological parameter $A_L$, which rescales the aforementioned lensing amplitude, is affected by the SNeIa systematic we are considering.  Anticipating our results, we find that the inferred properties of DE are relatively robust to a possible redshift-dependence of intrinsic SNeIa luminosities, particularly when SNeIa data are combined with additional low-redshift data. However, as one might expect, in general this possible additional systematic broadens the uncertainties on the inferred DE parameters.

The paper is then organized as follows. We start by describing our treatment of a possible redshift-dependence in the intrinsic SNeIa luminosities in Section~\ref{snia}. We then describe our analysis method in Section~\ref{analysis}. In Section~\ref{results}, we describe the four classes of dark energy models we study and examine the effects of this possible systematic by comparing the inferred dark energy properties with and without systematics for each of the models we consider. Our conclusions are summarized in Section~\ref{sec:conclusions}. For the sake of conciseness, given that the results obtained using the JLA sample are qualitatively similar to those obtained using the Pantheon one, we only summarize the former in our Tables but not in our Figures.

\section{Type Ia Supernovae measurements}
\label{snia}

Type Ia Supernovae (SNeIa) result from the thermonuclear disruption of carbon-oxygen white dwarfs, reaching their unstable phase through interaction with a binary companion. As discussed earlier, their wide use as distance indicators relies on the possibility of treating them as standardizable candles. The trustworthiness of this standardization procedure rests upon on a number of empirical correlations between SNeIa luminosities, colors, light-curve shapes, and host galaxy mass. In particular, the standardization procedure assumes that SNeIa form a homogeneous class of objects whose variability can mostly be characterized by two parameters~\cite{Tripp:1997wt}: the time stretching of the light-curve, the SNeIa color at maximum brightness. These two parameters are usually referred to as $X_1$ and $C$ respectively.

By measuring the redshifts and apparent magnitudes of SNeIa, one can extract their luminosity distances:
\begin{equation}
d_L(z) = (1+z)  \int_0^z \frac{dz'}{H(z')}\,,
\end{equation}
where $H(z)$ is the expansion rate of the Universe. When working with SNeIa measurements, one typically deals with the so-called distance modulus:
\begin{equation}
\mu(z) = 5 \log_{10} \left [ \frac{H_0}{c} d_L(z)\right ]+25 \,.
\end{equation}
The assumption underlying the standardization procedure is that SNeIa with identical color at maximum brightness ($C$), light-curve shape as characterized by its time stretching ($X_1$), and galactic environment, have on average the same intrinsic luminosity independently of redshift. Under these assumptions, one can model the \textit{observed} distance moduli as follows:
\begin{equation}
\mu_{\rm obs}(z) = m_B^*-(M_B-\alpha X_1 +\beta C)\,,
\label{eq:simple}
\end{equation}
where $m_B^*$ is the observed B-band rest-frame peak magnitude, and we have already discussed the physical meaning of $X_1$ and $C$ earlier. On the other hand, $\alpha$, $\beta$, and $M_B$ are nuisance parameters, whose physical interpretation is simply that of being the absolute magnitude of the SNeIa in the B-band rest-frame ($M_B$), the amplitude of the stretch correction ($\alpha$), and the amplitude of the color correction ($\beta$).

As stated  in the introductory section, a number of studies have considered the impact on cosmological parameter inference of several possible redshift-dependent parametrizations of intrinsic SNeIa luminosities. Our approach here is the following: given the fact that we are considering exotic DE models (including modified gravity models) which go beyond the simple cosmological constant, it could perfectly be the case that within these models, regardless of complications due to astrophysical evolution, the intrinsic SNeIa luminosity dependence on the redshift might not be as simple as implied by Eq.~\eqref{eq:simple}. As we discussed earlier, this might indeed be the case in several models which lead to space-time variations of the fundamental constants (see e.g.~\cite{Brans:1961sx,Carroll:1998zi,Sandvik:2001rv,Damour:2002mi,Calabrese:2013lga,Nunes:2016plz,Burrage:2016bwy,Wright:2017rsu,Vagnozzi:2019kvw,Jimenez:2020bgw,Jimenez:2020ysu}).

In order to parametrize our ignorance on a possible redshift-dependence of intrinsic SNeIa luminosities in an approach which is as agnostic as possible, we here follow the model presented in~\cite{Tutusaus:2018ulu,Tutusaus:2017ibk}, where Eq.~\eqref{eq:simple} is modified to the following:
\begin{equation}
\mu_{\rm obs}(z) = m_B^* -(M_B-\alpha X_1 +\beta C +\Delta m_{\rm{evo}}(z))\,,
\label{eq:comp}
\end{equation}
where $\Delta m_{\rm{evo}}(z)$ accounts for this possible redshift dependence of the intrinsic magnitude $M_B$ and takes the following functional form:
\begin{equation}
\Delta m_{\rm{evo}}(z) \equiv \epsilon z^\delta\,,
\label{eq:comp1}
\end{equation}
Therefore, the possible systematic we consider is described by the two parameters $\epsilon$ and $\delta$. A lower/higher power $\delta$ models a luminosity evolution dominant at low/high redshift.

We shall here explore the impact of the possible systematic described by Eqs.~\eqref{eq:comp} and \eqref{eq:comp1} by exploiting two different SNeIa samples. The Pantheon sample is the most complete one, consisting of distance moduli measurements from 1048 SNeIa from a number of surveys within the redshift range $0.01< z <2.3$~\cite{Scolnic:2017caz}. The earlier JLA sample instead includes distance moduli measurements from 740 SNeIa within the redshift range $0.01<z< 1$~\cite{Betoule:2014frx}. There is a key difference in the way these two samples are analyzed, related to the dependence of the covariance matrices on $\alpha$ and $\beta$, the amplitudes of the stretch and color corrections, as in Eq.~\eqref{eq:simple}. For the JLA sample, the covariance matrix depends explicitly on these two parameters, which must be added as extra nuisance parameters in the Monte Carlo Markov Chain (MCMC) scans. For the Pantheon sample, the effects of $\alpha$ and $\beta$ parameters are already taken into account (\textit{i.e.}\ pre-marginalized) in the calculation of the covariance matrix, so that these two parameters do not need to be varied in the MCMC scans.

\begin{figure}
\centering
\includegraphics[width=0.7\textwidth]{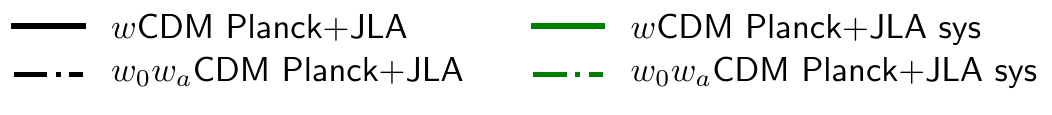}\\
\includegraphics[width=0.49\textwidth]{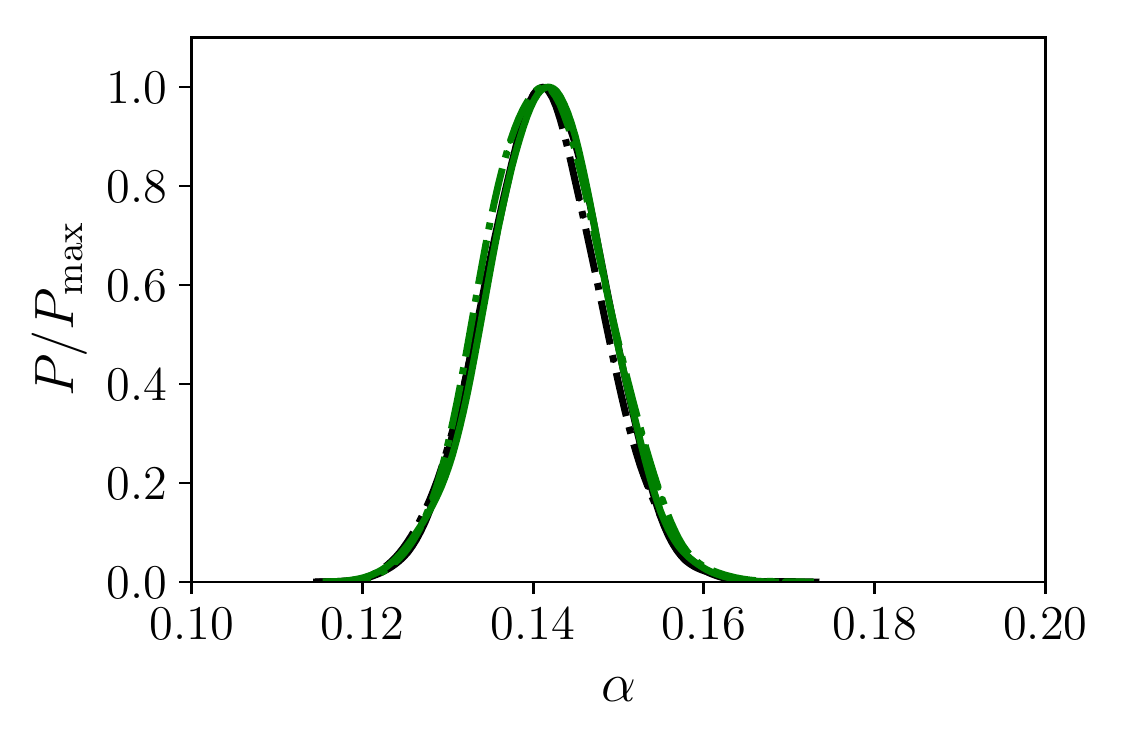}
\includegraphics[width=0.49\textwidth]{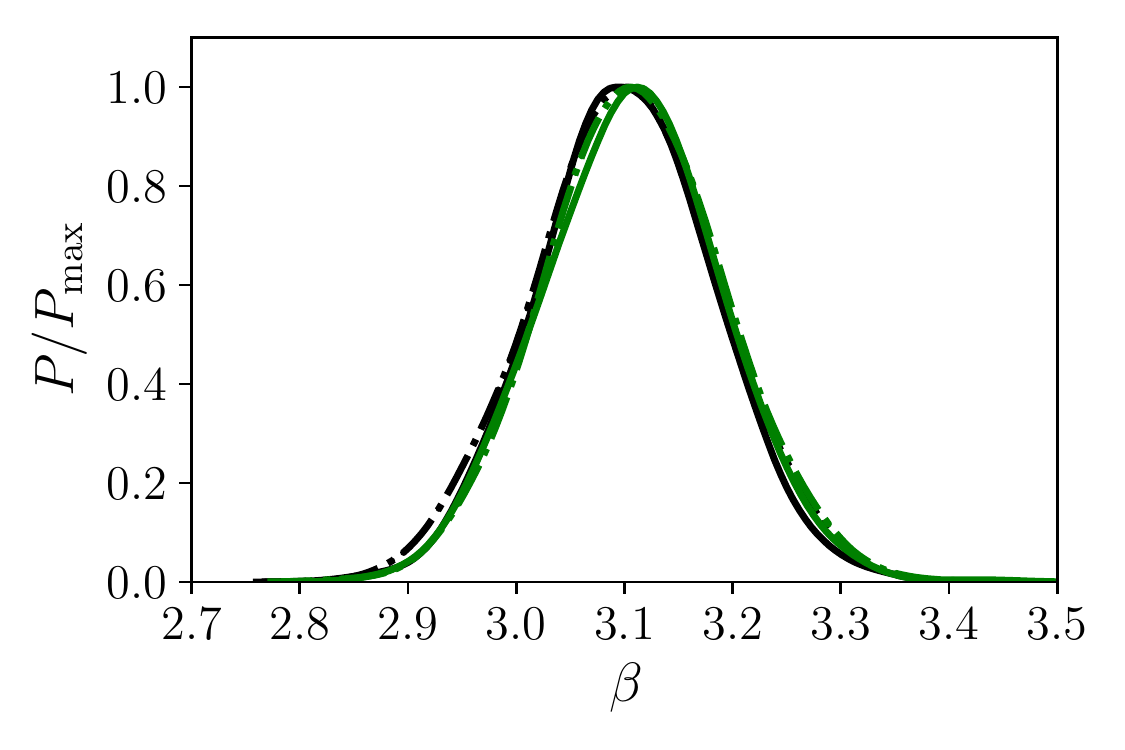}
\caption{1D marginalized posterior probability distributions for the amplitude of the stretch correction $\alpha$ (left panel) and the amplitude of the color correction $\beta$ (right panel), normalized to their maximum values. The posteriors have all been obtained combining \textit{Planck} CMB temperature and polarization anisotropy measurements with SNeIa distance moduli from the JLA sample, assuming a dark energy (DE) component with a constant equation of state (EoS) $w \neq -1$ (solid curves), and a DE component with a time-varying EoS $w(z)$ parametrized by the so-called CPL parametrization (dot-dashed curves). The black curves are obtained assuming no redshift-evolution of the intrinsic SNeIa luminosities, whereas the green curves account for a possible redshift evolution of the latter, as parametrized by Eqs.~\eqref{eq:comp} and \eqref{eq:comp1}.
}
\label{fig:beta}
\end{figure}

Nonetheless, one might be concerned that the introduction of systematics in the form of redshift-dependent intrinsic SNeIa luminosities might make the pre-marginalized Pantheon covariance matrix no longer safely utilizable. This could indeed be the case if introducing this systematic leads to important shifts in the inferred central values and uncertainties of $\alpha$ and $\beta$. To convince ourselves that this is not the case, we have analyzed the JLA sample for two reference DE models: the $w$CDM and $w_0w_a$CDM models, which we will describe in more detail in Sec.~\ref{ssec:wcdm} and Sec.~\ref{ssec:dde} respectively. We have found that, for both DE models and regardless of whether the systematic characterized by $\Delta m_{\rm{evo}}(z)$ is included or not, the posterior distributions of $\alpha$ and $\beta$ are essentially unaltered. For the $w$CDM case with (without) systematics we find $\alpha=0.142\pm 0.007$ ($0.141\pm 0.007$) and $\beta=3.11\pm 0.08$ ($3.10 \pm 0.08$). For the $w_0w_a$CDM case with (without) systematics we find $\alpha=1.414 \pm 0.007$ ($1.410 \pm 0.007$) and $\beta=3.11\pm 0.08$  ($3.10\pm 0.08$). These results are shown in Fig.~\ref{fig:beta}, where we plot the posterior distributions for $\alpha$ (left panel) and $\beta$ (right panel): the fact that in both panels the difference between the four curves is barely discernible by eye lends us confidence about the reliability of using the pre-marginalized Pantheon covariance matrix even in the presence of a possible redshift-dependence of intrinsic SNeIa luminosities. It is worth noting that these results agree with previous findings in the literature, see e.g.~\cite{Tutusaus:2018ulu}.~\footnote{One could in principle take a step further and relax the assumption of the stretch and color corrections being independent of redshift and environment, or equivalently that the associated nuisance parameters $\alpha$ and $\beta$ take the same values for all SNeIa in the sample. It is certainly worth examining the impact of relaxing these assumptions, which in turn would require introducing a significant amount of new nuisance parameters. As this is beyond the scope of the present work, we leave this investigation to a future project.}

\section{Datasets and analysis methods}
\label{analysis}

In the previous Section, we have already described the Pantheon~\cite{Scolnic:2017caz} and JLA~\cite{Betoule:2014frx} SNeIa samples, and justified our treatment of the covariance matrix of the former in the presence of the systematic we consider. Here, we describe the additional cosmological datasets and the analysis methods adopted. The following cosmological measurements are taken into account:
\begin{itemize}
\item \textbf{Planck}: Cosmic Microwave Background (CMB) measurements from the \textit{Planck} 2018 legacy release~\cite{Aghanim:2018eyx,Aghanim:2019ame}. We consider separately the temperature and polarization measurements and the CMB lensing measurements from the temperature four-point correlation function~\cite{Aghanim:2018oex}. In particular, temperature and polarization measurements are always combined with the Pantheon and/or JLA samples, whereas lensing measurements are only considered in the full dataset combination which also includes the BAO and Cosmic Chronometer measurements described below.
\item \textbf{BAO}: Baryon Acoustic Oscillations (BAO) distance measurements from the 6dFGS~\cite{Beutler:2011hx}, SDSS-MGS~\cite{Ross:2014qpa}, and BOSS DR12~\cite{Alam:2016hwk} surveys.
\item \textbf{Cosmic Chronometers}: cosmic chronometers (CC) exploit the evolution of differential ages of passive galaxies at different redshifts to directly constrain the Hubble parameter~\cite{Jimenez:2001gg}. We use 30 CC measurements of $H(z)$ in the redshift range $0<z<2$, compiled in~\cite{Moresco:2012by,Moresco:2012jh,Moresco:2015cya,Moresco:2016mzx}.~\footnote{The modified CC likelihood code we use is publicly available at \url{https://github.com/sunnyvagnozzi/CosmoMC-patches/tree/master/Cosmic_clocks}.}
\end{itemize}

We consider four different dataset combinations. In the first two, we combine the two different SNeIa compilations (one at any given time) with CMB temperature and polarization measurements: we refer to these dataset combinations  as \textit{Planck + Pantheon} and \textit{Planck + JLA}. We remark that these combinations do not include CMB lensing. To these two dataset combinations we then add CMB lensing, BAO, and Cosmic Chronometer measurements, and refer to the resulting dataset combinations as \textit{all + Pantheon} and \textit{all + JLA}, respectively. Each of these four dataset combinations is analyzed with and without systematics due to a possible redshift-dependence of intrinsic SNeIa luminosities: when systematics are included, we refer to the resulting dataset combinations as \textit{all + Pantheon sys} and \textit{all + JLA sys}, respectively. Therefore, we have in total eight possible scenarios for each of the DE models we consider, which will be described in the following.

Our baseline cosmology is described by the six parameters of the concordance $\Lambda$CDM model: the baryon and cold DM physical energy densities $\Omega_bh^2$ and $\Omega_ch^2$, the ratio between the sound horizon and the angular diameter distance at decoupling $\theta_s$, the optical depth to reionization $\tau$, and the amplitude and tilt of the primordial power spectrum of scalar fluctuations $A_s$ and $n_s$.

As anticipated earlier, we consider four classes of DE models, described by one or two extra parameters, for a total of seven or eight parameters. We start by considering the simplest case where the DE equation of state (EoS) $w$ is allowed to vary instead of being fixed to $w=-1$ as in the cosmological constant case. We refer to the resulting model, described by seven parameters, as the $w$CDM model, see Sec.~\ref{ssec:wcdm} for the results. We then consider a dynamical DE scenario, where the DE EoS is time-dependent, with its redshift evolution described by the widely used Chevallier-Polarski-Linder (CPL) parametrization. We refer to the resulting model, described by eight parameters, as the $w_0w_a$CDM model, see Sec.~\ref{ssec:dde} for the results. Afterwards, we consider three interacting dark energy (IDE) models, featuring energy exchange between DM and DE. The energy exchange rate is the same in all three models, but what is different is the number of free parameters and the allowed ranges thereof. We first consider a seven-parameter IDE model with one extra parameter, the DM-DE coupling strength $\xi$, which is required to satisfy $\xi<0$: we refer to this model as coupled vacuum or $\xi\Lambda$CDM model. We then consider two eight-parameter IDE models where in addition to $\xi$, also the DE EoS $w$ is allowed to vary. For stability reasons, we consider separately the case where $w<-1$ and $\xi>0$, which we refer to as coupled phantom or $\xi p$CDM model, and the case where $w>-1$ and $\xi<0$, which we refer to as coupled quintessence or $\xi q$CDM model. The results for the three IDE models, as well as the way the parameters $\xi$ and $w$ enter the cosmological equations (in particular modifying the continuity equations for DM and DE), are discussed in Sec.~\ref{ssec:ide}. We then consider an 8-parameter model-agnostic parametrization of modified gravity (MG), where the two free extra parameters $E_{11}$ and $E_{22}$ parametrize the deviations of the equations describing the metric perturbations from the canonical General Relativity case. The results for MG, as well as the way the two extra parameters $E_{11}$ and $E_{22}$ enter the cosmological equations (in particular modifying the Poisson and lensing equations, and mimicking anisotropic stress), are discussed in Sec.~\ref{sec:MG}. In addition, given that the MG results are to some extent affected by the mild disagreement on the amount of lensing between CMB temperature anisotropies and the reconstructed lensing potential, we consider a case where DE is described by a cosmological constant, but we treat the lensing amplitude $A_L$ (which rescales the amplitude of lensing in the CMB temperature power spectrum) as a free parameter, see Sec.~\ref{sec:alens} for the results.

\begin{table}[!b]
\begin{center}
\begin{tabular}{|c|c|c|c|c|c|c|}
\hline
Parameter   & $w$CDM & $w_0w_a$CDM & $\xi\Lambda$CDM & $\xi p$CDM & $\xi q$CDM & MG \\
\hline
$\Omega_bh^2$         & $[0.005,0.1]$& $[0.005,0.1]$& $[0.005,0.1]$& $[0.005,0.1]$& $[0.005,0.1]$& $[0.005,0.1]$ \\
$\Omega_ch^2$         & $[0.001,0.99]$& $[0.001,0.99]$& $[0.001,0.99]$& $[0.001,0.99]$& $[0.001,0.99]$& $[0.001,0.99]$ \\
$100\theta_s$             & $[0.5,10]$ & $[0.5,10]$& $[0.5,10]$& $[0.5,10]$& $[0.5,10]$& $[0.5,10]$\\
$\tau$                       & $[0.01,0.8]$& $[0.01,0.8]$& $[0.01,0.8]$& $[0.01,0.8]$& $[0.01,0.8]$& $[0.01,0.8]$ \\
$n_s$               & $[0.7,1.3]$& $[0.7,1.3]$& $[0.7,1.3]$& $[0.7,1.3]$& $[0.7,1.3]$& $[0.7,1.3]$ \\
$\log[10^{10}A_{s}]$      & $[1.7, 5.0]$& $[1.7, 5.0]$& $[1.7, 5.0]$& $[1.7, 5.0]$& $[1.7, 5.0]$& $[1.7, 5.0]$ \\
$w$                    & $[-2,1]$& $[-2,1]$ &-& $[-3,-1]$& $[-1,1]$&-\\
$w_0$                    & $[-2,1]$& $[-2,1]$ &-& $[-3,-1]$& $[-1,1]$&-\\
$w_a$                    &-& $[-2,2]$&-&-&-&- \\
$\xi$                  &-&-  & $[-1,0]$& $[0,0.5]$& $[-1,0]$ &-\\
$E_{11}$                   &-&- &-&-&-& $[-1,3]$ \\
$E_{22}$                 &- &-&- &-&- & $[-1.4,5]$ \\
\hline 
\end{tabular}
\end{center}
\caption{Flat prior ranges on the cosmological parameters adopted in our analyses. Each of the six columns corresponds to one of the six different dark energy models considered in this paper and described in the text.}
\label{tab:priors}
\end{table}
Unless otherwise specified, we adopt flat priors on all cosmological parameters. In Table~\ref{tab:priors}, we show the prior ranges adopted on the cosmological parameters, for each of the DE models we consider. For the two SNeIa systematic parameters $\epsilon$ and $\beta$, we set flat priors within $[-1;1]$ and $[0.2;2]$ respectively, in agreement with earlier works in~\cite{Tutusaus:2017ibk,Tutusaus:2018ulu}, in order to avoid degeneracies with $M_B$. We scan the parameter spaces in question using MCMC methods, through the MCMC sampler \texttt{CosmoMC}~\cite{Lewis:2002ah}, suitably interfaced with the Boltzmann solver \texttt{CAMB}~\cite{Lewis:1999bs}, and which supports the new 2018 Planck likelihood code. We monitor the convergence of the generated chains using the Gelman-Rubin parameter $R-1$~\cite{Gelman:1992zz}, and require $R-1<0.01$ for the chains to be considered converged.

\section{Results}
\label{results}

Here, we describe the results of our analyses for each of the four classes of dark energy models described above, as well as the case where we leave the lensing amplitude parameter $A_L$ free.

\subsection{Constant dark energy equation of state $w$}
\label{ssec:wcdm}

\begin{figure}
\centering
\begin{tabular}{cc}
\multicolumn{2}{c}{\includegraphics[width=.6\textwidth]{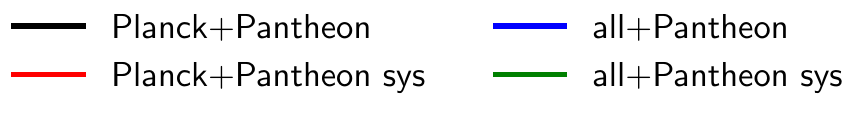}}\\
\includegraphics[width=.49\textwidth]{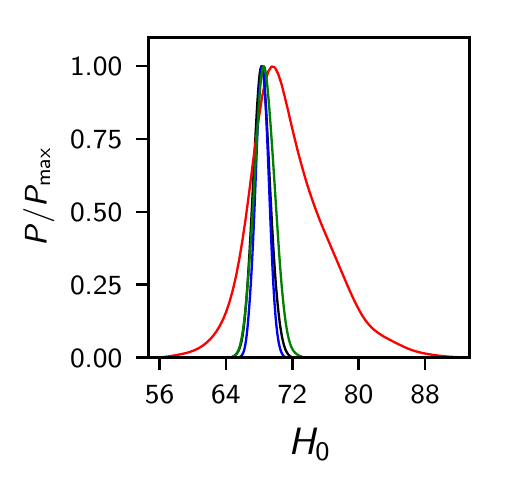}
&\includegraphics[width=.49\textwidth]{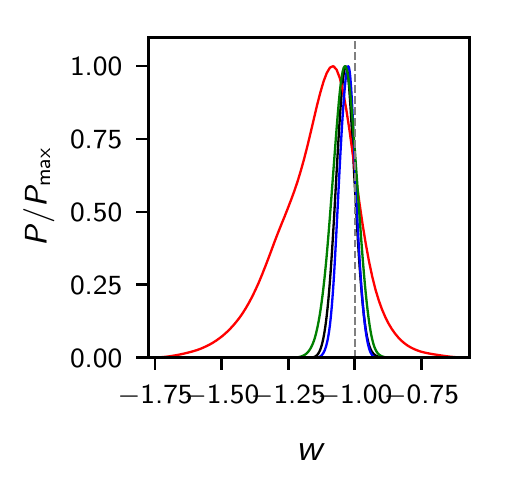}\\
\end{tabular}
\caption{1D marginalized posterior probability distributions for the Hubble constant $H_0$ (left panel) and the dark energy equation of state $w$ (right panel), normalized to their maximum values. The dashed grey vertical line at $w=-1$ indicates the standard $\Lambda$CDM value. The posteriors have been derived assuming the $w$CDM cosmology described in Sec.~\ref{ssec:wcdm}. We have used the following dataset combinations/treatment of systematics: \textit{Planck} CMB temperature and polarization anisotropies in combination with SNeIa distance moduli measurements from the Pantheon sample, with (red curves) and without (black curves) a possible systematic due to redshift evolution of the intrinsic SNeIa luminosities, as parametrized by Eqs.~\eqref{eq:comp}, \eqref{eq:comp1}; additionally considering CMB lensing, BAO, and Cosmic Chronometer measurements, with (green curves) and without (blue curves) this systematic. Qualitatively similar findings hold when using the JLA SNeIa sample instead of the Pantheon one: the results are not shown here, but summarized in Tables~\ref{tab1} and~\ref{tab1b}.}
\label{fig:wCDMP_bi}
\end{figure}

\begin{figure}
\centering
	\includegraphics[width=.7\textwidth]{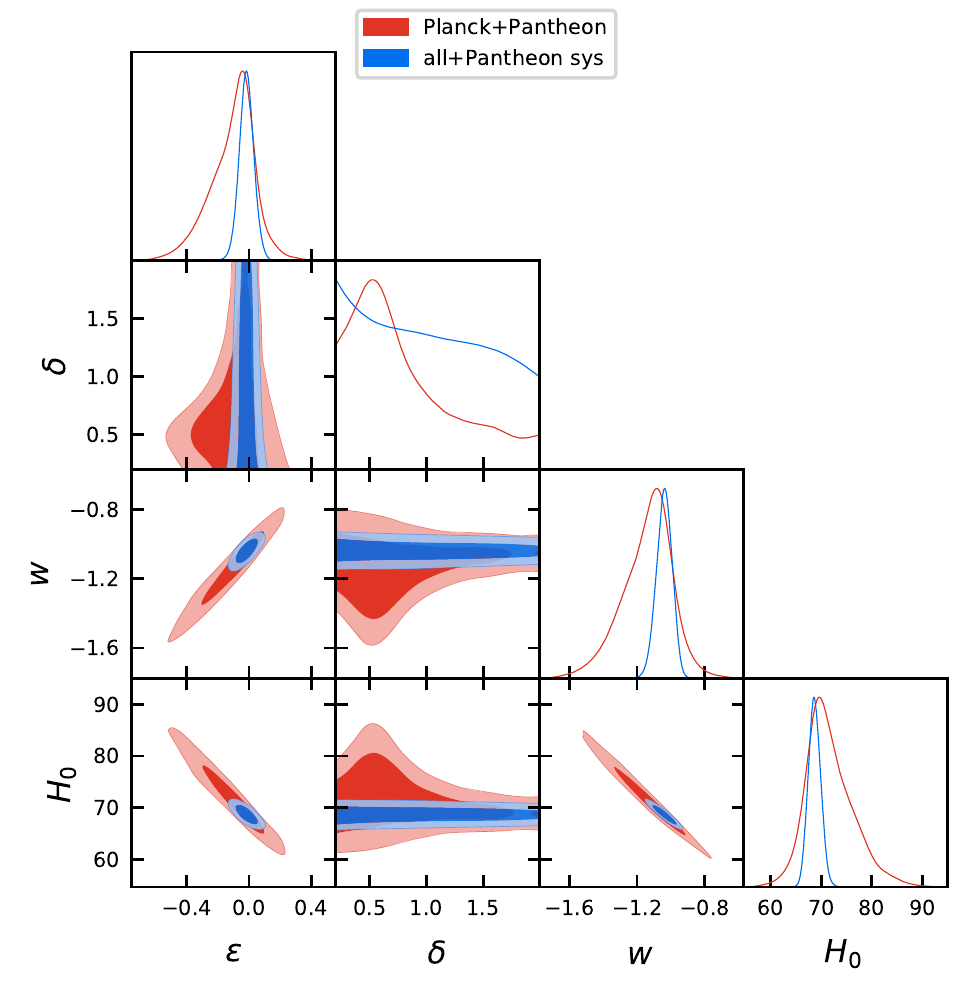}\\
	\caption{Triangular plot showing 2D joint and 1D marginalized posterior probability distributions for the Hubble constant $H_0$, the dark energy equation of state $w$, and the two parameters $\epsilon$ and $\delta$ controlling the redshift-dependence of the intrinsic SNeIa luminosities as parametrized by Eqs.~\eqref{eq:comp}, \eqref{eq:comp1}. The posteriors have been obtained combining \textit{Planck} CMB temperature and polarization anisotropies with SNeIa distance moduli measurements from the Pantheon sample (red contours), and additionally considering CMB lensing, BAO, and cosmic chronometer measurements (blue contours), in both cases with systematics characterizing a possible redshift-dependence of the intrinsic SNeIa luminosities included. Qualitatively similar findings hold when using the JLA SNeIa sample instead of the Pantheon one, with results not shown here.}
	\label{fig:wCDMJ_tri}
\end{figure}

We begin by considering the simplest DE scenario beyond the cosmological constant, where the DE EoS $w$ is treated as a constant free parameter. We refer to this model as the $w$CDM model. Considering first the Pantheon sample, the results are shown in Figs.~\ref{fig:wCDMP_bi} and~\ref{fig:wCDMJ_tri} as well as in Tabs.~\ref{tab1} and~\ref{tab1b}. The results including the JLA sample are instead shown in Tabs.~\ref{tab1} and~\ref{tab1b}. It is worth noticing that, when only combining SNeIa data with \textit{Planck} data, the effect of the SNeIa systematics is very noticeable: the DE EoS is shifted towards more phantom values, with the corresponding uncertainties being about four times larger than when SNeIa systematics are not taken into account.

It is worth noting that both the mean value of the Hubble constant, as well as its uncertainty, are larger when SNeIa systematics are included. The reason, as is clear from Fig.~\ref{fig:wCDMJ_tri}, is not only due to the existence of a strong anti-correlation between $H_0$ and $w$, but also to the mutual correlations between $H_0$, $w$, and $\epsilon$. As a result of these correlations, a larger positive (negative) value of $\epsilon$ will imply smaller (larger) distance moduli: this effect can be compensated by values of the DE EoS $w>-1$ ($w<-1$), which in turn implies a lower (higher) value of $H_0$. Notice that the mean value of $\epsilon$ is negative although compatible with zero at $1\sigma$, being $\epsilon=-0.11^{+0.16}_{-0.11}$ at $68\%$ confidence level (CL). Instead, the parameter $\delta$ associated to the redshift evolution of the intrinsic SNeIa luminosities shows no particular correlation with any of the parameters, and is poorly constrained.

At face value, the effects of SNeIa systematics on $H_0$ we have just described would appear to alleviate the so-called ``\textit{Hubble tension}'', \textit{i.e.}\ the $4.4\sigma$ disagreement between estimates of $H_0$ from \textit{Planck} CMB data assuming an underlying $\Lambda$CDM model~\cite{Aghanim:2018eyx}, and from a local distance ladder approach as done by the SH$0$ES team~\cite{Riess:2019cxk}. As discussed at length in the literature, the Hubble tension might be pointing towards new physics, a possibility which has been the subject of significant study in the literature: see e.g.~\cite{DiValentino:2016hlg,Bernal:2016gxb,Karwal:2016vyq,Zhao:2017urm,Sola:2017znb,Buen-Abad:2017gxg,Khosravi:2017hfi,Benetti:2017juy,Mortsell:2018mfj,Nunes:2018xbm,Poulin:2018zxs,Kumar:2018yhh,Banihashemi:2018oxo,DEramo:2018vss,Guo:2018ans,Graef:2018fzu,Banihashemi:2018has,Poulin:2018cxd,Kreisch:2019yzn,Raveri:2019mxg,Agrawal:2019lmo,Li:2019san,Yang:2019jwn,Keeley:2019esp,Lin:2019qug,Li:2019ypi,Rossi:2019lgt,DiValentino:2019exe,Archidiacono:2019wdp,Kazantzidis:2019dvk,Desmond:2019ygn,Nesseris:2019fwr,Vagnozzi:2019ezj,Pan:2019hac,Visinelli:2019qqu,Cai:2019bdh,Xiao:2019ccl,Smith:2019ihp,Ghosh:2019tab,Sola:2019jek,Escudero:2019gvw,Yan:2019gbw,Sakstein:2019fmf,Anchordoqui:2019amx,Frusciante:2019puu,Akarsu:2019hmw,Ye:2020btb,Krishnan:2020obg,DAgostino:2020dhv,Benevento:2020fev,Desmond:2020wep,Akarsu:2020yqa,Haridasu:2020xaa,Alestas:2020mvb,Braglia:2020iik,Ballardini:2020iws,Elizalde:2020mfs} for a selection of works examining this possibility. However, as pointed out in a number of recent works (e.g.~\cite{Lemos:2018smw,Aylor:2018drw,Knox:2019rjx}), it is important to check that proposed solutions are consistent with BAO distance measurements. Therefore, we defer a full assessment of whether the systematics we are examining can conclusively address the Hubble tension to our discussion concerning the full dataset combination (\textit{i.e.}\ the \textit{all}+\textit{Pantheon sys} or \textit{all}+\textit{JLA sys} dataset combinations).

Adding all the other cosmological datasets (CMB lensing, BAO, and Cosmic Chronometers) to our analysis results in the uncertainties on cosmological parameters being reduced. However, these uncertainties are still larger than those obtained for the case where CMB temperature and polarization data alone are combined with SNeIa data without taking this systematic into account. In fact, comparing the \textit{Planck}+\textit{Pantheon} to the \textit{all}+\textit{Pantheon sys} cases (see Fig.~\ref{fig:wCDMJ_tri} for a clear visual comparison), we see that in the latter case the uncertainties are broadened by about $40\%$. For instance, for these two cases we find $w=-1.035 \pm 0.035$ and $w=-1.040 \pm 0.046$ respectively. It is also worth noting that, once the full dataset combination is considered, the Hubble tension remains at the $\approx 3\sigma$ level, and hence is unsolved.

The results obtained considering the JLA SNeIa sample instead of the Pantheon one are shown in Tabs.~\ref{tab1} and~\ref{tab1b}. When combining JLA data with CMB temperature and polarization anisotropies alone, the uncertainties on the parameters are also $3$-$4$ times larger when allowing for a redshift-dependence in the intrinsic SNeIa luminosities, as we observed earlier for the Pantheon sample. However, the shifts in $H_0$ and $w$ are less noticeable than in the Pantheon case. When considering the full dataset combination, the uncertainties are again enlarged by about $40\%$, as we found for the Pantheon case. It is worth underlying that, as we had already noted earlier in Fig.~\ref{fig:beta}, the amplitudes of the stretch and color corrections [$\alpha$ and $\beta$ in Eq.~\eqref{eq:simple}] are basically unaffected by the introduction of the systematic we are considering, validating the robustness of our analysis, and in particular the choice of using the pre-marginalized covariance matrix when analyzing the Pantheon sample.

A final word concerning the difference in the parameter uncertainties when using the two SNeIa samples is in order. Neglecting a redshift-dependence in the intrinsic SNeIa luminosities, the uncertainties inferred when using the Pantheon sample are always smaller than that obtained using the JLA sample, regardless of whether the \textit{Planck} or \textit{all} dataset combinations are used. This is in agreement with previous works~\cite{Scolnic:2017caz,Jones:2017udy}. However, once the possibility of redshift-dependent intrinsic SNeIa luminosities is taken into account, the difference between the two samples becomes barely noticeable, except for a slightly larger uncertainty on both $\epsilon$ and $\delta$ for the JLA case (compare the second and fourth columns of Tabs.~\ref{tab1} and~\ref{tab1b}).

\begin{center}                              
\begin{table*} 
\scalebox{1}{
\begin{tabular}{cccccccccccccccc}       
\hline\hline                                                                                                                    
Parameters & Planck   & Planck & Planck& Planck \\ 
 & +Pantheon &+Pantheon sys & +JLA  & +JLA sys \\ \hline

$w$ & $    -1.035\pm0.035$ &  $    -1.14^{+0.16}_{-0.12}$ & $ -1.038\pm0.051   $ & $  -1.06^{+0.18}_{-0.11}  $ \\

$H_0 $[km/s/Mpc] & $   68.3\pm1.0$&  $   71.7^{+3.5}_{-5.2}$ & $ 68.4\pm1.6  $ & $69.1^{+3.0}_{-5.7}   $ \\

$\Omega_m$ & $    0.307\pm0.010$ &  $    0.282\pm0.037$ & $ 0.307^{+0.014}_{-0.016}   $ & $ 0.305^{+0.046}_{-0.036}   $ \\

$\alpha$ & $    - $ &  $    -$ & $  0.1414\pm0.0066  $ & $ 0.1415\pm0.0066   $ \\

$\beta$ & $    - $ &  $    -$ & $  3.107\pm0.081  $ & $ 3.111\pm0.081   $ \\

$\epsilon$ & $    - $ &  $    -0.11^{+0.16}_{-0.11}$ & $  -  $ & $ -0.02^{+0.18}_{-0.10}   $ \\

$\delta$ & $    - $ &  $    <0.934$ & $  -  $ & $ <1.19   $ \\

\hline\hline                                                  
\end{tabular} 
}
\caption{Constraints on selected cosmological parameters of the $w$CDM cosmology described in Sec.~\ref{ssec:wcdm}. Constraints are reported as 68\% CL intervals or 68\% upper/lower limits. The results have been obtained combining \textit{Planck} CMB temperature and polarization anisotropies with SNeIa distance moduli measurements from the Pantheon sample (first and second columns), or using the JLA sample instead of the Pantheon one (third and fourth columns). We illustrate separately the cases with (second and fourth columns) and without (first and third columns) a possible systematic due to redshift evolution of the intrinsic SNeIa luminosities, as parametrized by Eqs.~\eqref{eq:comp}, \eqref{eq:comp1}.}
\label{tab1}                                              
\end{table*}                                                    
\end{center}

\begin{center}                              
\begin{table*} 
\scalebox{1}{
\begin{tabular}{cccccccccccccccc}       
\hline\hline                                                                                                                    
Parameters & all   & all & all& all \\ 
 & +Pantheon &+Pantheon sys & +JLA  & +JLA sys \\ 
  \hline

$w$& $    -1.028\pm0.031$ &  $    -1.040\pm0.046$ & $ -1.029\pm0.037   $ & $ -1.022^{+0.049}_{-0.042}   $ \\

$H_0 $[km/s/Mpc] & $   68.36\pm0.82$&  $   68.7\pm1.2$ & $ 68.40\pm0.97 $ & $ 68.2^{+1.1}_{-1.3}  $ \\

$\Omega_m$ & $    0.3054\pm0.0076$ &  $    0.303\pm0.011$ & $ 0.3051\pm0.0086   $ & $ 0.306\pm0.011   $ \\

$\alpha$ & $    - $ &  $    -$ & $  0.1413\pm0.0065  $ & $ 0.1415\pm0.0065   $ \\

$\beta$ & $    - $ &  $    -$ & $  3.106\pm0.081  $ & $ 3.109\pm0.082   $ \\

$\epsilon$ & $    - $ &  $    -0.016\pm0.048$ & $  -  $ & $ 0.016^{+0.064}_{-0.057}   $ \\

$\delta$ & $    - $ &  $    <1.33$ & $  -  $ & $ unconstrained   $ \\

\hline\hline                                                  
\end{tabular} 
}
\caption{As in Table~\ref{tab1}, but considering \textit{Planck} CMB temperature and polarization anisotropies and lensing measurements in addition to late-time Baryon Acoustic Oscillations and Cosmic Chronometer distance and expansion history measurements, in combination with SNeIa distance moduli measurements from the Pantheon sample (first and second columns), or using the JLA sample instead of the Pantheon one (third and fourth columns).}
\label{tab1b}                                              
\end{table*}                                                    
\end{center}

\subsection{Dynamical Dark Energy: a redshift-dependent dark energy equation of state $w(z)$}
\label{ssec:dde}

The next logical step in exploring the robustness of inferred constraints on DE properties against the possibility of redshift-dependent intrinsic SNeIa luminosities is to allow for a time-dependent DE EoS $w(z)$. Several parametrizations of time-varying dark energy, typically based on one or two additional free parameters, have been proposed in the literature: see for example~\cite{Efstathiou:1999tm,Jassal:2004ej,Barboza:2008rh,Ma:2011nc,Pantazis:2016nky,Escamilla-Rivera:2016qwv,Yang:2017alx,Pan:2017zoh,Vagnozzi:2018jhn,Yang:2018qmz}, and see also~\cite{Perkovic:2020mph} for a recent study examining the theoretical viability conditions for these parametrizations. It is also worth pointing out that the community has recently also been moving towards non-parametric reconstructions of the DE EoS, see e.g.~\cite{Hee:2016nho,Shafieloo:2018gin,Wang:2018fng,Gerardi:2019obr,DiazRivero:2019ukx}. Here, we shall instead focus on the simplest and undoubtedly most commonly adopted parametrization of a time-varying DE component, the so-called Chevallier-Polarski-Linder (CPL) parametrization. Within the CPL parametrization, the evolution of the DE EoS as a function of redshift is expressed in terms of two free parameters, $w_0$ and $w_a$, and takes the following form~\cite{Chevallier:2000qy,Linder:2002et}:
\begin{equation}\label{eq:cpl}
w(z) = w_0 + w_a\frac{z}{1+z}\,.
\end{equation}
This expression can be identified with the first two terms in a Taylor expansion of the DE EoS in powers of the scale factor $a=1/(1+z)$ around the present time. The truncation is well-justified as long as the DE EoS is smooth and does not oscillate in time. We refer to this model as the $w_0w_a$CDM model.

\begin{figure}
\centering
\begin{tabular}{ccc}
\multicolumn{3}{c}{\includegraphics[width=.6\textwidth]{legend_1d_Pantheon.pdf}}\\
\includegraphics[width=.32\textwidth]{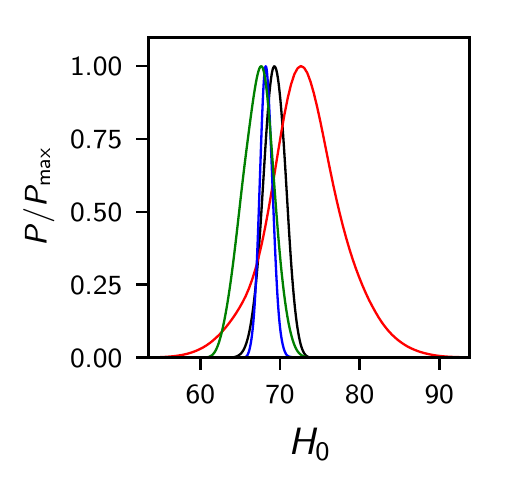}
&\includegraphics[width=.32\textwidth]{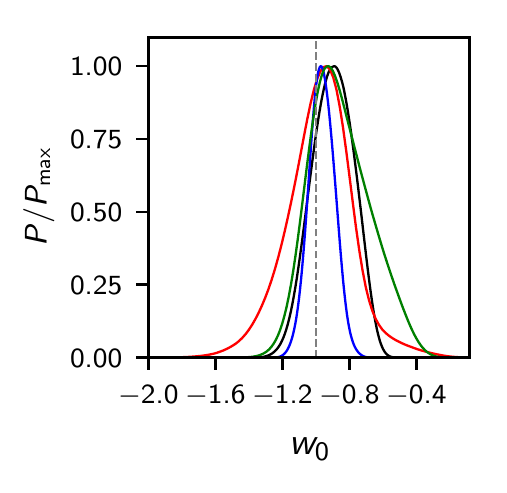}
&\includegraphics[width=.32\textwidth]{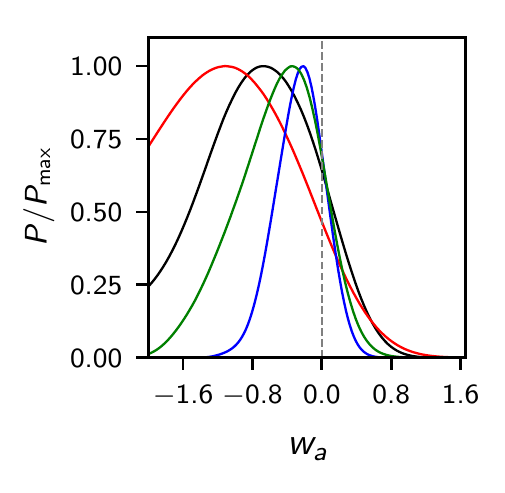}\\
\end{tabular}
\caption{As in Fig.~\ref{fig:wCDMP_bi}, for the parameters of the CPL dynamical dark energy $w_0w_a$CDM cosmology described in Sec.~\ref{ssec:dde}, with the qualitatively similar JLA results summarized in Tables~\ref{tabwa} and~\ref{tab1wa}.}
\label{fig:w0waCDMJLA_bi}
\end{figure}

\begin{figure}
\centering
	\includegraphics[width=.7\textwidth]{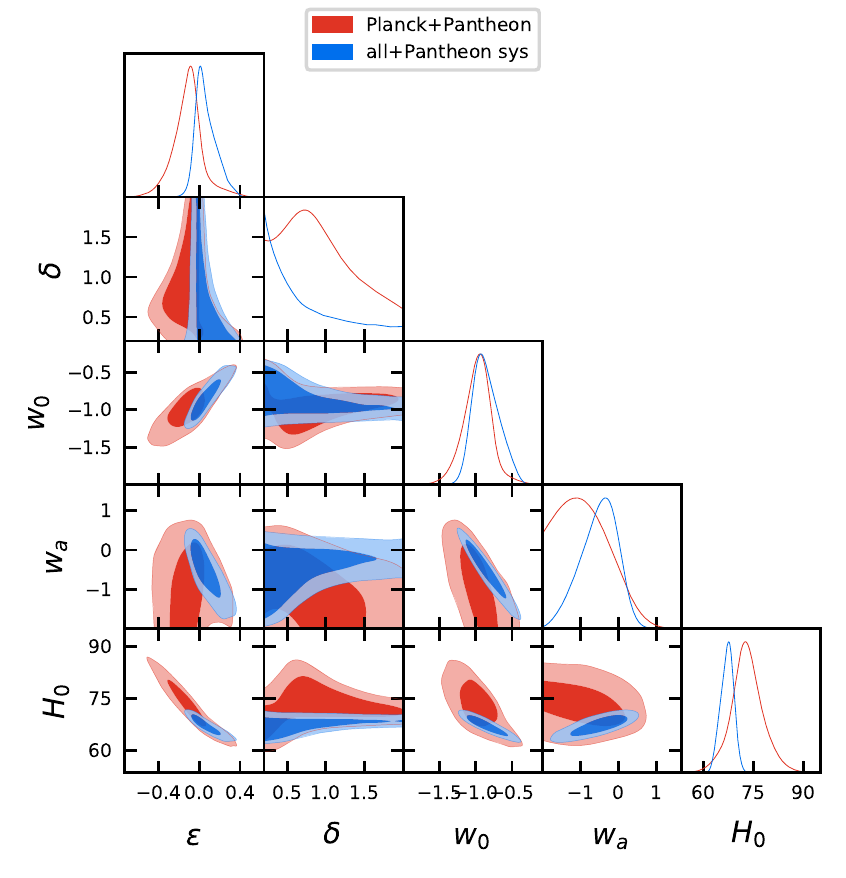}\\
	\caption{As in Fig.~\ref{fig:wCDMJ_tri}, for the parameters of the CPL dynamical dark energy $w_0w_a$CDM cosmology described in Sec.~\ref{ssec:dde}, with the qualitatively similar JLA results not shown here.}
	\label{fig:w0waCDMJLA_tri}
\end{figure}

The results for the $w_0w_a$CDM model are presented in Figs.~\ref{fig:w0waCDMJLA_bi} and~\ref{fig:w0waCDMJLA_tri}, and in Tabs.~\ref{tabwa} and~\ref{tab1wa}.  Considering the \textit{Planck}+\textit{Pantheon} dataset combination, we see that the uncertainties on the DE EoS parameters $w_0$ and $w_a$ can be increased by up to $30\%$ when allowing for the systematic we consider. The Hubble constant is also shifted to much larger values, with the corresponding uncertainty being about four times larger. As for the $w$CDM model we studied earlier, in the $w_0w_a$CDM there are also strong degeneracies between $w_0$, $H_0$, and $\epsilon$, which are now further complicated due to the presence of the extra parameter $w_a$. While the parameter $w_a$ is perfectly consistent with $w_a=0$ when not accounting for systematics, there is a $\approx 2\sigma$ preference for non-zero $w_a$, and hence for evolving DE, when accounting for these effects. However, once we consider the \textit{all}+\textit{Pantheon} dataset combination, this preference disappears completely.

If we inspect Tab.~\ref{tab1wa} we see that even for the full dataset combination, the systematic SNeIa we are considering still leaves a clear imprint on the inferred values of the DE EoS: the uncertainties are broadened by up to a factor of three, a factor that is comparable to the reduction in uncertainty brought upon by the addition of CMB lensing, BAO, and CC measurements to the starting \textit{Planck}+\textit{Pantheon} dataset combination. In other words, the improvement in constraints on DE parameters brought upon by adding CMB lensing, BAO, and CC measurements is basically removed if one allows for a redshift-dependence in the intrinsic SNeIa luminosities. In addition, it is clear that the Hubble tension remains unsolved.

When considering the JLA sample instead of the Pantheon one, we find qualitatively similar conclusions, apart for a significant difference regarding the Hubble constant. In this case, accounting for the systematic we are studying always shifts $H_0$ to lower values, further intensifying the Hubble tension. We explain this finding as follows. While constraints from the Pantheon and JLA samples are known to be in good agreement once combined with CMB and BAO measurements, there are nonetheless shifts in the constraints on $w_0$ and $w_a$ between the two, with up to $\sim 0.5\sigma$ significance~\cite{Jones:2017udy}. These shifts go in the direction of preferring more negative values of $w_0$ and less negative values of $w_a$, as we observe both in Tab.~\ref{tabwa} and Tab.~\ref{tab1wa} in our work, as well as Tab.~9 in~\cite{Jones:2017udy}. These shifts are further intensified once we allow for the systematics we are considering, because these systematics are redshift-dependent and the Pantheon sample covers a larger redshift range. When considering the mutual degeneracies between $H_0$, $w_0$, and $\epsilon$ (and to a minor extent $w_a$), as is clear from Fig.~\ref{fig:w0waCDMJLA_tri}, the net effect of these shifts is that JLA is expected to prefer a lower $H_0$ value compared to Pantheon when systematics are included. Finally, as already noted earlier in Fig.~\ref{fig:beta}, the amplitudes of the stretch and color corrections [$\alpha$ and $\beta$ in Eq.~\eqref{eq:simple}] are basically unaffected by the introduction of the systematic we are considering. 
\begin{center}                              
\begin{table*}
\scalebox{1}{
\begin{tabular}{cccccccccccccccc}       
\hline\hline                                                                                                                    
Parameters & Planck   & Planck & Planck& Planck \\ 
 & +Pantheon &+Pantheon sys & +JLA  & +JLA sys \\ \hline

$w_0$ & $    -0.90\pm0.13$ &  $    -0.98^{+0.19}_{-0.17}$ & $ -0.83^{+0.15}_{-0.10}   $ & $  -0.78^{+0.21}_{-0.29}  $ \\

$w_a$ & $    -0.69\pm0.61$ &  $    -0.90^{+0.44}_{-0.94}$ & $ <-0.797   $ & $  <-0.839  $ \\

$H_0 $[km/s/Mpc] & $   69.3\pm1.4$&  $   73.1\pm4.9$ & $ 70.1\pm1.7  $ & $68.9\pm5.8   $ \\

$\Omega_m$ & $    0.299^{+0.012}_{-0.014}$ &  $    0.271^{+0.029}_{-0.038}$ & $ 0.292^{+0.013}_{-0.017}   $ & $ 0.308^{+0.038}_{-0.059}   $ \\

$\alpha$ & $    - $ &  $    -$ & $  0.1412\pm0.0066  $ & $ 0.1413\pm0.0066   $ \\

$\beta$ & $    - $ &  $    -$ & $  3.106\pm0.081  $ & $ 3.109\pm0.081   $ \\

$\epsilon$ & $    - $ &  $    -0.11\pm0.15$ & $  -  $ & $ -0.05^{+0.17}_{-0.22}   $ \\

$\delta$ & $    - $ &  $    <1.12$ & $  -  $ & $ <1.01   $ \\

\hline\hline                                                  
\end{tabular} 
}
\caption{As in Table~\ref{tab1}, for the parameters of the CPL dynamical dark energy $w_0w_a$CDM cosmology described in Sec.~\ref{ssec:dde}.}
\label{tabwa}                                              
\end{table*}                                                    
\end{center}

\begin{center}                              
\begin{table*}
\scalebox{1}{
\begin{tabular}{cccccccccccccccc}       
\hline\hline                                                                                                                    
Parameters & all   & all & all& all \\ 
 & +Pantheon &+Pantheon sys & +JLA  & +JLA sys \\ 
  \hline

$w_0$ & $    -0.964\pm0.077$ &  $    -0.85^{+0.15}_{-0.21}$ & $ -0.92\pm0.10   $ & $  -0.70\pm0.19  $ \\

$w_a$ & $    -0.25^{+0.30}_{-0.26}$ &  $    -0.52^{+0.57}_{-0.40}$ & $ -0.39^{+0.36}_{-0.31}   $ & $  -0.91\pm0.52  $ \\

$H_0 $[km/s/Mpc] & $   68.28\pm0.81$&  $   67.2^{+2.1}_{-1.8}$ & $ 68.0\pm1.1 $ & $ 65.7\pm2.0  $ \\

$\Omega_m$ & $    0.3067\pm0.0076$ &  $    0.318^{+0.016}_{-0.021}$ & $ 0.309\pm0.010   $ & $ 0.331\pm0.020   $ \\

$\alpha$ & $    - $ &  $    -$ & $  0.1410\pm0.0066  $ & $ 0.1413\pm0.0066   $ \\

$\beta$ & $    - $ &  $    -$ & $  3.102\pm0.080  $ & $ 3.106\pm0.082   $ \\

$\epsilon$ & $    - $ &  $    0.07^{+0.07}_{-0.12}$ & $  -  $ & $ 0.15^{+0.10}_{-0.13}   $ \\

$\delta$ & $    - $ &  $    <0.923$ & $  -  $ & $ <0.923   $ \\

\hline\hline                                                  
\end{tabular} 
}
\caption{As in Table~\ref{tab1b}, for the parameters of the CPL dynamical dark energy $w_0w_a$CDM cosmology described in Sec.~\ref{ssec:dde}.}
\label{tab1wa}                                              
\end{table*}                                                    
\end{center}

\subsection{Interacting Dark Energy}
\label{ssec:ide}

In order to continue our study on the robustness of the inferred properties of DE against the possibility of redshift-dependent intrinsic SNeIa luminosities, we will now study three interacting dark energy (IDE) models. In particular, we consider a class of models where the DM and DE continuity equations are coupled as follows:
\begin{eqnarray}
&&\dot{\rho}_c + 3H \rho_c = Q\,,\\
&&\dot{\rho}_x + 3H (1+w) \rho_x =-Q\,,
\end{eqnarray}
where $\rho_c$ and $\rho_x$ are the DM and DE energy densities respectively, the DE EoS $w$ is assumed to be constant, and the coupling function $Q$ governs the interaction rate between the two dark components. A number of IDE models, and corresponding coupling functions, have been studied in the literature, see Refs.~\cite{Amendola:1999er,Mangano:2002gg,Farrar:2003uw,Pettorino:2004zt,Barrow:2006hia,Amendola:2006dg,He:2008tn,Pettorino:2008ez,Valiviita:2008iv,Baldi:2008ay,Gavela:2009cy,Majerotto:2009np,Gavela:2010tm,Baldi:2010td,Chimento:2011pk,Clemson:2011an,Amendola:2011ie,Bettoni:2012xv,Chimento:2012zz,Pettorino:2012ts,Chimento:2012aea,Chimento:2013se,Chimento:2013qja,Pettorino:2013oxa,Chimento:2013ira,Chimento:2013rya,Nunes:2014qoa,Richarte:2014yva,Casas:2015qpa,Murgia:2016ccp,Nunes:2016dlj,Kumar:2016zpg,Pan:2016ngu,Sharov:2017iue,Benisty:2017eqh,Kumar:2017dnp,Guo:2017hea,DiValentino:2017iww,Yang:2017zjs,Yang:2017ccc,Costa:2018aoy,Yang:2018pej,Yang:2018euj,Yang:2018uae,Li:2018ydj,Martinelli:2019dau,Paliathanasis:2019hbi,Kumar:2019wfs,Pan:2019jqh,Li:2019loh,Anagnostopoulos:2019myt,Yang:2019vni,Yang:2019uzo,Pan:2019gop,Nakamura:2019phn,DiValentino:2019ffd,Benetti:2019lxu,Kase:2019veo,Yang:2019uog,Aljaf:2019ilr,Cheng:2019bkh,Pan:2020zza,Yang:2020uga,Lucca:2020zjb,Hogg:2020rdp,Amendola:2020ldb,Benisty:2020nql,Gomez-Valent:2020mqn} for a list of both seminal and more recent works on the subject or Ref.~\cite{Wang:2016lxa} for a comprehensive review. These works have considered both theoretically motivated functions for $Q$, based on physical models for DM and DE (the latter often in the form of a scalar field), as well as more phenomenological functional forms.

In the following, we shall assume the following functional form for $Q$, which is purely phenomenological but has been widely studied in the literature:
\begin{eqnarray}
Q = 3H\xi \rho_x\,,
\label{model}
\end{eqnarray}
where $\xi$ is a dimensionless parameter characterizing the strength of the coupling. Notice that $\xi>0$ corresponds to an energy flow from DE to DM, and conversely for $\xi<0$. While the form of the coupling in Eq.~(\ref{model}) might raise the question as to how microscopic DM-DE interactions should know about the overall expansion rate $H$, one can instead argue that such a choice is instead a natural one from a thermodynamical perspective. In fact, this particular form implies that one can use the scale factor as time variable, and eliminate $H$ entirely, so that the conservation equations end up knowing only about the change in scale factor (or correspondingly volume), and nothing about the specific cosmology or theory of gravity. In other words, the presence of $H$ reflects the fact that a change in density depends on a change in volume, as one expects from thermodynamics considerations, and the functional form of the conservation equation is unaffected by the specific theory of gravity.

Following considerations concerning early-time instabilities (see e.g.~\cite{Valiviita:2008iv,Gavela:2009cy}), for the specific model given by Eq.~(\ref{model}), the signs of $(1+w)$ and $\xi$ should be opposite. In other words, within this model, an energy flow from DE to DM can only occur for phantom DE, and a flow in the reverse direction for quintessence-like DE. We shall therefore consider three sub-classes of this particular model, following the notation of the earlier~\cite{DiValentino:2019jae}. We begin by fixing the DE EoS to $w=-0.999$, which in turn requires $\xi<0$. We refer to the resulting model as coupled vacuum model, or $\xi\Lambda$CDM model. The rationale behind this choice, explained in e.g.~\cite{DiValentino:2019ffd}, is that for $w$ sufficiently close to $-1$ the effect of DE perturbations is basically unnoticeable: consequently, in the coupled vacuum model one is essentially only capturing the effect of the DM-DE coupling $\xi$, while at the same time ensuring the absence of instabilities which would be present for $w=-1$. The coupled vacuum model therefore provides a rather accurate surrogate for a $\Lambda$CDM+$\xi$ cosmology, and has seven free parameters. We then consider two eight-parameter models, where both $w$ and $\xi$ are free parameters. We first consider a coupled phantom model, where $w<-1$ and $\xi>0$, which we refer to as $\xi p$CDM model. Similarly, we consider a coupled quintessence model, where $w>-1$ and $\xi<0$, which we refer to as $\xi q$CDM model. We remark that all three the $\xi\Lambda$CDM, $\xi p$CDM  and $\xi q$CDM models are described by the same exchange rate function $Q$ given in Eq.~(\ref{model}).

\begin{figure}[!ht]
\centering
\includegraphics[width=.6\textwidth]{legend_1d_Pantheon.pdf}\\
\includegraphics[width=.49\textwidth]{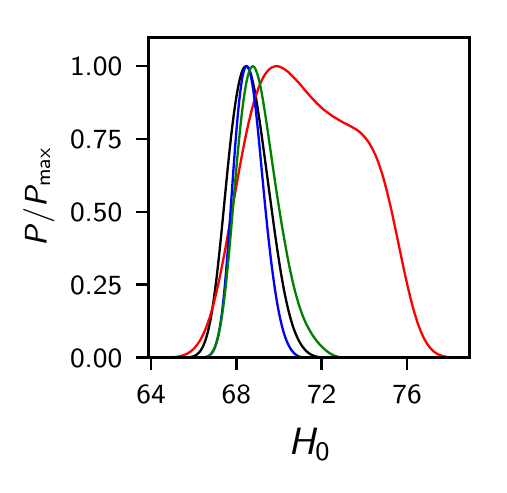}
\includegraphics[width=.49\textwidth]{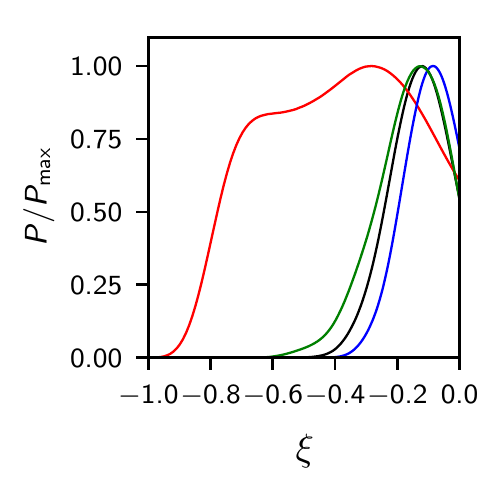}\\
\caption{As in Fig.~\ref{fig:wCDMP_bi}, for the parameters of the coupled vacuum $\xi\Lambda$CDM cosmology described in Sec.~\ref{ssec:ide}, with the qualitatively similar JLA results summarized in Tables~\ref{tabxi} and~\ref{tab1xi}.}
\label{fig:coupling_bi}
\end{figure}

\begin{figure}
\centering
	\includegraphics[width=.7\textwidth]{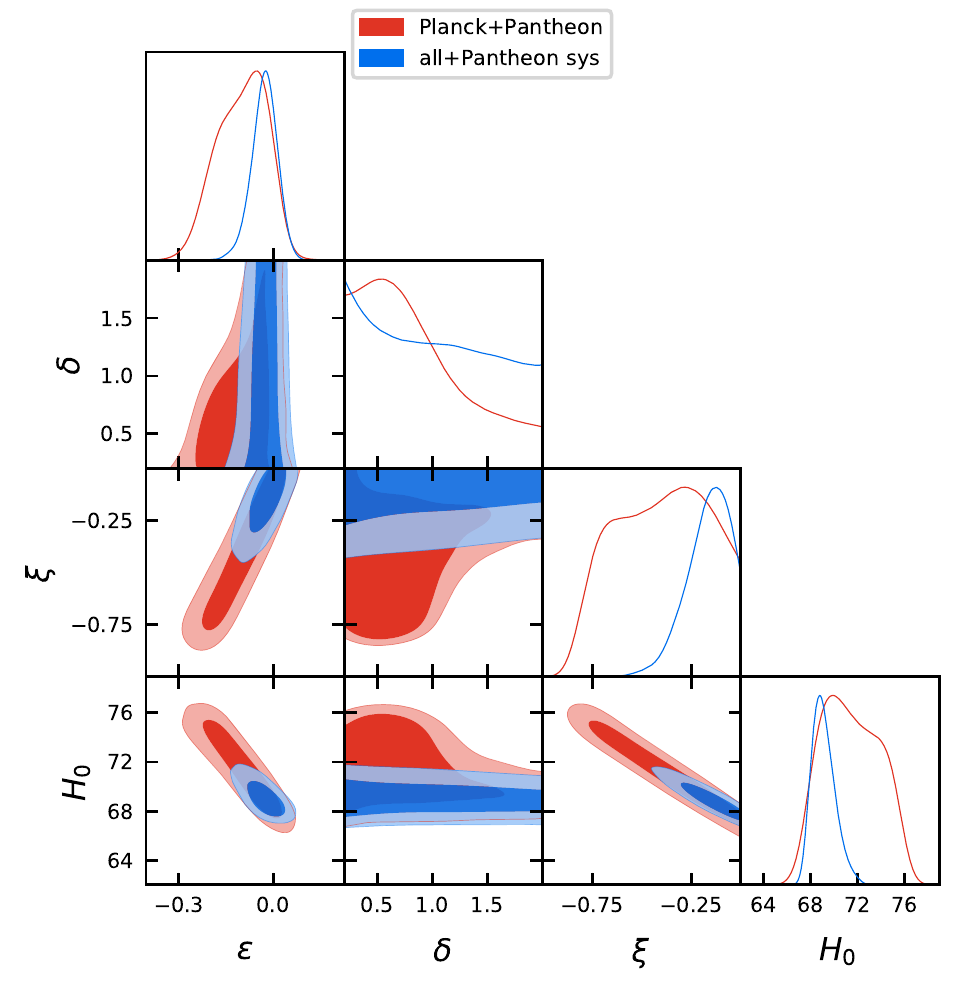}
	\caption{As in Fig.~\ref{fig:wCDMJ_tri}, for the parameters of the coupled vacuum $\xi\Lambda$CDM cosmology described in Sec.~\ref{ssec:ide}, with the qualitatively similar JLA results not shown here.}
	\label{fig:coupling_tri}
\end{figure}

We start by commenting on the results (shown in Figs.~\ref{fig:coupling_bi} and~\ref{fig:coupling_tri}, and Tabs.~\ref{tabxi} and~\ref{tab1xi}) for the coupled vacuum $\xi\Lambda$CDM model where $w$ is fixed. When only combining SNeIa data with CMB temperature and polarization anisotropies  data, we see that the effect of SNeIa systematics is rather large: for the \textit{Planck}+\textit{Pantheon} combination, accounting for a possible redshift-dependence in the intrinsic SNeIa luminosities leads to an approximately $2\sigma$ preference for $\xi<0$, whereas for the \textit{Planck}+\textit{JLA} combination the constraint on $\xi$ is considerably loosened. For both dataset combinations, the value of $H_0$ is also higher, as we found previously for the $w$CDM model but not for the $w_0w_a$CDM one.

The effects we described above can be understood in terms of the mutual correlations between $H_0$, $\xi$ and $\epsilon$, see Fig.~\ref{fig:coupling_tri}. Note that $H_0$ and $\epsilon$ are negatively correlated: a larger positive (negative) value of $\epsilon$ will imply smaller (larger) distance moduli: this can be compensated by a value of $\xi>0$ ($\xi<0$), which in turn implies a lower (higher) value of $H_0$. These degeneracies are indeed closely related to those we found within the $w$CDM model. The reason is easy to understand: in this IDE model, the DE energy density $\rho_x$ scales as $\rho_x \propto (1+z)^{-3(1+w+\xi/3)}$, as if the DE component had an effective EoS $(w+\xi/3)$. The mean values of the matter energy density are much lower (with corresponding uncertainties about twice as large) when taking SNeIa systematics into account. The reason is that $\xi$ and $\Omega_m$ are positively correlated. In the $\xi\Lambda$CDM coupled vacuum model, the matter energy density in the past is higher than in the standard $\Lambda$CDM model since the background evolution of the matter energy density gets an extra contribution proportional to the DE component. Consequently, the amount of energy density associated to the total matter currently needed will be smaller and smaller as the coupling $\xi$ becomes more and more negative. 

When we consider the full dataset combinations (\textit{all}+\textit{Pantheon} and \textit{all}+\textit{JLA}), we recover the patterns already seen earlier for the $w$CDM and $w_0w_a$CDM models. In other words, the bounds on the cosmological parameters are more constraining, and the effect of a possible redshift-dependence in the intrinsic SNeIa luminosities becomes much less noticeable at the level of inferred central values of the parameters. What persists is a degradation in the limits on $\xi$, even if less significant than in the case when SNeIa data is combined with CMB temperature and polarization anisotropy measurements alone.

\begin{center}                              
\begin{table*} 
\scalebox{1}{
\begin{tabular}{cccccccccccccccc}       
\hline\hline                                                                        Parameters & Planck & Planck & Planck& Planck \\ 
 & +Pantheon &+Pantheon sys & +JLA  & +JLA sys \\ \hline

$\xi$ & $    -0.15^{+0.11}_{-0.06}$ &  $    -0.41^{+0.29}_{-0.22}$ & $ -0.19^{+0.17}_{-0.07}   $ & $  >-0.493  $ \\

$H_0 $[km/s/Mpc] & $   68.6^{+0.8}_{-1.0}$&  $   71.4^{+2.4}_{-2.0}$ & $ 69.1^{+1.0}_{-1.5}  $ & $71.0^{+2.0}_{-3.3}   $ \\

$\Omega_m$ & $    0.271^{+0.032}_{-0.022}$ &  $    0.18^{+0.10}_{-0.07}$ & $ 0.255^{+0.048}_{-0.026}   $ & $ 0.20^{+0.11}_{-0.06}   $ \\

$\alpha$ & $    - $ &  $    -$ & $  0.1417\pm0.0065  $ & $ 0.1416\pm0.0065   $ \\

$\beta$ & $    - $ &  $    -$ & $  3.114\pm0.081  $ & $ 3.112\pm0.082   $ \\

$\epsilon$ & $    - $ &  $    -0.098^{+0.094}_{-0.071}$ & $  -  $ & $ -0.07^{+0.11}_{-0.09}   $ \\

$\delta$ & $    - $ &  $    <1.00$ & $  -  $ & $ <1.08   $ \\

\hline\hline                                                  
\end{tabular} 
}
\caption{As in Table~\ref{tab1}, for the parameters of the coupled vacuum $\xi\Lambda$CDM cosmology described in Sec.~\ref{ssec:ide}.}
\label{tabxi}                                              
\end{table*}                                                    
\end{center}

\begin{center}                              
\begin{table*} 
\scalebox{1}{
\begin{tabular}{cccccccccccccccc}       
\hline\hline                                                                                                                    
Parameters & all   & all & all& all \\ 
 & +Pantheon &+Pantheon sys & +JLA  & +JLA sys \\ 
  \hline

$\xi$& $    -0.12^{+0.11}_{-0.04}$ &  $    -0.17^{+0.14}_{-0.06}$ & $ -0.14^{+0.13}_{-0.04}   $ & $ >-0.178   $ \\

$H_0 $[km/s/Mpc] & $   68.61^{+0.61}_{-0.77}$&  $   69.1^{+0.8}_{-1.2}$ & $ 68.79^{+0.68}_{-0.98}  $ & $ 68.8^{+0.7}_{-1.1}  $ \\

$\Omega_m$ & $    0.276^{+0.027}_{-0.016}$ &  $    0.259^{+0.040}_{-0.021}$ & $ 0.270^{+0.034}_{-0.018}   $ & $ 0.268^{+0.038}_{-0.018}   $ \\

$\alpha$ & $    - $ &  $    -$ & $  0.1416\pm0.0066  $ & $ 0.1416\pm0.0067   $ \\

$\beta$ & $    - $ &  $    -$ & $  3.111\pm0.080  $ & $ 3.110\pm0.081   $ \\

$\epsilon$ & $    - $ &  $    -0.028^{+0.044}_{-0.036}$ & $  -  $ & $ -0.001^{+0.060}_{-0.053}   $ \\

$\delta$ & $    - $ &  $    <1.33$ & $  -  $ & $ unconstrained   $ \\

\hline\hline                                                  
\end{tabular} 
}
\caption{As in Table~\ref{tab1b}, for the parameters of the coupled vacuum $\xi\Lambda$CDM cosmology described in Sec.~\ref{ssec:ide}.}
\label{tab1xi}                                              
\end{table*}                                                    
\end{center}


\begin{figure}
\centering
\includegraphics[width=.6\textwidth]{legend_1d_Pantheon.pdf}\\
\includegraphics[width=.32\textwidth]{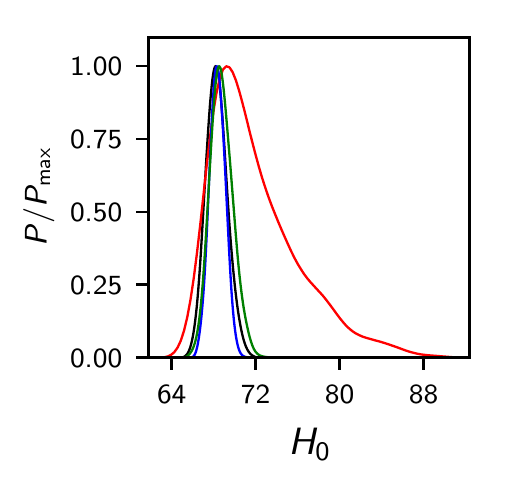}
\includegraphics[width=.32\textwidth]{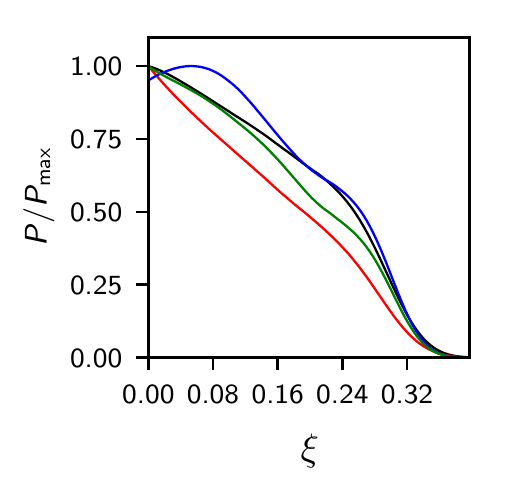}
\includegraphics[width=.32\textwidth]{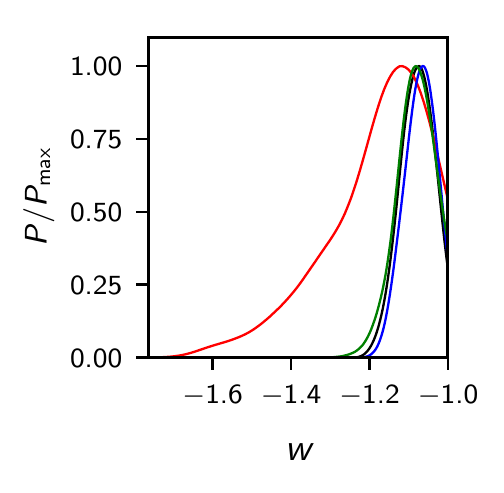}\\

\caption{As in Fig.~\ref{fig:wCDMP_bi}, for the parameters of the coupled phantom $\xi p$CDM cosmology described in Sec.~\ref{ssec:ide}, with the qualitatively similar JLA results summarized in Tables~\ref{tabxip} and~\ref{tab1xip}.}
\label{fig:couplingneg_bi}
\end{figure}

\begin{figure}
\centering
	\includegraphics[width=.7\textwidth]{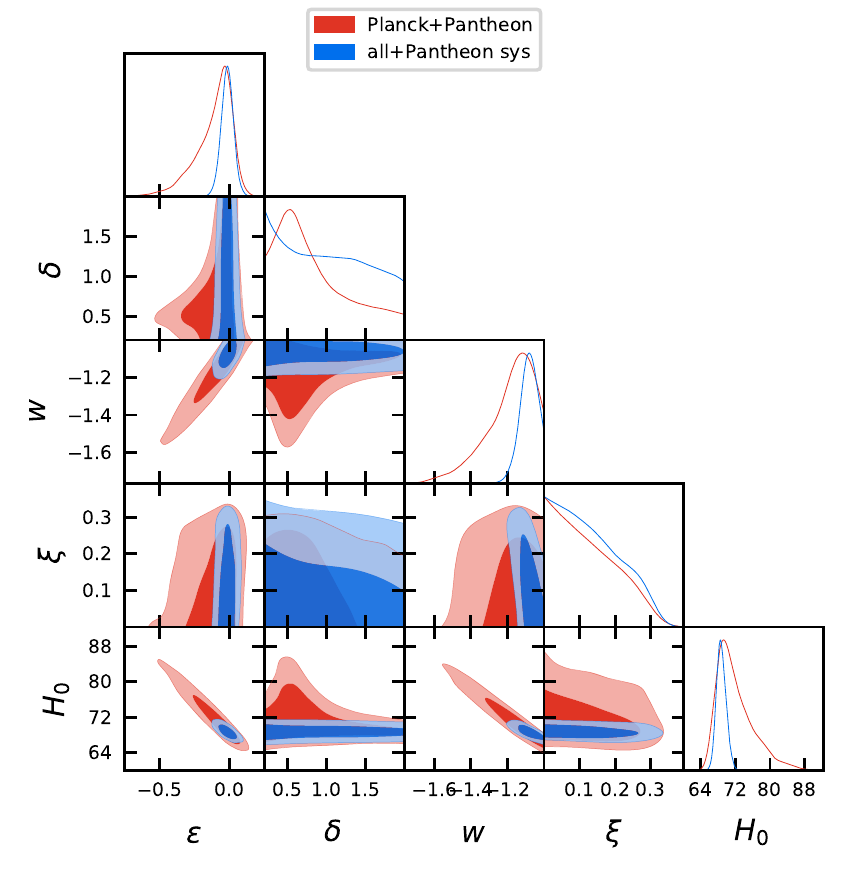}
	
	\caption{As in Fig.~\ref{fig:wCDMJ_tri}, for the parameters of the coupled phantom $\xi p$CDM cosmology described in Sec.~\ref{ssec:ide}, with the qualitatively similar JLA results not shown here.}
	\label{fig:couplingneg_tri}
\end{figure}

We now move on to the coupled phantom $\xi p$CDM model, where to avoid instabilities we consider $w<-1$ and $\xi>0$, with results shown in Figs.~\ref{fig:couplingneg_bi} and~\ref{fig:couplingneg_tri}, and Tabs.~\ref{tabxip} and~\ref{tab1xip}. For this model, we first of all notice that the upper limits on $\xi$ are almost unaffected by the SNeIa systematic we are considering, regardless of whether we are combining the SNeIa data with CMB temperature and polarization anisotropy data alone, or with the full dataset combination including CMB lensing, BAO, and Cosmic Chronometer measurements. However, we also see a strong correlation between $w$, $H_0$, and $\epsilon$. To some extent, the model is similar to the $w$CDM scenario previously analyzed: a larger negative value of $\epsilon$ will imply larger distance moduli, which can be compensated by values $w<-1$, in turn implying higher values of $H_0$, see Fig.~\ref{fig:couplingneg_tri}.

It is also worth noticing from Tabs.~\ref{tabxip} and~\ref{tab1xip} that the mean values of $\epsilon$ we infer are negative. For the \textit{Planck}+\textit{Pantheon} dataset combination we find $w= -1.081^{+0.050}_{-0.038}$ ($w=-1.19^{+0.18}_{-0.05}$) without (with) systematics, and similarly $w=-1.092^{+0.060}_{-0.044}$ ($w >-1.17$) without (with) systematics for the \textit{Planck}+\textit{JLA} dataset combination. This results in the uncertainties on $w$ being increased significantly by the introduction of this systematic (by a factor of four or more). Similar considerations hold for $H_0$, whose inferred value shifts to larger values, and the matter density $\Omega_m$, whose uncertainty is doubled once this systematic is introduced. Considering the full dataset combination, the preference for a larger value of $H_0$  completely disappears, although its uncertainty remains larger than in the no-systematics case (as is the case for $w$, but not for $\xi$ and $\Omega_m$).

\begin{center}                              
\begin{table*} 
\scalebox{1}{
\begin{tabular}{cccccccccccccccc}       
\hline\hline                                                                                                                    
Parameters & Planck   & Planck & Planck& Planck \\ 
 & +Pantheon &+Pantheon sys & +JLA  & +JLA sys \\ \hline

$\xi$ & $    <0.180$ &  $    <0.163$ & $  <0.187  $ & $ <0.190   $ \\

$w$ & $    -1.081^{+0.050}_{-0.038}$ &  $    -1.19^{+0.18}_{-0.05}$ & $ -1.092^{+0.060}_{-0.044}   $ & $ >-1.17   $ \\

$H_0 $[km/s/Mpc] & $   68.3^{+0.9}_{-1.1}$&  $   71.9^{+2.1}_{-5.1}$ & $ 68.6\pm1.4  $ & $ 70.5^{+1.6}_{-4.7}  $ \\

$\Omega_m$ & $    0.336^{+0.020}_{-0.025}$ &  $    0.305^{+0.042}_{-0.036}$ & $ 0.334^{+0.022}_{-0.027}   $ & $ 0.320^{+0.046}_{-0.037}   $ \\

$\alpha$ & $    - $ &  $    -$ & $  0.1415\pm0.0066  $ & $ 0.1415\pm0.0066   $ \\

$\beta$ & $    - $ &  $    -$ & $  3.111\pm0.082  $ & $ 3.112\pm0.081   $ \\

$\epsilon$ & $    - $ &  $    -0.11^{+0.15}_{-0.07}$ & $  -  $ & $  -0.06^{+0.16}_{-0.07}  $ \\

$\delta$ & $    - $ &  $    <0.990$ & $  -  $ & $  <1.21  $ \\

\hline\hline                                                  
\end{tabular} 
}
\caption{As in Table~\ref{tab1}, for the parameters of the coupled phantom $\xi p$CDM cosmology described in Sec.~\ref{ssec:ide}.}
\label{tabxip}                                              
\end{table*}                                                    
\end{center}

\begin{center}                              
\begin{table*} 
\scalebox{1}{
\begin{tabular}{cccccccccccccccc}       
\hline\hline                                                                                                                    
Parameters & all   & all & all& all \\ 
 & +Pantheon &+Pantheon sys & +JLA  & +JLA sys \\ 
  \hline

$\xi$ & $    <0.177$ &  $    <0.170$ & $ <0.186   $ & $  <0.191  $ \\

$w$ & $    -1.073^{+0.049}_{-0.033}$ &  $    -1.086^{+0.057}_{-0.039}$ & $ -1.080^{+0.054}_{-0.036}   $ & $  -1.076^{+0.067}_{-0.027}  $ \\

$H_0 $[km/s/Mpc] & $   68.36\pm0.77$&  $   68.7^{+1.0}_{-1.2}$ & $ 68.5^{+0.9}_{-1.0}  $ & $  68.4^{+1.0}_{-1.2} $ \\

$\Omega_m$ & $    0.334^{+0.018}_{-0.024}$ &  $    0.330^{+0.019}_{-0.026}$ & $ 0.334^{+0.020}_{-0.026}   $ & $  0.336^{+0.021}_{-0.027}  $ \\

$\alpha$ & $    - $ &  $    -$ & $  0.1415\pm0.0066  $ & $ 0.1415\pm0.0066   $ \\

$\beta$ & $    - $ &  $    -$ & $  3.108\pm0.081  $ & $ 3.109\pm0.081   $ \\

$\epsilon$ & $    - $ &  $    -0.018\pm0.045$ & $  -  $ & $ 0.012\pm0.062   $ \\

$\delta$ & $    - $ &  $    <1.33$ & $  -  $ & $ unconstrained   $ \\

\hline\hline                                                  
\end{tabular} 
}
\caption{As in Table~\ref{tab1b}, for the parameters of the coupled phantom $\xi p$CDM cosmology described in Sec.~\ref{ssec:ide}.}
\label{tab1xip}                                              
\end{table*}                                                    
\end{center}

\begin{figure}
\centering
\includegraphics[width=.6\textwidth]{legend_1d_Pantheon.pdf}\\
\includegraphics[width=.32\textwidth]{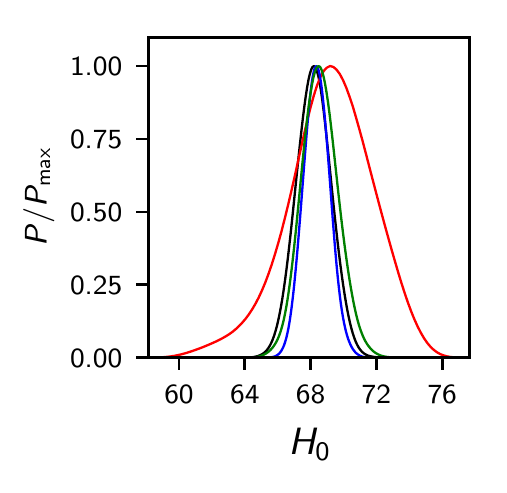}
\includegraphics[width=.32\textwidth]{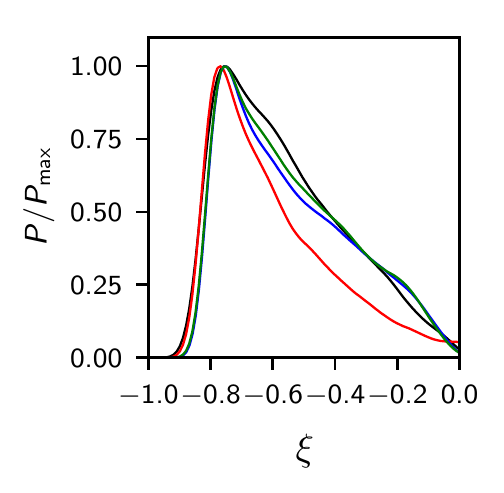}
\includegraphics[width=.32\textwidth]{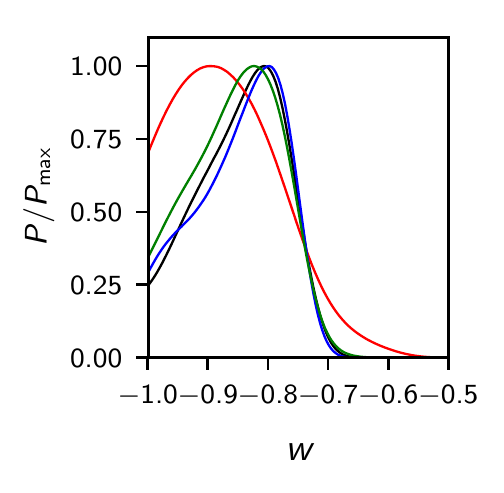}\\

\caption{As in Fig.~\ref{fig:wCDMP_bi}, for the parameters of the coupled quintessence $\xi q$CDM cosmology described in Sec.~\ref{ssec:ide}, with the qualitatively similar JLA results summarized in Tables~\ref{tabxiq} and~\ref{tab1xiq}.}
\label{fig:couplingpos_bi}
\end{figure}

\begin{figure}
\centering
	\includegraphics[width=.7\textwidth]{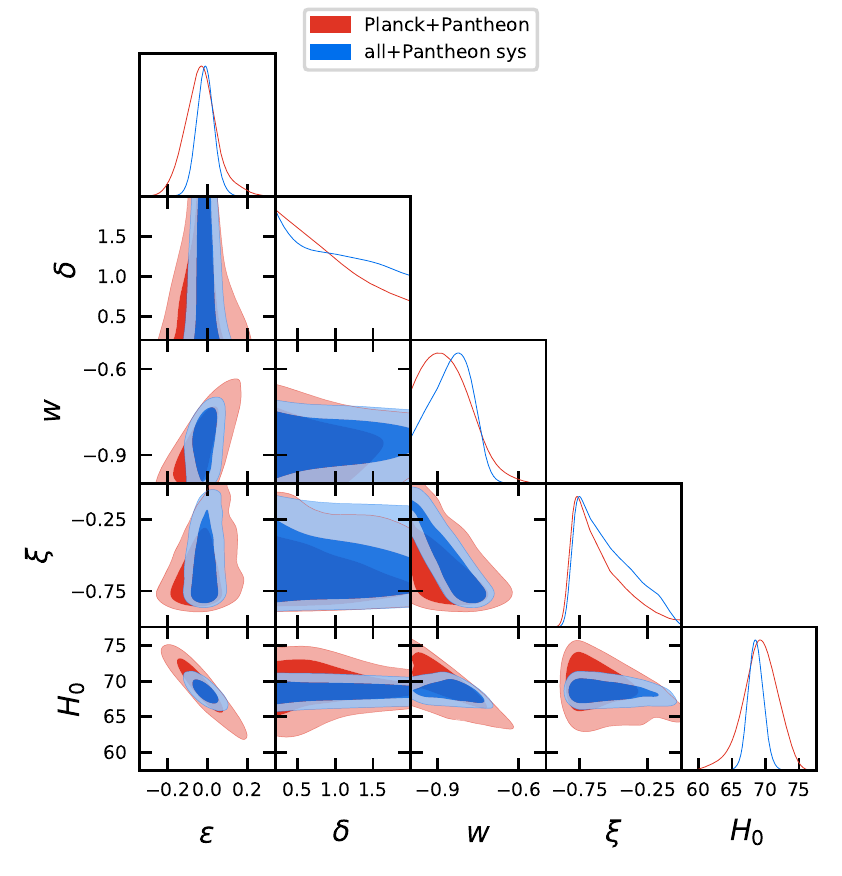}

	\caption{As in Fig.~\ref{fig:wCDMJ_tri}, for the parameters of the coupled quintessence $\xi q$CDM cosmology described in Sec.~\ref{ssec:ide}, with the qualitatively similar JLA results not shown here.}
	\label{fig:couplingpos_tri}
\end{figure}

Finally, we discuss the results obtained for the coupled quintessence $\xi q$CDM model, where the absence of instabilities requires $w>-1$ and $\xi<0$, with results shown in Figs.~\ref{fig:couplingpos_bi} and~\ref{fig:couplingpos_tri}, and Tabs.~\ref{tabxiq} and~\ref{tab1xiq}. Notice that, independently from the SNeIa sample used, the dataset combination considered, and whether or not a possible redshift-dependence in the intrinsic luminosities thereof is considered, we always recover a preference for a non-zero value of $\xi$. In fact, we find $\xi=-0.55^{+0.11}_{-0.27}$ ($\xi=-0.55^{+0.11}_{-0.27}$) for the \textit{all}+\textit{Pantheon} dataset combination without (with) SNeIa systematics, and $\xi=-0.54^{+0.12}_{-0.27}$ ($\xi=-0.55^{+0.11}_{-0.27}$) for the \textit{all}+\textit{JLA} dataset combination without (with) SNeIa systematics.

The correlations among the parameters are similar to those previously described: since the DE EoS is now required to be in the \textit{quintessence} region, \textit{i.e.}\ $w>-1$, this will imply a lower value of $H_0$, even in the presence of a redshift-dependence in the intrinsic SNeIa luminosities. However, in some cases, for instance when SNeIa data are combined with CMB temperature and polarization anisotropy measurements alone, the strong correlation between $H_0$ and $\epsilon$ (see Fig.~\ref{fig:couplingpos_tri}) shifts the mean value of $H_0$ towards larger quantities and/or increases the uncertainties on $H_0$ by a factor of two. 

As we already found for the $w$CDM and $w_0w_a$CDM models, also in the three IDE cases considered here it is worth noticing that the amplitudes of the stretch and color corrections [$\alpha$ and $\beta$ in Eq.~\eqref{eq:simple}] are basically unaffected by the introduction of the SNeIa systematic we are considering (see the JLA results in Tabs.~\ref{tabxi}, \ref{tab1xi}, \ref{tabxip}, \ref{tab1xip}, \ref{tabxiq}, and~\ref{tab1xiq}), thus validating the choice of using the pre-marginalized covariance matrix when analyzing the Pantheon sample and more in general the robustness of our analysis.

\begin{center}                              
\begin{table*} 
\scalebox{1}{
\begin{tabular}{cccccccccccccccc}       
\hline\hline                                                                                                                    
Parameters & Planck   & Planck & Planck& Planck \\ 
 & +Pantheon &+Pantheon sys & +JLA  & +JLA sys \\ \hline

$\xi$ & $    -0.57^{+0.11}_{-0.26}$ &  $    -0.60^{+0.08}_{-0.24}$ & $  -0.59^{+0.10}_{-0.24}  $ & $ -0.58^{+0.10}_{-0.26}  $ \\

$w$ & $    -0.843^{+0.087}_{-0.061}$ &  $    -0.86^{+0.05}_{-0.12}$ & $  -0.840^{+0.081}_{-0.073}  $ & $ -0.82^{+0.09}_{-0.12}   $ \\

$H_0 $[km/s/Mpc] & $   68.2\pm1.1$&  $   69.2^{+2.8}_{-2.3}$ & $ 68.4\pm1.5  $ & $ 67.8\pm3.0  $ \\

$\Omega_m$ & $    0.148^{+0.039}_{-0.091}$ &  $    0.133^{+0.027}_{-0.084}$ & $  0.138^{+0.034}_{-0.085}  $ & $ 0.146^{+0.034}_{-0.096}   $ \\

$\alpha$ & $    - $ &  $    -$ & $  0.1416\pm0.0066  $ & $ 0.1414\pm0.0066   $ \\

$\beta$ & $    - $ &  $    -$ & $  3.109\pm0.080  $ & $ 3.110\pm0.080   $ \\

$\epsilon$ & $    - $ &  $    -0.032^{+0.071}_{-0.084}$ & $  -  $ & $ 0.02^{+0.10}_{-0.09}   $ \\

$\delta$ & $    - $ &  $    <1.15$ & $  -  $ & $ unconstrained   $ \\

\hline\hline                                                  
\end{tabular} 
}
\caption{As in Table~\ref{tab1}, for the parameters of the coupled quintessence $\xi q$CDM cosmology described in Sec.~\ref{ssec:ide}.}
\label{tabxiq}                                              
\end{table*}                                                    
\end{center}

\begin{center}                              
\begin{table*} 
\scalebox{1}{
\begin{tabular}{cccccccccccccccc}       
\hline\hline                                                                                                                    
Parameters & all   & all & all& all \\ 
 & +Pantheon &+Pantheon sys & +JLA  & +JLA sys \\ 
  \hline

$\xi$ & $    -0.55^{+0.11}_{-0.27}$ &  $    -0.55^{+0.11}_{-0.27}$ & $  -0.54^{+0.12}_{-0.27}  $ & $  -0.55^{+0.11}_{-0.27}  $ \\

$w$ & $    -0.843^{+0.093}_{-0.055}$ &  $    -0.850^{+0.088}_{-0.067}$ & $ -0.848^{+0.088}_{-0.063}   $ & $ -0.840^{+0.091}_{-0.069}   $ \\

$H_0 $[km/s/Mpc] & $   68.36\pm0.81$&  $   68.6\pm1.1$ & $ 68.40\pm0.97  $ & $ 68.3^{+1.1}_{-1.2}  $ \\

$\Omega_m$ & $    0.149^{+0.040}_{-0.094}$ &  $    0.149^{+0.039}_{-0.093}$ & $ 0.153^{+0.044}_{-0.091}   $ & $  0.150^{+0.041}_{-0.092}  $ \\

$\alpha$ & $    - $ &  $    -$ & $  0.1415\pm0.0066  $ & $ 0.1415\pm0.0066   $ \\

$\beta$ & $    - $ &  $    -$ & $  3.107\pm0.081  $ & $ 3.111\pm0.081   $ \\

$\epsilon$ & $    - $ &  $    -0.012\pm0.045$ & $  -  $ & $ 0.016\pm0.062   $ \\

$\delta$ & $    - $ &  $    <1.33$ & $  -  $ & $ unconstrained   $ \\

\hline\hline                                                  
\end{tabular} 
}
\caption{As in Table~\ref{tab1b}, for the parameters of the coupled quintessence $\xi q$CDM cosmology described in Sec.~\ref{ssec:ide}.}
\label{tab1xiq}                                              
\end{table*}                                                    
\end{center}

\subsection{Modified Gravity Scenarios}
\label{sec:MG}

Modifications to General Relativity, which we shall refer to, generically, as modified gravity (MG), are a viable possibility for explaining cosmic acceleration. The phenomenology of MG theories is typically quite different from that of the DE models we have studied so far, in that MG models usually affect the growth of structure significantly. We cannot do justice to the broad literature on MG models sourcing cosmic acceleration: for an inevitably incomplete list, see for instance~\cite{Nojiri:2006ri,Amendola:2006we,Hu:2007nk,Cognola:2007zu,Saridakis:2009bv,Lim:2010yk,Deffayet:2010qz,Clifton:2011jh,Chamseddine:2013kea,Addazi:2016oob,Sebastiani:2016ras,Barvinsky:2017pmm,Renk:2017rzu,Dutta:2017fjw,Calza:2018ohl,Casalino:2018tcd,Saridakis:2018unr,Kase:2018aps,Odintsov:2019evb,Hogas:2019ywm}. In the following, we shall explore how the constraints on a particular model-agnostic parametrization of deviations from General Relativity are altered when accounting for a possible redshift-dependence in the intrinsic SNeIa luminosities.

A relatively simple and model-agnostic way of parametrizing effects due to MG is to parametrize its effects on the metric perturbations, governed by the two gravitational potentials $\Psi$ and $\Phi$: a non-vanishing anisotropic stress, proportional to $(\Psi-\Phi)$, is in fact a very common signature of several MG models. One commonly adopted parametrization relies on a set of functions usually denoted by $\mu(k,a)$, $\eta(k,a)$, and $\Sigma(k,a)$, depending on the wavenumber $k$ and the scale factor $a$. Below, we describe how these three functions modify the relevant equations for the metric perturbations.
\begin{itemize}
    \item $\mu(k,a)$ modifies the Poisson equation for the metric perturbation $\Psi$:
    \begin{equation}
    k^2 \Psi = -4 \pi a^2 G \mu(k,a)\rho\Delta\,,
    \label{mu}
    \end{equation}
    where $\rho$ and $\Delta$ are respectively the matter energy density and density contrast.
    \item $\Sigma(a,k)$ modifies the equation for the lensing potential:
    \begin{equation}
    -k^2 (\Psi +\Phi) = 8 \pi a^2 G \Sigma(k,a)\rho\Delta\,.
    \end{equation}
    \item Finally, $\eta(a,k)$ mimics the effect of an  anisotropic stress:
    \begin{equation}
    \eta(k,a)=\frac{\Phi}{\Psi}\,.
    \end{equation}
\end{itemize}
This phenomenological parametrization of the effects of MG has been used in a vast number of papers, most notably by the \textit{Planck} collaboration~\cite{Ade:2015rim,Aghanim:2018eyx}, see also e.g.~\cite{Bertschinger:2008zb,Martinelli:2010wn,Baker:2014zva} for other important works exploring this and similar parametrizations.

The three functions $\mu$, $\eta$, and $\Sigma$ are both time- and scale-dependent, and are mutually dependent. When $\mu=\eta=\Sigma=1$, one recovers General Relativity. Only two of these functions are needed to describe the effects of MG: any choice of two of these functions will fully parametrize the deviations of the perturbations from a smooth DE model. Here, we  parametrize $\mu$ and  $\eta$  as follows~\cite{Ade:2015rim,Aghanim:2018eyx,DiValentino:2015bja}:
\begin{eqnarray}
\mu(k,a)&= &1+f_1(a)\frac{1+c_1(\lambda H/k)^2}{1+(\lambda H /k)^2}\,,\\
\eta(k,a)&=&1+f_2(a)\frac{1+c_2(\lambda H/k)^2}{1+(\lambda H/k)^2}\,,
\end{eqnarray}
where $c_1$ and $c_2$ are constants, while the functions $f_i(a)$ describe the time-dependence of the deviations from General Relativity. On the other hand, $\Sigma$ can be recovered by $\Sigma \equiv (\mu/2)(1+\eta)$. Following earlier work~\cite{DiValentino:2015bja}, we assume a time-dependence for these functions proportional to the DE energy density $\Omega_x(a)$, motivated by the somewhat natural expectation that the modifications induced to the clustering and to the anisotropic stress in these extended gravitational theories should be proportional to their effective energy density (as is the case for the matter and radiation components). Therefore, $f_i(a)\equiv E_{ii} \Omega_x(a)$, where the coefficients $E_{ii}$ are constants. In other words, we are assuming that the background evolution is the same as in $\Lambda$CDM, with only the evolution of perturbations being affected by the MG parametrization in question. As the CMB measurements are not expected to be very sensitive to scale-dependent effects, we shall make the simplifying assumption $c_1=c_2=1$. Overall, we are therefore utilizing the following parametrization:
\begin{eqnarray}
\mu(k,a) &= &1+ E_{11} \Omega_x(a)\,,\\
\eta(k,a)&=&1+ E_{22} \Omega_x(a)\,.
\label{etaka}
\end{eqnarray}
Values of $E_{11}$ and $E_{22}$ other than zero imply deviations from General Relativity. It is worth clarifying that, within the MG parametrization we are considering, $E_{11}$ and $E_{22}$ are the free parameters, whereas $\mu$, $\eta$, and $\Sigma$ are the derived ones, see Tab.~\ref{tab:priors}.

\begin{figure}
\centering
\includegraphics[width=.6\textwidth]{legend_1d_Pantheon.pdf}\\
\includegraphics[width=.45\textwidth]{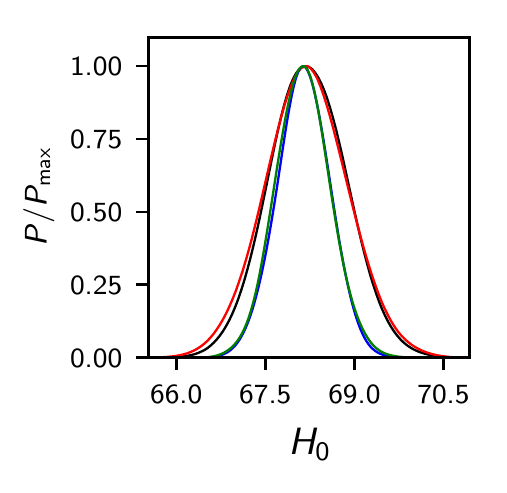}
\includegraphics[width=.45\textwidth]{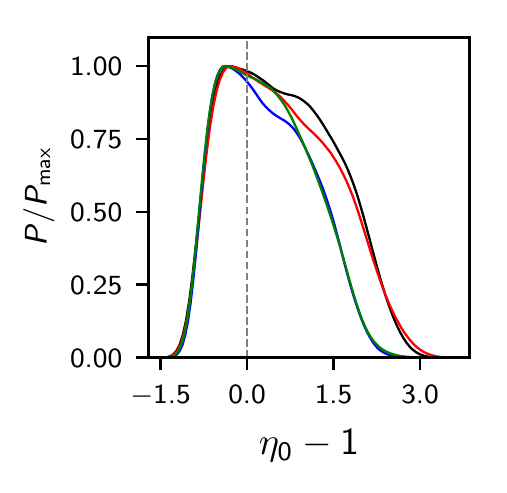}
\includegraphics[width=.45\textwidth]{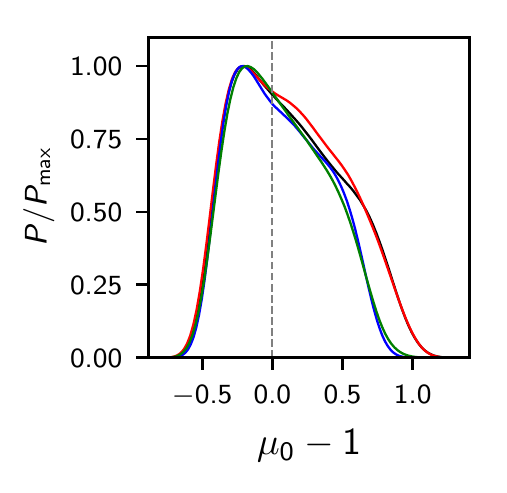}
\includegraphics[width=.45\textwidth]{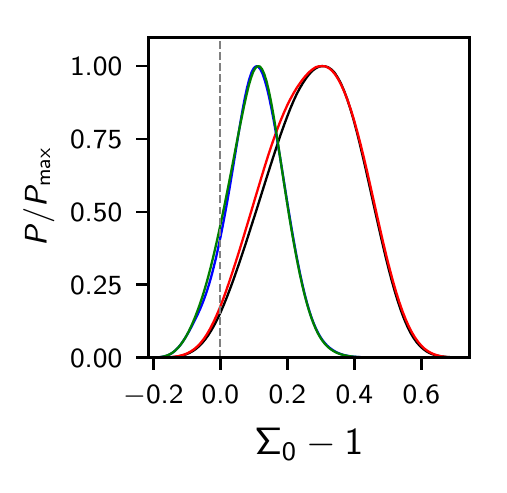}

\caption{As in Fig.~\ref{fig:wCDMP_bi}, for the parameters of the model-agnostic MG parametrization described in Sec.~\ref{sec:MG}, with the qualitatively similar JLA results summarized in Tables~\ref{tabmg} and~\ref{tab1mg}. Note that all three $\eta_0$, $\mu_0$, and $\Sigma_0$ are derived parameters, whereas the free parameters are $E_{11}$ and $E_{22}$.}
\label{mg}
\end{figure}

\begin{figure}
\centering
	\includegraphics[width=1.0\textwidth]{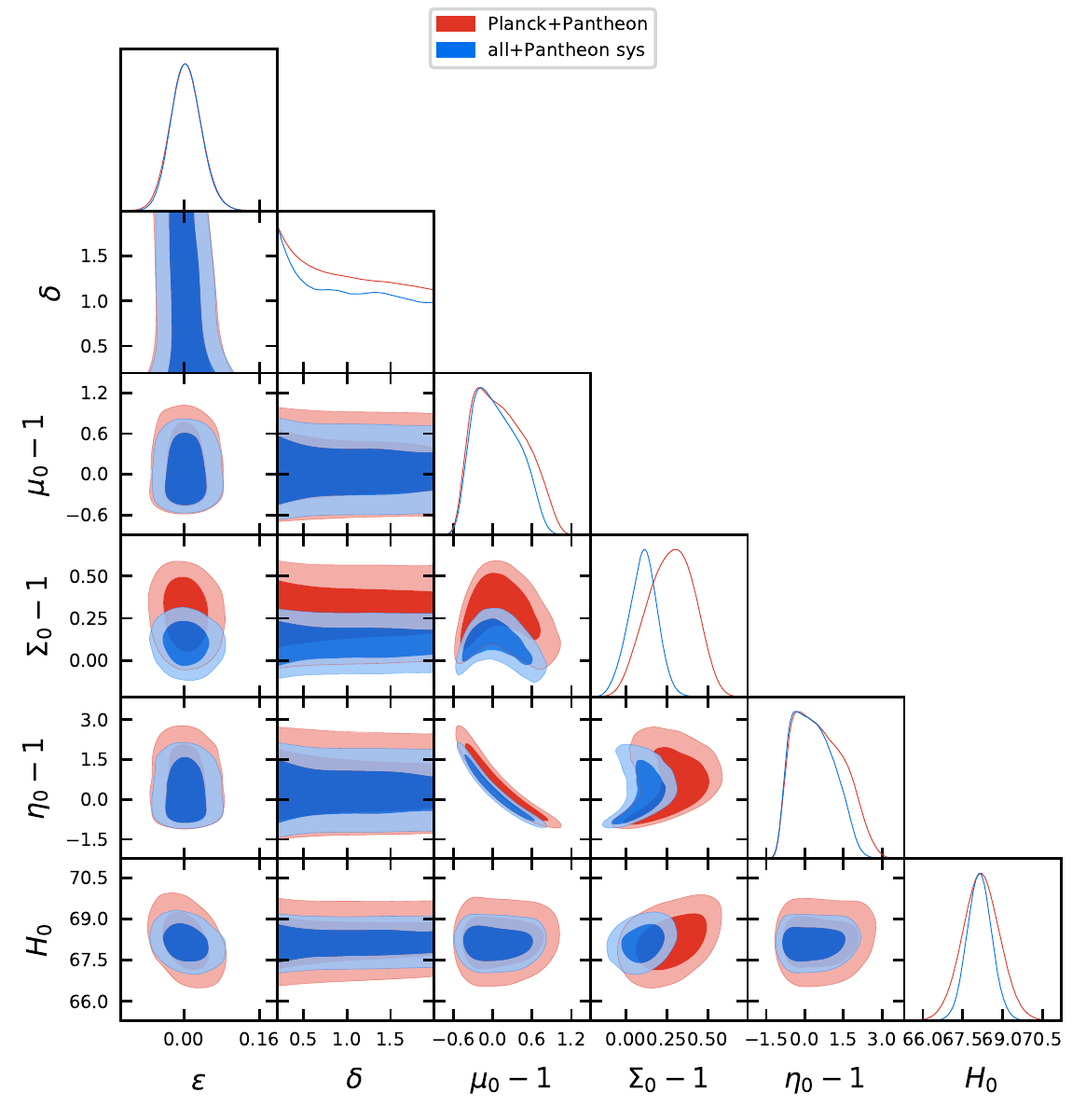}
	\caption{As in Fig.~\ref{fig:wCDMJ_tri}, for the model-agnostic MG parametrization described in Sec.~\ref{sec:MG}, with the qualitatively similar JLA results not shown here. Note that all three $\eta_0$, $\mu_0$, and $\Sigma_0$ are derived parameters, whereas the free parameters are $E_{11}$ and $E_{22}$.}
	\label{mg1}
\end{figure}

The results for the our model-agnostic parametrization of MG are shown in Figs.~\ref{mg} and~\ref{mg1}, and in Tabs.~\ref{tabmg} and~\ref{tab1mg}, where $\eta_0$, $\mu_0$, and $\Sigma_0$ denote the current values of these parameters, \textit{i.e.}\ at $a=1$. Notice that the MG parameters $\eta_0$ and $\mu_0$ are both perfectly consistent with their General Relativity expectations (\textit{i.e.}\ $\eta_0=\mu_0=1$) when only combining SNeIa data with CMB temperature and polarization anisotropy measurements. However, we find a $2\sigma$ preference for $\Sigma_0>1$. These results are qualitatively unaffected if we model a possible redshift-dependence of intrinsic SNeIa luminosities, both for the Pantheon and JLA samples. This is due to the lack of strong degeneracies between the MG parameters and the two parameters controlling SNeIa systematics, $\epsilon$ and $\delta$. Given our assumption that the background evolution is the same as in $\Lambda$CDM, another important ingredient towards reaching this conclusion is the lack of significant degeneracies between $\Omega_x$ and $\epsilon$.

In principle, one could argue that this result could have been expected. In fact, SNeIa are probing the background expansion rate, whereas our parametrization of MG alters the growth of structure. However, we believe one should not immediately jump to this conclusion. In fact, if $\epsilon$ and $\delta$ were strongly degenerate with parameters such as $\Omega_m$ or $H_0$, changes in the inferred values of the latter within a MG scenario could be transferred to the other parameters due to mutual degeneracies with $\epsilon$ and $\delta$, a possibility which cannot be logically excluded a priori. Our analysis, nonetheless, shows that this does not occur.

The deviations from General Relativity of the mean values of $\eta_0$, $\mu_0$, and $\Sigma_0$ are much larger (about twice as large) when combining SNeIa data with CMB temperature and polarization anisotropy measurements only. Once CMB lensing, BAO, and Cosmic Chronometer measurements are included in the analyses, the deviations of these mean values from their standard expectations are significantly reduced. However, the uncertainties on the MG parameters are mildly dependent on the dataset combinations used except for $\Sigma_0$, whose uncertainties are much smaller when the full dataset combination is taken into account, due to the impact of CMB lensing data in improving the constraints on this parameter, as previously noticed in~\cite{DiValentino:2015bja}.

\begin{center}                              
\begin{table*} 
\scalebox{1}{
\begin{tabular}{cccccccccccccccc}       
\hline\hline                                                                                                                    
Parameters & Planck & Planck & Planck& Planck \\ 
 & +Pantheon &+Pantheon sys & +JLA & +JLA sys \\ \hline

$\mu_0-1$ & $    0.12^{+0.28}_{-0.52}$ &  $    0.12^{+0.29}_{-0.51}$ & $ 0.12^{+0.29}_{-0.49}   $ & $ 0.12^{+0.30}_{-0.52}   $ \\

$\eta_0-1$ & $    0.6^{+0.7}_{-1.2}$ &  $    0.6^{+0.7}_{-1.2}$ & $  0.6^{+0.7}_{-1.2}  $ & $ 0.6^{+0.7}_{-1.3}   $ \\

$\Sigma_0-1$ & $    0.28^{+0.15}_{-0.13}$ &  $    0.28^{+0.15}_{-0.13}$ & $  0.28^{+0.15}_{-0.13}  $ & $ 0.27^{+0.15}_{-0.13}   $ \\

$H_0 $[km/s/Mpc] & $   68.22\pm0.66$&  $   68.19\pm0.70$ & $ 68.22\pm0.66  $ & $ 68.12\pm0.70  $ \\

$\Omega_m$ & $    0.3041\pm0.0086$ &  $    0.3044\pm0.0092$ & $ 0.3041\pm0.0087   $ & $  0.3054\pm0.0092  $ \\

$\alpha$ & $    - $ &  $    -$ & $  0.1412\pm0.0066  $ & $ 0.1416\pm0.0067   $ \\

$\beta$ & $    - $ &  $    -$ & $  3.104\pm0.080  $ & $ 3.110\pm0.081   $ \\

$\epsilon$ & $    - $ &  $    0.005\pm0.034$ & $  -  $ & $ 0.028\pm0.051   $ \\

$\delta$ & $    - $ &  $    <1.35$ & $  -  $ & $  unconstrained  $ \\

\hline\hline                                                  
\end{tabular} 
}
\caption{As in Table~\ref{tab1}, for the parameters of the  model-agnostic MG parametrization described in Sec.~\ref{sec:MG}. Note that all three $\eta_0$, $\mu_0$, and $\Sigma_0$ are derived parameters, whereas the free parameters are $E_{11}$ and $E_{22}$.}
\label{tabmg}                                              
\end{table*}                                                    
\end{center}

\begin{center}                              
\begin{table*} 
\scalebox{1}{
\begin{tabular}{cccccccccccccccc}       
\hline\hline                                                                                                                    
Parameters & all   & all & all& all \\ 
 & +Pantheon &+Pantheon sys & +JLA  & +JLA sys \\ 
  \hline

$\mu_0-1$ & $    0.06^{+0.26}_{-0.43}$ &  $    0.06^{+0.27}_{-0.42}$ & $ 0.06^{+0.27}_{-0.43}   $ & $ 0.06^{+0.26}_{-0.40}   $ \\

$\eta_0-1$ & $    0.3^{+0.6}_{-1.0}$ &  $    0.3^{+0.6}_{-1.0}$ & $  0.3^{+0.6}_{-1.0}  $ & $  0.29^{+0.58}_{-0.95}  $ \\

$\Sigma_0-1$ & $    0.106^{+0.089}_{-0.080}$ &  $    0.103^{+0.090}_{-0.082}$ & $ 0.105\pm0.088   $ & $  0.104\pm0.086  $ \\

$H_0 $[km/s/Mpc] & $   68.14\pm0.45$&  $   68.13\pm0.46$ & $ 68.15\pm0.47  $ & $  68.10\pm0.46 $ \\

$\Omega_m$ & $    0.3047\pm0.0059$ &  $    0.3048\pm0.0061$ & $  0.3046\pm0.0061  $ & $ 0.3052\pm0.0060   $ \\

$\alpha$ & $    - $ &  $    -$ & $  0.1413\pm0.0066  $ & $ 0.1415\pm0.0067   $ \\

$\beta$ & $    - $ &  $    -$ & $  3.104\pm0.080  $ & $ 3.111\pm0.081   $ \\

$\epsilon$ & $    - $ &  $    0.005^{+0.0030}_{0.034}$ & $  -  $ & $ 0.027\pm0.050   $ \\

$\delta$ & $    - $ &  $    <1.35$ & $  -  $ & $  unconstrained  $ \\

\hline\hline                                                  
\end{tabular} 
}
\caption{As in Table~\ref{tab1b}, for the parameters of the model-agnostic MG parametrization described in Sec.~\ref{sec:MG}. Note that all three $\eta_0$, $\mu_0$, and $\Sigma_0$ are derived parameters, whereas the free parameters are $E_{11}$ and $E_{22}$.}
\label{tab1mg}                                              
\end{table*}                                                    
\end{center}

\subsection{The lensing amplitude anomaly}
\label{sec:alens}

The hints we previously found for $\Sigma_0-1 \neq 0$ are inherently connected to the fact that the \textit{Planck} temperature anisotropy spectrum prefers a higher amount of lensing than that expected within the baseline $\Lambda$CDM model, once all other parameters are fixed~\cite{DiValentino:2015bja}. This preference, which is also partly responsible for \textit{Planck} temperature measurements favouring a closed Universe, is usually quantified through the purely phenomenological $A_L$ parameter (sometimes referred to as $A_{\rm lens}$), first introduced in~\cite{Calabrese:2008rt}, which rescales the lensing amplitude in the CMB power spectra. A value of $A_L \neq 1$ reflects the disagreement between the amplitude of lensing as inferred by the smoothing of the CMB temperature anisotropy power spectrum, and the amplitude of the CMB lensing power spectrum as reconstructed from the temperature four-point function using the same temperature maps. There is an indication for $A_L>1$ at the $\sim 2.8\sigma$ level from Planck CMB measurements~\cite{Aghanim:2018eyx}, which has motivated  plenty of discussions in the literature, see e.g.~\cite{Handley:2019tkm,DiValentino:2019qzk,Efstathiou:2020wem}. It is as of today yet unclear whether this preference is due to a statistical fluctuation, although a re-analysis of \textit{Planck} High Frequency maps with access to a larger sky fraction appears to support this interpretation~\cite{Efstathiou:2019mdh}.

Since the amplitude of lensing-induced smoothing parametrized by $A_L$ is expected to be degenerate with certain models of DE or MG~\cite{DiValentino:2015bja}, we extend the baseline $\Lambda$CDM model and add $A_L$ as a free parameter to test whether the preference for $A_L>1$ is affected by the possibility of a redshift-dependence of intrinsic SNeIa luminosities. We refer to the resulting model as the $\Lambda$CDM+$A_L$ model. We remark that, within this model, DE is described by a cosmological constant, whereas $A_L$ mimics possible effects due to MG.

\begin{figure}
\centering
\includegraphics[width=.6\textwidth]{legend_1d_Pantheon.pdf}\\
\includegraphics[width=.45\textwidth]{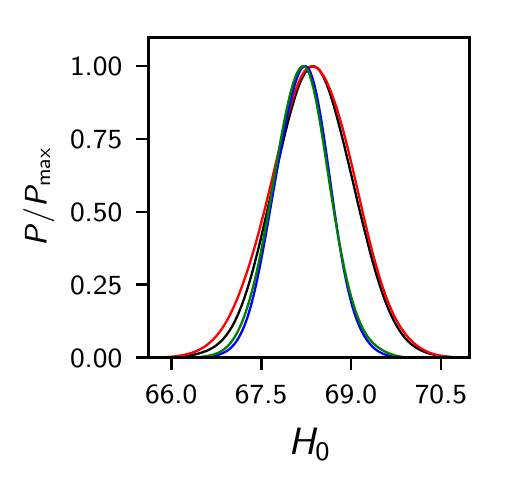}
\includegraphics[width=.45\textwidth]{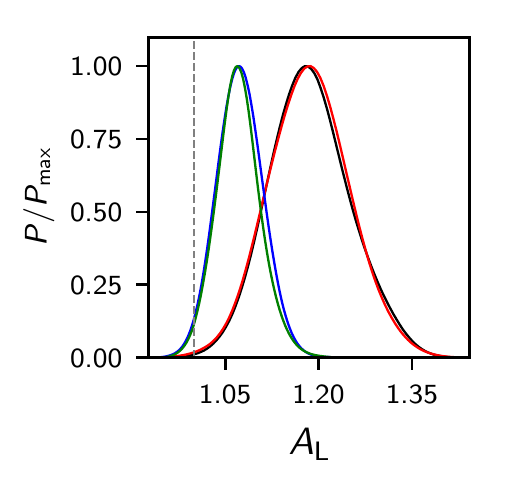}\\

\caption{As in Fig.~\ref{fig:wCDMP_bi}, for the $\Lambda$CDM+$A_L$ model described in Sec.~\ref{sec:alens}, with the qualitatively similar JLA results summarized in Tables~\ref{tabalens} and~\ref{tab1alens}.}
\label{Alens}
\end{figure}

\begin{figure}
\centering
	\includegraphics[width=.7\textwidth]{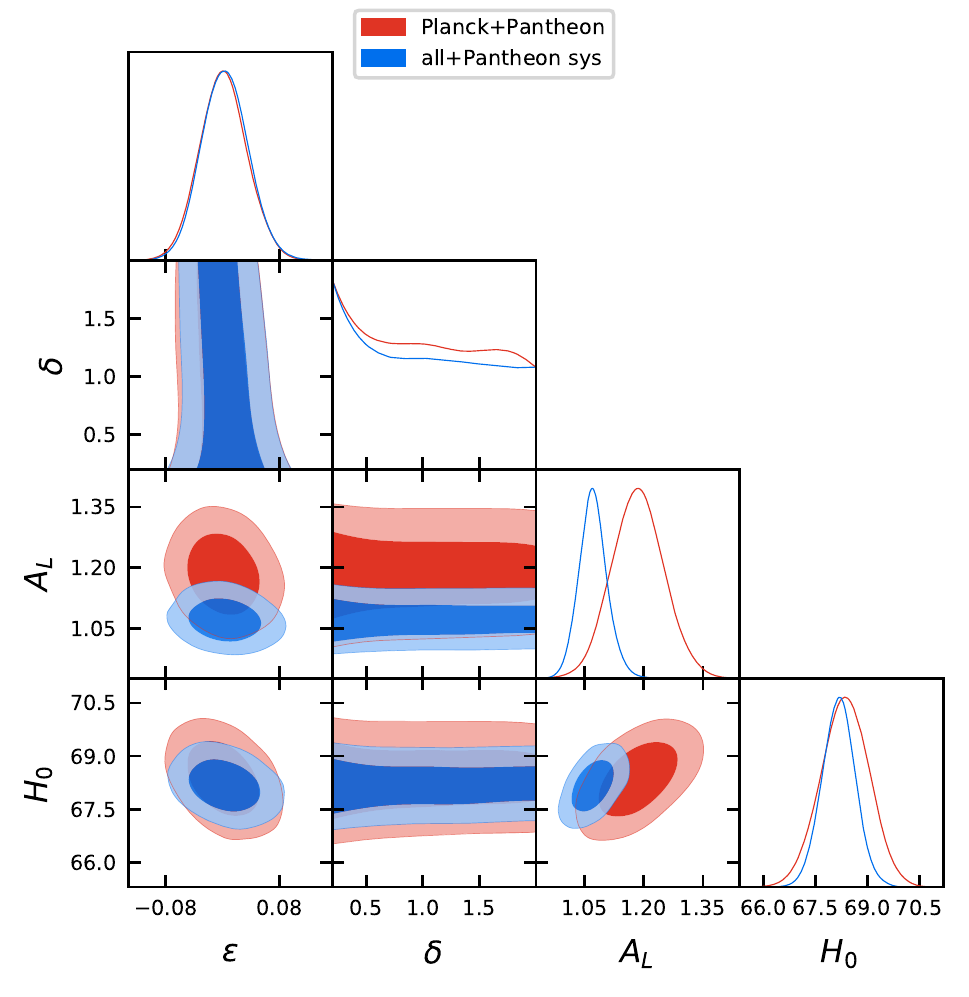}
	\caption{As in Fig.~\ref{fig:wCDMJ_tri}, for the $\Lambda$CDM+$A_L$ model described in Sec.~\ref{sec:alens}, with the qualitatively similar JLA results not shown here.}
	\label{Alens1}
\end{figure}

Our findings are summarized in Figs.~\ref{Alens} and~\ref{Alens1}, and in Tabs.~\ref{tabalens} and~\ref{tab1alens}. From both the tables and the figures we see that the preference for $A_L>1$ remains at a $\gtrsim 2\sigma$ significance level for all the cases we examine, regardless of whether or not the SNeIa systematics are considered. As for the MG case we studied earlier, in principle one could argue that this result could have been expected, given that SNeIa are probing the background expansion rate. However, if $\epsilon$ and $\delta$ were strongly degenerate with parameters such as $\Omega_m$ or $H_0$, changes in the inferred values of the latter within a MG scenario could be transferred to the other parameters due to mutual degeneracies with $\epsilon$ and $\delta$. This possibility cannot be excluded a priori.

In fact, a careful inspection of Fig.~\ref{Alens1} shows a correlation between $\epsilon$, controlling the amplitude of SNeIa systematics, and $A_L$. This correlation is only present for the case when SNeIa data is combined with CMB temperature and polarization anisotropy data, and not when the full dataset combination is considered. The reason why this degeneracy exists in first place is that $\epsilon$ and $H_0$ are negatively correlated, whereas $A_L$ and $H_0$ are positively correlated. Therefore, a mutual mild anti-correlation is inherited in the ($\epsilon$,$A_L$) plane. Nonetheless, the result of this degeneracy is not that of shifting the value of $A_L$, but just to slightly broaden its uncertainty with respect to the no-systematics case, by about 10\% for the \textit{Planck}+\textit{Pantheon} dataset combination, and by about 3\% for the \textit{Planck}+\textit{JLA} dataset combination. As we explained earlier, when one considers the full \textit{all}+\textit{Pantheon} or \textit{all}+\textit{JLA} dataset combinations, these already mild effects disappear completely.

As already noticed in~\cite{DiValentino:2015bja}, the hints for $\Sigma_0-1>0$ we have found in the previous section are directly connected to the lensing anomaly as characterized by $A_L>1$. These two effects are unmodified by the possible SNeIa systematic we consider. Broadly speaking, an important signature of several MG models is to modify (usually enhance) the lensing amplitude in the CMB power spectra. The very same effect can be obtained by increasing $A_L$, and therefore an indication for $\Sigma_0-1>0$ is generally expected to translate into an indication for $A_L>1$.

\begin{center}                              
\begin{table*} 
\scalebox{1}{
\begin{tabular}{cccccccccccccccc}       
\hline\hline                                                                                                                    
Parameters & Planck   & Planck & Planck& Planck \\ 
 & +Pantheon &+Pantheon sys & +JLA  & +JLA sys \\ \hline

$A_L$ & $    1.186^{+0.060}_{-0.071}$ &  $    1.185\pm 0.066$ & $ 1.186\pm 0.066   $ & $ 1.181\pm0.068   $ \\

$H_0 $[km/s/Mpc] & $   68.38\pm0.65$&  $   68.36\pm0.70$ & $ 68.36\pm0.69  $ & $ 68.31\pm0.71  $ \\

$\Omega_m$ & $    0.3020\pm0.0085$ &  $    0.3024^{+0.0084}_{-0.0095}$ & $ 0.3023\pm0.0090   $ & $  0.3030\pm0.0093  $ \\

$\alpha$ & $    - $ &  $    -$ & $  0.1413\pm0.0066  $ & $ 0.1416\pm0.0066   $ \\

$\beta$ & $    - $ &  $    -$ & $  3.104\pm0.080  $ & $ 3.108\pm0.081   $ \\

$\epsilon$ & $    - $ &  $    0.001\pm0.034$ & $  -  $ & $ 0.025\pm0.051   $ \\

$\delta$ & $    - $ &  $    <1.37$ & $  -  $ & $  unconstrained  $ \\

\hline\hline                                                  
\end{tabular} 
}
\caption{As in Table~\ref{tab1}, for the the $\Lambda$CDM+$A_L$ model described in Sec.~\ref{sec:alens}.}
\label{tabalens}                                              
\end{table*}                                                    
\end{center}

\begin{center}                              
\begin{table*} 
\scalebox{1}{
\begin{tabular}{cccccccccccccccc}       
\hline\hline                                                                                            Parameters & all   & all & all& all \\ 
 & +Pantheon &+Pantheon sys & +JLA  & +JLA sys \\ 
  \hline

$A_L$ & $    1.074\pm 0.037$ &  $    1.072\pm 0.036$ & $ 1.072\pm 0.036   $ & $ 1.072\pm0.037   $ \\

$H_0 $[km/s/Mpc] & $   68.19\pm0.46$&  $   68.17\pm0.48$ & $ 68.17\pm0.48  $ & $  68.16\pm0.48 $ \\

$\Omega_m$ & $    0.3041\pm0.0060$ &  $    0.3043\pm0.0063$ & $  0.3043\pm0.0062  $ & $ 0.3045\pm0.0063   $ \\

$\alpha$ & $    - $ &  $    -$ & $  0.1413\pm0.0066  $ & $ 0.1415\pm0.0066   $ \\

$\beta$ & $    - $ &  $    -$ & $  3.103\pm0.080  $ & $ 3.108\pm0.081   $ \\

$\epsilon$ & $    - $ &  $    0.004\pm 0.033$ & $  -  $ & $ 0.027\pm0.050   $ \\

$\delta$ & $    - $ &  $    <1.36$ & $  -  $ & $  unconstrained  $ \\

\hline\hline                                                  
\end{tabular} 
}
\caption{As in Table~\ref{tab1b}, for the $\Lambda$CDM+$A_L$ model described in Sec.~\ref{sec:alens}.}
\label{tab1alens}                                              
\end{table*}                                                    
\end{center}

\section{Conclusions}
\label{sec:conclusions}

Type Ia Supernovae (SNeIa) used as distance indicators were instrumental to the discovery of cosmic acceleration in 1998. Cosmic acceleration is usually attributed to a form of dark energy (DE) whose energy density constitutes 70\% of the Universe's energy budget. The simplest DE candidate in the form of a cosmological constant, albeit fully consistent with data, is theoretically problematic. As a consequence, several alternative models have been proposed, making use of new particles/fields or modifications to Einstein's General Relativity on ultra-large scales as in modified gravity (MG) models. The use of SNeIa as distance indicators rests upon the possibility of using them as standardizable candles, with the standardization procedure being independent of host galaxy redshifts and environments. In other words, two different SNeIa found in different hosts, with the same colour, light-curve stretch, and host stellar mass, should on average have the same intrinsic luminosity regardless of their redshift.

However, in recent years, several studies have established correlations between host galaxy properties and standardized SNeIa luminosities~\cite{Gallagher:2008zi,Kelly:2009iy,Uddin:2017rmc}, raising the question of whether or not intrinsic SNeIa luminosities might evolve with redshift~\cite{Kim:2019npy,Kang:2019azh,Rose:2020shp}. While the evidence for cosmic acceleration is robust to the possibility of a redshift-dependence of intrinsic SNeIa luminosities~\cite{Rose:2020shp}, the question remains of whether the DE properties one infers from SNeIa data are robust to this possible systematic. This work has therefore been devoted to testing the \textit{soundness} of inferred properties/limits on dark energy and MG models, obtained using SNeIa data, and relaxing the assumption of intrinsic SNeIa luminosities being redshift-independent. We have adopted a simple phenomenological power-law parametrization to account for this effect in the modelling of observed SNeIa distance moduli as in Eqs.~\eqref{eq:comp}, \eqref{eq:comp1}.

Using two different SNeIa compilations, the \textit{Pantheon} and \textit{JLA} samples, we have then considered various DE and MG models and examined whether the inferred properties thereof are robust to the inclusion of the systematic in question. We have considered models where the DE equation of state is a free parameter, either constant or time-varying, as well as models where DE and dark matter interact, and finally a model-agnostic parametrization of MG models at the level of the equations governing the evolution of metric perturbations. We remark that besides astrophysical complications, non-standard DE/MG models (for instance leading to time-varying fundamental constant, e.g.~\cite{Brans:1961sx,Carroll:1998zi,Sandvik:2001rv,Damour:2002mi,Calabrese:2013lga,Nunes:2016plz,Burrage:2016bwy,Wright:2017rsu,Vagnozzi:2019kvw,Jimenez:2020bgw,Jimenez:2020ysu}) can also lead to redshift-dependent intrinsic SNeIa luminosities, and hence including the possibility of this effect when studying models of DE beyond the cosmological constant might simply be a matter of theoretical consistency (see for instance the discussion in~\cite{Zumalacarregui:2020cjh}).

Our main findings can be summarized as follow. The main effect of the SNeIa systematic we have considered has been that of broadening the uncertainties on the inferred DE parameters. This has been true for all the DE models we have tested, with the exception of the model-agnostic MG parametrization, whose inferred properties are instead extremely robust to this systematic. When only combining SNeIa data with CMB temperature and polarization anisotropy measurements from the \textit{Planck} satellite, the effect of SNeIa systematics is more noticeable: besides in some cases broadening the uncertainties by up to a factor of $4$, including the systematics has been found to lead to ${\cal O}(\sigma)$ shifts in the central values of the inferred parameters, due to degeneracies between DE parameters and parameters characterizing the SNeIa systematics. However, robust conclusions regarding models of cosmic acceleration should rely on the analysis of additional measurements of the late-time expansion history such as BAO measurements. When considering a full dataset combination which also includes CMB lensing, BAO, and Cosmic Chronometer measurements, we find that the aforementioned shifts in the central values are considerably reduced. The only remaining effect of the SNeIa systematic we consider is then a small, typically $\lesssim 40\%$, broadening of the uncertainties on the DE parameters.

Our main conclusion, therefore, is that when a compilation of robust cosmological measurements is studied in combination with SNeIa data, the corresponding inferred dark energy parameters (for the dark energy models we studied) are robust to the possibility of intrinsic SNeIa luminosities being redshift-dependent, with the main effect of the latter being that of broadening the parameter uncertainties. In other words, we confirm the \textit{soundness} of dark energy properties. Moving forward, it will be important to settle the ongoing debate around whether or not intrinsic SNeIa luminosities evolve with redshift. Upcoming surveys such as the Vera Rubin Observatory~\cite{Ivezic:2008fe} will help addressing this question by detecting millions of SNeIa. In addition, it would definitely be worth exploring more physical parametrizations of observed SNeIa distance moduli which can better account for post-standardization residuals as well as correlations of the latter with host galaxy properties. An example is the recent two-component color model recently proposed by~\cite{Brout:2020msh}. We leave the exploration of these and related issues to future work.

\acknowledgments
S.V.\ thanks Subir Sarkar for important remarks. E.D.V.\ acknowledges support from the European Research Council in the form of a Consolidator Grant with number 681431. S.G.\ acknowledges financial support by the ``Juan de la Cierva-Incorporaci\'on'' program (IJC2018-036458-I) of the Spanish MINECO, and from the Spanish grants FPA2017-85216-P (AEI/FEDER, UE), PROMETEO/2018/165 (Generalitat Valenciana) and the Red Consolider MultiDark FPA2017-90566-REDC. O.M.\ is supported by the Spanish grants FPA2017-85985-P, PROMETEO/2019/083 and by the European Union Horizon 2020 research and innovation program (grant agreements No. 690575 and 67489) S.V.\ acknowledges support from the Isaac Newton Trust and the Kavli Foundation through a Newton-Kavli Fellowship, and acknowledges a College Research Associateship at Homerton College, University of Cambridge. This work makes use of the publicly available \texttt{CosmoMC}~\cite{Lewis:2002ah} and \texttt{CAMB}~\cite{Lewis:1999bs} codes and of the Planck data release 2018 Likelihood Code.

\bibliography{bibliography}

\providecommand{\href}[2]{#2}\begingroup\raggedright\begin{thebibliography}{100}

\bibitem{Alam:2016hwk}
{\bf BOSS} Collaboration, S.~Alam et~al., {\it {The clustering of galaxies in
  the completed SDSS-III Baryon Oscillation Spectroscopic Survey: cosmological
  analysis of the DR12 galaxy sample}},  {\em Mon.\ Not.\ Roy.\ Astron.\ Soc.}
  {\bf 470} (2017), no.~3 2617--2652,
  [\href{http://arxiv.org/abs/1607.03155}{{\tt arXiv:1607.03155}}].

\bibitem{Scolnic:2017caz}
D.~M. Scolnic et~al., {\it {The Complete Light-curve Sample of
  Spectroscopically Confirmed SNe Ia from Pan-STARRS1 and Cosmological
  Constraints from the Combined Pantheon Sample}},  {\em Astrophys. J.} {\bf
  859} (2018), no.~2 101, [\href{http://arxiv.org/abs/1710.00845}{{\tt
  arXiv:1710.00845}}].

\bibitem{Aghanim:2018eyx}
{\bf Planck} Collaboration, N.~Aghanim et~al., {\it {Planck 2018 results. VI.
  Cosmological parameters}},  \href{http://arxiv.org/abs/1807.06209}{{\tt
  arXiv:1807.06209}}.

\bibitem{Sahni:2004ai}
V.~Sahni, {\it {Dark matter and dark energy}},  {\em Lect. Notes Phys.} {\bf
  653} (2004) 141--180, [\href{http://arxiv.org/abs/astro-ph/0403324}{{\tt
  astro-ph/0403324}}].

\bibitem{Wetterich:1987fm}
C.~Wetterich, {\it {Cosmology and the Fate of Dilatation Symmetry}},  {\em
  Nucl. Phys.} {\bf B302} (1988) 668--696,
  [\href{http://arxiv.org/abs/1711.03844}{{\tt arXiv:1711.03844}}].

\bibitem{Ratra:1987rm}
B.~Ratra and P.~J.~E. Peebles, {\it {Cosmological Consequences of a Rolling
  Homogeneous Scalar Field}},  {\em Phys. Rev.} {\bf D37} (1988) 3406.

\bibitem{Caldwell:1997ii}
R.~R. Caldwell, R.~Dave, and P.~J. Steinhardt, {\it {Cosmological imprint of an
  energy component with general equation of state}},  {\em Phys. Rev. Lett.}
  {\bf 80} (1998) 1582--1585,
  [\href{http://arxiv.org/abs/astro-ph/9708069}{{\tt astro-ph/9708069}}].

\bibitem{Peebles:1998qn}
P.~J.~E. Peebles and A.~Vilenkin, {\it {Quintessential inflation}},  {\em Phys.
  Rev.} {\bf D59} (1999) 063505,
  [\href{http://arxiv.org/abs/astro-ph/9810509}{{\tt astro-ph/9810509}}].

\bibitem{Kamenshchik:2001cp}
A.~{\relax Yu}. Kamenshchik, U.~Moschella, and V.~Pasquier, {\it {An
  Alternative to quintessence}},  {\em Phys. Lett.} {\bf B511} (2001) 265--268,
  [\href{http://arxiv.org/abs/gr-qc/0103004}{{\tt gr-qc/0103004}}].

\bibitem{Bento:2002ps}
M.~C. Bento, O.~Bertolami, and A.~A. Sen, {\it {Generalized Chaplygin gas,
  accelerated expansion and dark energy matter unification}},  {\em Phys. Rev.}
  {\bf D66} (2002) 043507, [\href{http://arxiv.org/abs/gr-qc/0202064}{{\tt
  gr-qc/0202064}}].

\bibitem{Freese:2002sq}
K.~Freese and M.~Lewis, {\it {Cardassian expansion: A Model in which the
  universe is flat, matter dominated, and accelerating}},  {\em Phys. Lett.}
  {\bf B540} (2002) 1--8, [\href{http://arxiv.org/abs/astro-ph/0201229}{{\tt
  astro-ph/0201229}}].

\bibitem{Li:2004rb}
M.~Li, {\it {A Model of holographic dark energy}},  {\em Phys. Lett.} {\bf
  B603} (2004) 1, [\href{http://arxiv.org/abs/hep-th/0403127}{{\tt
  hep-th/0403127}}].

\bibitem{Barbieri:2005gj}
R.~Barbieri, L.~J. Hall, S.~J. Oliver, and A.~Strumia, {\it {Dark energy and
  right-handed neutrinos}},  {\em Phys. Lett.} {\bf B625} (2005) 189--195,
  [\href{http://arxiv.org/abs/hep-ph/0505124}{{\tt hep-ph/0505124}}].

\bibitem{Cicoli:2012tz}
M.~Cicoli, F.~G. Pedro, and G.~Tasinato, {\it {Natural Quintessence in String
  Theory}},  {\em JCAP} {\bf 1207} (2012) 044,
  [\href{http://arxiv.org/abs/1203.6655}{{\tt arXiv:1203.6655}}].

\bibitem{Rinaldi:2014yta}
M.~Rinaldi, {\it {Higgs Dark Energy}},  {\em Class. Quant. Grav.} {\bf 32}
  (2015) 045002, [\href{http://arxiv.org/abs/1404.0532}{{\tt
  arXiv:1404.0532}}].

\bibitem{Hlozek:2014lca}
R.~Hlozek, D.~Grin, D.~J.~E. Marsh, and P.~G. Ferreira, {\it {A search for
  ultralight axions using precision cosmological data}},  {\em Phys. Rev.} {\bf
  D91} (2015), no.~10 103512, [\href{http://arxiv.org/abs/1410.2896}{{\tt
  arXiv:1410.2896}}].

\bibitem{Rinaldi:2015iza}
M.~Rinaldi, {\it {Dark energy as a fixed point of the Einstein Yang-Mills Higgs
  Equations}},  {\em JCAP} {\bf 1510} (2015) 023,
  [\href{http://arxiv.org/abs/1508.04576}{{\tt arXiv:1508.04576}}].

\bibitem{Nunes:2016aup}
R.~C. Nunes and S.~Pan, {\it {Cosmological consequences of an adiabatic matter
  creation process}},  {\em Mon. Not. Roy. Astron. Soc.} {\bf 459} (2016),
  no.~1 673--682, [\href{http://arxiv.org/abs/1603.02573}{{\tt
  arXiv:1603.02573}}].

\bibitem{Sola:2016ecz}
J.~Solà~Peracaula, J.~de~Cruz~Pérez, and A.~Gómez-Valent, {\it {Dynamical
  dark energy vs. $\Lambda$ = const in light of observations}},  {\em EPL} {\bf
  121} (2018), no.~3 39001, [\href{http://arxiv.org/abs/1606.00450}{{\tt
  arXiv:1606.00450}}].

\bibitem{Capozziello:2017buj}
S.~Capozziello, R.~D'Agostino, and O.~Luongo, {\it {Cosmic acceleration from a
  single fluid description}},  {\em Phys. Dark Univ.} {\bf 20} (2018) 1--12,
  [\href{http://arxiv.org/abs/1712.04317}{{\tt arXiv:1712.04317}}].

\bibitem{Benisty:2018qed}
D.~Benisty and E.~I. Guendelman, {\it {Unified dark energy and dark matter from
  dynamical spacetime}},  {\em Phys. Rev.} {\bf D98} (2018), no.~2 023506,
  [\href{http://arxiv.org/abs/1802.07981}{{\tt arXiv:1802.07981}}].

\bibitem{Visinelli:2018utg}
L.~Visinelli and S.~Vagnozzi, {\it {Cosmological window onto the string
  axiverse and the supersymmetry breaking scale}},  {\em Phys. Rev.} {\bf D99}
  (2019), no.~6 063517, [\href{http://arxiv.org/abs/1809.06382}{{\tt
  arXiv:1809.06382}}].

\bibitem{Benisty:2018oyy}
D.~Benisty, E.~Guendelman, and Z.~Haba, {\it {Unification of dark energy and
  dark matter from diffusive cosmology}},  {\em Phys. Rev.} {\bf D99} (2019),
  no.~12 123521, [\href{http://arxiv.org/abs/1812.06151}{{\tt
  arXiv:1812.06151}}]. [Erratum: Phys. Rev.D101,no.4,049901(2020)].

\bibitem{DAgostino:2019wko}
R.~D'Agostino, {\it {Holographic dark energy from nonadditive entropy:
  cosmological perturbations and observational constraints}},  {\em Phys. Rev.
  D} {\bf 99} (2019), no.~10 103524,
  [\href{http://arxiv.org/abs/1903.03836}{{\tt arXiv:1903.03836}}].

\bibitem{Heckman:2019dsj}
J.~J. Heckman, C.~Lawrie, L.~Lin, J.~Sakstein, and G.~Zoccarato, {\it
  {Pixelated Dark Energy}},  {\em Fortsch. Phys.} {\bf 67} (2019), no.~11
  1900071, [\href{http://arxiv.org/abs/1901.10489}{{\tt arXiv:1901.10489}}].

\bibitem{Riess:1998cb}
{\bf Supernova Search Team} Collaboration, A.~G. Riess et~al., {\it
  {Observational evidence from supernovae for an accelerating universe and a
  cosmological constant}},  {\em Astron. J.} {\bf 116} (1998) 1009--1038,
  [\href{http://arxiv.org/abs/astro-ph/9805201}{{\tt astro-ph/9805201}}].

\bibitem{Perlmutter:1998np}
{\bf Supernova Cosmology Project} Collaboration, S.~Perlmutter et~al., {\it
  {Measurements of $\Omega$ and $\Lambda$ from 42 high redshift supernovae}},
  {\em Astrophys. J.} {\bf 517} (1999) 565--586,
  [\href{http://arxiv.org/abs/astro-ph/9812133}{{\tt astro-ph/9812133}}].

\bibitem{Tripp:1997wt}
R.~Tripp, {\it {A Two-parameter luminosity correction for type Ia supernovae}},
   {\em Astron. Astrophys.} {\bf 331} (1998) 815--820.

\bibitem{Gallagher:2008zi}
J.~S. Gallagher, P.~M. Garnavich, N.~Caldwell, R.~P. Kirshner, S.~W. Jha,
  W.~Li, M.~Ganeshalingam, and A.~V. Filippenko, {\it {Supernovae in Early-Type
  Galaxies: Directly Connecting Age and Metallicity with Type Ia Luminosity}},
  {\em Astrophys. J.} {\bf 685} (2008) 752--766,
  [\href{http://arxiv.org/abs/0805.4360}{{\tt arXiv:0805.4360}}].

\bibitem{Kelly:2009iy}
P.~L. Kelly, M.~Hicken, D.~L. Burke, K.~S. Mandel, and R.~P. Kirshner, {\it
  {Hubble Residuals of Nearby Type Ia Supernovae Are Correlated with Host
  Galaxy Masses}},  {\em Astrophys. J.} {\bf 715} (2010) 743--756,
  [\href{http://arxiv.org/abs/0912.0929}{{\tt arXiv:0912.0929}}].

\bibitem{Uddin:2017rmc}
S.~A. Uddin, J.~Mould, C.~Lidman, V.~Ruhlmann-Kleider, and B.~R. Zhang, {\it
  {The influence of Host Galaxies in Type Ia Supernova Cosmology}},  {\em
  Astrophys. J.} {\bf 848} (2017), no.~1 56,
  [\href{http://arxiv.org/abs/1709.05830}{{\tt arXiv:1709.05830}}].

\bibitem{Kim:2019npy}
Y.-L. Kim, Y.~Kang, and Y.-W. Lee, {\it {Environmental Dependence of Type Ia
  Supernova Luminosities from the YONSEI Supernova Catalog}},  {\em J. Korean
  Astron. Soc.} {\bf 52} (2019), no.~5 181--205,
  [\href{http://arxiv.org/abs/1908.10375}{{\tt arXiv:1908.10375}}].

\bibitem{Kang:2019azh}
Y.~Kang, Y.-W. Lee, Y.-L. Kim, C.~Chung, and C.~H. Ree, {\it {Early-type Host
  Galaxies of Type Ia Supernovae. II. Evidence for Luminosity Evolution in
  Supernova Cosmology}},  \href{http://arxiv.org/abs/1912.04903}{{\tt
  arXiv:1912.04903}}.

\bibitem{Rose:2020shp}
B.~M. Rose, D.~Rubin, A.~Cikota, S.~Deustua, S.~Dixon, A.~Fruchter, D.~O.
  Jones, A.~G. Riess, and D.~M. Scolnic, {\it {No Evidence for Type Ia
  Supernova Luminosity Evolution: Evidence for Dark Energy is Robust}},
  \href{http://arxiv.org/abs/2002.12382}{{\tt arXiv:2002.12382}}.

\bibitem{Rigault:2014kaa}
M.~Rigault et~al., {\it {Confirmation of a Star Formation Bias in Type Ia
  Supernova Distances and its Effect on Measurement of the Hubble Constant}},
  {\em Astrophys. J.} {\bf 802} (2015), no.~1 20,
  [\href{http://arxiv.org/abs/1412.6501}{{\tt arXiv:1412.6501}}].

\bibitem{Rigault:2018ffm}
{\bf Nearby Supernova Factory} Collaboration, M.~Rigault et~al., {\it {Strong
  Dependence of Type Ia Supernova Standardization on the Local Specific Star
  Formation Rate}},  \href{http://arxiv.org/abs/1806.03849}{{\tt
  arXiv:1806.03849}}.

\bibitem{Rigault:2013gux}
{\bf Nearby Supernova factory} Collaboration, M.~Rigault et~al., {\it {Evidence
  of Environmental Dependencies of Type Ia Supernovae from the Nearby Supernova
  Factory indicated by Local H$\alpha$}},  {\em Astron. Astrophys.} {\bf 560}
  (2013) A66, [\href{http://arxiv.org/abs/1309.1182}{{\tt arXiv:1309.1182}}].
  [Corrigendum: Astron. Astrophys.612,C1(2018)].

\bibitem{Childress:2014vka}
M.~J. Childress, C.~Wolf, and H.~J. Zahid, {\it {Ages of Type Ia Supernovae
  Over Cosmic Time}},  {\em Mon. Not. Roy. Astron. Soc.} {\bf 445} (2014),
  no.~2 1898--1911, [\href{http://arxiv.org/abs/1409.2951}{{\tt
  arXiv:1409.2951}}].

\bibitem{Jones:2018vbn}
D.~O. Jones et~al., {\it {Should Type Ia Supernova Distances be Corrected for
  their Local Environments?}},  {\em Astrophys. J.} {\bf 867} (2018), no.~2
  108, [\href{http://arxiv.org/abs/1805.05911}{{\tt arXiv:1805.05911}}].

\bibitem{Timmes:2003xx}
F.~X. Timmes, E.~F. Brown, and J.~W. Truran, {\it {On variations in the peak
  luminosity of type ia supernovae}},  {\em Astrophys. J.} {\bf 590} (2003)
  L83--L86, [\href{http://arxiv.org/abs/astro-ph/0305114}{{\tt
  astro-ph/0305114}}].

\bibitem{Travaglio:2005yt}
C.~Travaglio, W.~Hillebrandt, and M.~Reinecke, {\it {Metallicity effect in
  multi-dimensional SNIa nucleosynthesis}},  {\em Astron. Astrophys.} {\bf 443}
  (2005) 1007, [\href{http://arxiv.org/abs/astro-ph/0507510}{{\tt
  astro-ph/0507510}}].

\bibitem{Moreno-Raya:2015jqq}
M.~E. Moreno-Raya, M.~Moll{\'a}, {\'A}.~R. L{\'o}pez-S{\'a}nchez, L.~Galbany,
  J.~V{\'i}lchez, A.~Carnero, and I.~Dom{\'i}nguez, {\it {On the dependence of
  the type Ia SNe luminosities on the metallicity of their host galaxies}},
  \href{http://arxiv.org/abs/1511.05348}{{\tt arXiv:1511.05348}}.

\bibitem{Moreno-Raya:2016rlw}
M.~E. Moreno-Raya, A.~R. Lopez-Sanchez, M.~Moll{\'a}, L.~Galbany, J.~M.
  Vilchez, and A.~Carnero, {\it {Using the local gas-phase oxygen abundances to
  explore a metallicity-dependence in SNe Ia luminosities}},
  \href{http://arxiv.org/abs/1607.05526}{{\tt arXiv:1607.05526}}.

\bibitem{Campbell:2016zzh}
H.~Campbell, M.~Fraser, and G.~Gilmore, {\it {How SN Ia host-galaxy properties
  affect cosmological parameters}},  {\em Mon. Not. Roy. Astron. Soc.} {\bf
  457} (2016), no.~4 3470--3491, [\href{http://arxiv.org/abs/1602.02596}{{\tt
  arXiv:1602.02596}}].

\bibitem{2018A&A...615A..68R}
M.~{Roman}, D.~{Hardin}, M.~{Betoule}, P.~{Astier}, C.~{Balland}, R.~S.
  {Ellis}, S.~{Fabbro}, J.~{Guy}, I.~{Hook}, D.~A. {Howell}, C.~{Lidman},
  A.~{Mitra}, A.~{M{\"o}ller}, A.~M. {Mour{\~a}o}, J.~{Neveu},
  N.~{Palanque-Delabrouille}, C.~J. {Pritchet}, N.~{Regnault},
  V.~{Ruhlmann-Kleider}, C.~{Saunders}, and M.~{Sullivan}, {\it {Dependence of
  Type Ia supernova luminosities on their local environment}},  {\em Astron.
  Astrophys.} {\bf 615} (July, 2018) A68,
  [\href{http://arxiv.org/abs/1706.07697}{{\tt arXiv:1706.07697}}].

\bibitem{Brout:2020msh}
D.~Brout and D.~Scolnic, {\it {It's Dust: Solving the Mysteries of the
  Intrinsic Scatter and Host-Galaxy Dependence of Standardized Type Ia
  Supernova Brightnesses}},  \href{http://arxiv.org/abs/2004.10206}{{\tt
  arXiv:2004.10206}}.

\bibitem{Nordin:2008aa}
J.~Nordin, A.~Goobar, and J.~Jonsson, {\it {Quantifying systematic
  uncertainties in supernova cosmology}},  {\em JCAP} {\bf 0802} (2008) 008,
  [\href{http://arxiv.org/abs/0801.2484}{{\tt arXiv:0801.2484}}].

\bibitem{Ferramacho:2008ap}
L.~D. Ferramacho, A.~Blanchard, and Y.~Zolnierowski, {\it {Constraints on
  C.D.M. cosmology from galaxy power spectrum, CMB and SNIa evolution}},  {\em
  Astron. Astrophys.} {\bf 499} (2009) 21,
  [\href{http://arxiv.org/abs/0807.4608}{{\tt arXiv:0807.4608}}].

\bibitem{Linden:2009vh}
S.~Linden, J.-M. Virey, and A.~Tilquin, {\it {Cosmological Parameter Extraction
  and Biases from Type Ia Supernova Magnitude Evolution}},  {\em Astron.
  Astrophys.} {\bf 50} (2009) 1095--1105,
  [\href{http://arxiv.org/abs/0907.4495}{{\tt arXiv:0907.4495}}].

\bibitem{Tutusaus:2017ibk}
I.~Tutusaus, B.~Lamine, A.~Dupays, and A.~Blanchard, {\it {Is cosmic
  acceleration proven by local cosmological probes?}},  {\em Astron.
  Astrophys.} {\bf 602} (2017) A73,
  [\href{http://arxiv.org/abs/1706.05036}{{\tt arXiv:1706.05036}}].

\bibitem{Tutusaus:2018ulu}
I.~Tutusaus, B.~Lamine, and A.~Blanchard, {\it {Model-independent cosmic
  acceleration and redshift-dependent intrinsic luminosity in type-Ia
  supernovae}},  {\em Astron. Astrophys.} {\bf 625} (2019) A15,
  [\href{http://arxiv.org/abs/1803.06197}{{\tt arXiv:1803.06197}}].

\bibitem{LHuillier:2018rsv}
B.~L'Huillier, A.~Shafieloo, E.~V. Linder, and A.~G. Kim, {\it {Model
  Independent Expansion History from Supernovae: Cosmology versus
  Systematics}},  {\em Mon. Not. Roy. Astron. Soc.} {\bf 485} (2019), no.~2
  2783--2790, [\href{http://arxiv.org/abs/1812.03623}{{\tt arXiv:1812.03623}}].

\bibitem{Martinelli:2019krf}
M.~Martinelli and I.~Tutusaus, {\it {CMB tensions with low-redshift $H_0$ and
  $S_8$ measurements: impact of a redshift-dependent type-Ia supernovae
  intrinsic luminosity}},  {\em Symmetry} {\bf 11} (2019), no.~8 986,
  [\href{http://arxiv.org/abs/1906.09189}{{\tt arXiv:1906.09189}}].

\bibitem{Sapone:2020wwz}
D.~Sapone, S.~Nesseris, and C.~A.~P. Bengaly, {\it {Is there any measurable
  redshift dependence on the SN Ia absolute magnitude?}},
  \href{http://arxiv.org/abs/2006.05461}{{\tt arXiv:2006.05461}}.

\bibitem{Nielsen:2015pga}
J.~T. Nielsen, A.~Guffanti, and S.~Sarkar, {\it {Marginal evidence for cosmic
  acceleration from Type Ia supernovae}},  {\em Sci. Rep.} {\bf 6} (2016)
  35596, [\href{http://arxiv.org/abs/1506.01354}{{\tt arXiv:1506.01354}}].

\bibitem{Rubin:2016iqe}
D.~Rubin and B.~Hayden, {\it {Is the expansion of the universe accelerating?
  All signs point to yes}},  {\em Astrophys. J.} {\bf 833} (2016), no.~2 L30,
  [\href{http://arxiv.org/abs/1610.08972}{{\tt arXiv:1610.08972}}].

\bibitem{Dam:2017xqs}
L.~H. Dam, A.~Heinesen, and D.~L. Wiltshire, {\it {Apparent cosmic acceleration
  from type Ia supernovae}},  {\em Mon. Not. Roy. Astron. Soc.} {\bf 472}
  (2017), no.~1 835--851, [\href{http://arxiv.org/abs/1706.07236}{{\tt
  arXiv:1706.07236}}].

\bibitem{Colin:2018ghy}
J.~Colin, R.~Mohayaee, M.~Rameez, and S.~Sarkar, {\it {Evidence for anisotropy
  of cosmic acceleration}},  {\em Astron. Astrophys.} {\bf 631} (2019) L13,
  [\href{http://arxiv.org/abs/1808.04597}{{\tt arXiv:1808.04597}}].

\bibitem{Desgrange:2019npu}
C.~Desgrange, A.~Heinesen, and T.~Buchert, {\it {Dynamical spatial curvature as
  a fit to type Ia supernovae}},  {\em Int. J. Mod. Phys.} {\bf D28} (2019),
  no.~11 1950143, [\href{http://arxiv.org/abs/1902.07915}{{\tt
  arXiv:1902.07915}}].

\bibitem{Sherwin:2011gv}
B.~D. Sherwin et~al., {\it {Evidence for dark energy from the cosmic microwave
  background alone using the Atacama Cosmology Telescope lensing
  measurements}},  {\em Phys. Rev. Lett.} {\bf 107} (2011) 021302,
  [\href{http://arxiv.org/abs/1105.0419}{{\tt arXiv:1105.0419}}].

\bibitem{Nadathur:2020kvq}
S.~Nadathur, W.~J. Percival, F.~Beutler, and H.~Winther, {\it {Testing
  low-redshift cosmic acceleration with large-scale structure}},
  \href{http://arxiv.org/abs/2001.11044}{{\tt arXiv:2001.11044}}.

\bibitem{Hunt:2008wp}
P.~Hunt and S.~Sarkar, {\it {Constraints on large scale inhomogeneities from
  WMAP-5 and SDSS: confrontation with recent observations}},  {\em Mon. Not.
  Roy. Astron. Soc.} {\bf 401} (2010) 547,
  [\href{http://arxiv.org/abs/0807.4508}{{\tt arXiv:0807.4508}}].

\bibitem{Hunt:2015iua}
P.~Hunt and S.~Sarkar, {\it {Search for features in the spectrum of primordial
  perturbations using Planck and other datasets}},  {\em JCAP} {\bf 1512}
  (2015) 052, [\href{http://arxiv.org/abs/1510.03338}{{\tt arXiv:1510.03338}}].

\bibitem{Calabrese:2013lga}
E.~Calabrese, M.~Martinelli, S.~Pandolfi, V.~F. Cardone, C.~J. A.~P. Martins,
  S.~Spiro, and P.~E. Vielzeuf, {\it {Dark Energy coupling with
  electromagnetism as seen from future low-medium redshift probes}},  {\em
  Phys. Rev.} {\bf D89} (2014), no.~8 083509,
  [\href{http://arxiv.org/abs/1311.5841}{{\tt arXiv:1311.5841}}].

\bibitem{Wright:2017rsu}
B.~S. Wright and B.~Li, {\it {Type Ia supernovae, standardizable candles, and
  gravity}},  {\em Phys. Rev. D} {\bf 97} (2018), no.~8 083505,
  [\href{http://arxiv.org/abs/1710.07018}{{\tt arXiv:1710.07018}}].

\bibitem{Brans:1961sx}
C.~Brans and R.~H. Dicke, {\it {Mach's principle and a relativistic theory of
  gravitation}},  {\em Phys. Rev.} {\bf 124} (1961) 925--935.

\bibitem{Nunes:2016plz}
R.~C. Nunes, A.~Bonilla, S.~Pan, and E.~N. Saridakis, {\it {Observational
  Constraints on $f(T)$ gravity from varying fundamental constants}},  {\em
  Eur. Phys. J.} {\bf C77} (2017), no.~4 230,
  [\href{http://arxiv.org/abs/1608.01960}{{\tt arXiv:1608.01960}}].

\bibitem{Jimenez:2020bgw}
J.~B. Jimenez, D.~Bettoni, and P.~Brax, {\it {Charged Dark Matter and the $H_0$
  tension}},  \href{http://arxiv.org/abs/2004.13677}{{\tt arXiv:2004.13677}}.

\bibitem{Carroll:1998zi}
S.~M. Carroll, {\it {Quintessence and the rest of the world}},  {\em Phys. Rev.
  Lett.} {\bf 81} (1998) 3067--3070,
  [\href{http://arxiv.org/abs/astro-ph/9806099}{{\tt astro-ph/9806099}}].

\bibitem{Burrage:2016bwy}
C.~Burrage and J.~Sakstein, {\it {A Compendium of Chameleon Constraints}},
  {\em JCAP} {\bf 1611} (2016) 045,
  [\href{http://arxiv.org/abs/1609.01192}{{\tt arXiv:1609.01192}}].

\bibitem{Vagnozzi:2019kvw}
S.~Vagnozzi, L.~Visinelli, O.~Mena, and D.~F. Mota, {\it {Do we have any hope
  of detecting scattering between dark energy and baryons through cosmology?}},
   {\em Mon. Not. Roy. Astron. Soc.} {\bf 493} (2020), no.~1 1139--1152,
  [\href{http://arxiv.org/abs/1911.12374}{{\tt arXiv:1911.12374}}].

\bibitem{Jimenez:2020ysu}
J.~B. Jiménez, D.~Bettoni, D.~Figueruelo, and F.~A. Teppa~Pannia, {\it {On
  cosmological signatures of baryons-dark energy elastic couplings}},
  \href{http://arxiv.org/abs/2004.14661}{{\tt arXiv:2004.14661}}.

\bibitem{Sandvik:2001rv}
H.~B. Sandvik, J.~D. Barrow, and J.~Magueijo, {\it {A simple cosmology with a
  varying fine structure constant}},  {\em Phys. Rev. Lett.} {\bf 88} (2002)
  031302, [\href{http://arxiv.org/abs/astro-ph/0107512}{{\tt
  astro-ph/0107512}}].

\bibitem{Damour:2002mi}
T.~Damour, F.~Piazza, and G.~Veneziano, {\it {Runaway dilaton and equivalence
  principle violations}},  {\em Phys. Rev. Lett.} {\bf 89} (2002) 081601,
  [\href{http://arxiv.org/abs/gr-qc/0204094}{{\tt gr-qc/0204094}}].

\bibitem{Zumalacarregui:2020cjh}
M.~Zumalacarregui, {\it {Gravity in the Era of Equality: Towards solutions to
  the Hubble problem without fine-tuned initial conditions}},
  \href{http://arxiv.org/abs/2003.06396}{{\tt arXiv:2003.06396}}.

\bibitem{Betoule:2014frx}
{\bf SDSS} Collaboration, M.~Betoule et~al., {\it {Improved cosmological
  constraints from a joint analysis of the SDSS-II and SNLS supernova
  samples}},  {\em Astron. Astrophys.} {\bf 568} (2014) A22,
  [\href{http://arxiv.org/abs/1401.4064}{{\tt arXiv:1401.4064}}].

\bibitem{Aubourg:2014yra}
E.~Aubourg et~al., {\it {Cosmological implications of baryon acoustic
  oscillation measurements}},  {\em Phys. Rev.} {\bf D92} (2015), no.~12
  123516, [\href{http://arxiv.org/abs/1411.1074}{{\tt arXiv:1411.1074}}].

\bibitem{Chevallier:2000qy}
M.~Chevallier and D.~Polarski, {\it {Accelerating universes with scaling dark
  matter}},  {\em Int.\ J.\ Mod.\ Phys.\ D} {\bf 10} (2001) 213--224,
  [\href{http://arxiv.org/abs/gr-qc/0009008}{{\tt gr-qc/0009008}}].

\bibitem{Linder:2002et}
E.~V. Linder, {\it {Exploring the expansion history of the universe}},  {\em
  Phys.\ Rev.\ Lett.} {\bf 90} (2003) 091301,
  [\href{http://arxiv.org/abs/astro-ph/0208512}{{\tt astro-ph/0208512}}].

\bibitem{Aghanim:2019ame}
{\bf Planck} Collaboration, N.~Aghanim et~al., {\it {Planck 2018 results. V.
  CMB power spectra and likelihoods}},
  \href{http://arxiv.org/abs/1907.12875}{{\tt arXiv:1907.12875}}.

\bibitem{Aghanim:2018oex}
{\bf Planck} Collaboration, N.~Aghanim et~al., {\it {Planck 2018 results. VIII.
  Gravitational lensing}},  \href{http://arxiv.org/abs/1807.06210}{{\tt
  arXiv:1807.06210}}.

\bibitem{Beutler:2011hx}
F.~Beutler, C.~Blake, M.~Colless, D.~Jones, L.~Staveley-Smith, L.~Campbell,
  Q.~Parker, W.~Saunders, and F.~Watson, {\it {The 6dF Galaxy Survey: Baryon
  Acoustic Oscillations and the Local Hubble Constant}},  {\em Mon.\ Not.\
  Roy.\ Astron.\ Soc.} {\bf 416} (2011) 3017--3032,
  [\href{http://arxiv.org/abs/1106.3366}{{\tt arXiv:1106.3366}}].

\bibitem{Ross:2014qpa}
A.~J. Ross, L.~Samushia, C.~Howlett, W.~J. Percival, A.~Burden, and M.~Manera,
  {\it {The clustering of the SDSS DR7 main Galaxy sample -- I. A 4 per cent
  distance measure at $z = 0.15$}},  {\em Mon.\ Not.\ Roy.\ Astron.\ Soc.} {\bf
  449} (2015), no.~1 835--847, [\href{http://arxiv.org/abs/1409.3242}{{\tt
  arXiv:1409.3242}}].

\bibitem{Jimenez:2001gg}
R.~Jimenez and A.~Loeb, {\it {Constraining cosmological parameters based on
  relative galaxy ages}},  {\em Astrophys.\ J.} {\bf 573} (2002) 37--42,
  [\href{http://arxiv.org/abs/astro-ph/0106145}{{\tt astro-ph/0106145}}].

\bibitem{Moresco:2012by}
M.~Moresco, L.~Verde, L.~Pozzetti, R.~Jimenez, and A.~Cimatti, {\it {New
  constraints on cosmological parameters and neutrino properties using the
  expansion rate of the Universe to z\textasciitilde 1.75}},  {\em JCAP} {\bf
  07} (2012) 053, [\href{http://arxiv.org/abs/1201.6658}{{\tt
  arXiv:1201.6658}}].

\bibitem{Moresco:2012jh}
M.~Moresco et~al., {\it {Improved constraints on the expansion rate of the
  Universe up to z\textasciitilde 1.1 from the spectroscopic evolution of
  cosmic chronometers}},  {\em JCAP} {\bf 08} (2012) 006,
  [\href{http://arxiv.org/abs/1201.3609}{{\tt arXiv:1201.3609}}].

\bibitem{Moresco:2015cya}
M.~Moresco, {\it {Raising the bar: new constraints on the Hubble parameter with
  cosmic chronometers at z $\sim$ 2}},  {\em Mon.\ Not.\ Roy.\ Astron.\ Soc.}
  {\bf 450} (2015), no.~1 L16--L20,
  [\href{http://arxiv.org/abs/1503.01116}{{\tt arXiv:1503.01116}}].

\bibitem{Moresco:2016mzx}
M.~Moresco, L.~Pozzetti, A.~Cimatti, R.~Jimenez, C.~Maraston, L.~Verde,
  D.~Thomas, A.~Citro, R.~Tojeiro, and D.~Wilkinson, {\it {A 6\% measurement of
  the Hubble parameter at $z\sim0.45$: direct evidence of the epoch of cosmic
  re-acceleration}},  {\em JCAP} {\bf 05} (2016) 014,
  [\href{http://arxiv.org/abs/1601.01701}{{\tt arXiv:1601.01701}}].

\bibitem{Lewis:2002ah}
A.~Lewis and S.~Bridle, {\it {Cosmological parameters from CMB and other data:
  A Monte Carlo approach}},  {\em Phys. Rev.} {\bf D66} (2002) 103511,
  [\href{http://arxiv.org/abs/astro-ph/0205436}{{\tt astro-ph/0205436}}].

\bibitem{Lewis:1999bs}
A.~Lewis, A.~Challinor, and A.~Lasenby, {\it {Efficient computation of CMB
  anisotropies in closed FRW models}},  {\em Astrophys. J.} {\bf 538} (2000)
  473--476, [\href{http://arxiv.org/abs/astro-ph/9911177}{{\tt
  astro-ph/9911177}}].

\bibitem{Gelman:1992zz}
A.~Gelman and D.~B. Rubin, {\it {Inference from Iterative Simulation Using
  Multiple Sequences}},  {\em Statist. Sci.} {\bf 7} (1992) 457--472.

\bibitem{Riess:2019cxk}
A.~G. Riess, S.~Casertano, W.~Yuan, L.~M. Macri, and D.~Scolnic, {\it {Large
  Magellanic Cloud Cepheid Standards Provide a 1\% Foundation for the
  Determination of the Hubble Constant and Stronger Evidence for Physics beyond
  $\Lambda$CDM}},  {\em Astrophys. J.} {\bf 876} (2019), no.~1 85,
  [\href{http://arxiv.org/abs/1903.07603}{{\tt arXiv:1903.07603}}].

\bibitem{DiValentino:2016hlg}
E.~Di~Valentino, A.~Melchiorri, and J.~Silk, {\it {Reconciling Planck with the
  local value of $H_0$ in extended parameter space}},  {\em Phys. Lett.} {\bf
  B761} (2016) 242--246, [\href{http://arxiv.org/abs/1606.00634}{{\tt
  arXiv:1606.00634}}].

\bibitem{Bernal:2016gxb}
J.~L. Bernal, L.~Verde, and A.~G. Riess, {\it {The trouble with $H_0$}},  {\em
  JCAP} {\bf 1610} (2016) 019, [\href{http://arxiv.org/abs/1607.05617}{{\tt
  arXiv:1607.05617}}].

\bibitem{Karwal:2016vyq}
T.~Karwal and M.~Kamionkowski, {\it {Dark energy at early times, the Hubble
  parameter, and the string axiverse}},  {\em Phys. Rev.} {\bf D94} (2016),
  no.~10 103523, [\href{http://arxiv.org/abs/1608.01309}{{\tt
  arXiv:1608.01309}}].

\bibitem{Zhao:2017urm}
M.-M. Zhao, D.-Z. He, J.-F. Zhang, and X.~Zhang, {\it {Search for sterile
  neutrinos in holographic dark energy cosmology: Reconciling Planck
  observation with the local measurement of the Hubble constant}},  {\em Phys.
  Rev.} {\bf D96} (2017), no.~4 043520,
  [\href{http://arxiv.org/abs/1703.08456}{{\tt arXiv:1703.08456}}].

\bibitem{Sola:2017znb}
J.~Solà, A.~Gómez-Valent, and J.~de~Cruz~Pérez, {\it {The $H_0$ tension in
  light of vacuum dynamics in the Universe}},  {\em Phys. Lett.} {\bf B774}
  (2017) 317--324, [\href{http://arxiv.org/abs/1705.06723}{{\tt
  arXiv:1705.06723}}].

\bibitem{Buen-Abad:2017gxg}
M.~A. Buen-Abad, M.~Schmaltz, J.~Lesgourgues, and T.~Brinckmann, {\it
  {Interacting Dark Sector and Precision Cosmology}},  {\em JCAP} {\bf 1801}
  (2018) 008, [\href{http://arxiv.org/abs/1708.09406}{{\tt arXiv:1708.09406}}].

\bibitem{Khosravi:2017hfi}
N.~Khosravi, S.~Baghram, N.~Afshordi, and N.~Altamirano, {\it {$H_0$ tension as
  a hint for a transition in gravitational theory}},  {\em Phys. Rev.} {\bf
  D99} (2019), no.~10 103526, [\href{http://arxiv.org/abs/1710.09366}{{\tt
  arXiv:1710.09366}}].

\bibitem{Benetti:2017juy}
M.~Benetti, L.~L. Graef, and J.~S. Alcaniz, {\it {The $H_0$ and $\sigma_8$
  tensions and the scale invariant spectrum}},  {\em JCAP} {\bf 1807} (2018)
  066, [\href{http://arxiv.org/abs/1712.00677}{{\tt arXiv:1712.00677}}].

\bibitem{Mortsell:2018mfj}
E.~Mörtsell and S.~Dhawan, {\it {Does the Hubble constant tension call for new
  physics?}},  {\em JCAP} {\bf 1809} (2018) 025,
  [\href{http://arxiv.org/abs/1801.07260}{{\tt arXiv:1801.07260}}].

\bibitem{Nunes:2018xbm}
R.~C. Nunes, {\it {Structure formation in $f(T)$ gravity and a solution for
  $H_0$ tension}},  {\em JCAP} {\bf 1805} (2018) 052,
  [\href{http://arxiv.org/abs/1802.02281}{{\tt arXiv:1802.02281}}].

\bibitem{Poulin:2018zxs}
V.~Poulin, K.~K. Boddy, S.~Bird, and M.~Kamionkowski, {\it {Implications of an
  extended dark energy cosmology with massive neutrinos for cosmological
  tensions}},  {\em Phys. Rev.} {\bf D97} (2018), no.~12 123504,
  [\href{http://arxiv.org/abs/1803.02474}{{\tt arXiv:1803.02474}}].

\bibitem{Kumar:2018yhh}
S.~Kumar, R.~C. Nunes, and S.~K. Yadav, {\it {Cosmological bounds on dark
  matter-photon coupling}},  {\em Phys. Rev.} {\bf D98} (2018), no.~4 043521,
  [\href{http://arxiv.org/abs/1803.10229}{{\tt arXiv:1803.10229}}].

\bibitem{Banihashemi:2018oxo}
A.~Banihashemi, N.~Khosravi, and A.~H. Shirazi, {\it {Ups and Downs in Dark
  Energy: phase transition in dark sector as a proposal to lessen cosmological
  tensions}},  \href{http://arxiv.org/abs/1808.02472}{{\tt arXiv:1808.02472}}.

\bibitem{DEramo:2018vss}
F.~D'Eramo, R.~Z. Ferreira, A.~Notari, and J.~L. Bernal, {\it {Hot Axions and
  the $H_0$ tension}},  {\em JCAP} {\bf 1811} (2018) 014,
  [\href{http://arxiv.org/abs/1808.07430}{{\tt arXiv:1808.07430}}].

\bibitem{Guo:2018ans}
R.-Y. Guo, J.-F. Zhang, and X.~Zhang, {\it {Can the $H_0$ tension be resolved
  in extensions to $\Lambda$CDM cosmology?}},  {\em JCAP} {\bf 1902} (2019)
  054, [\href{http://arxiv.org/abs/1809.02340}{{\tt arXiv:1809.02340}}].

\bibitem{Graef:2018fzu}
L.~L. Graef, M.~Benetti, and J.~S. Alcaniz, {\it {Primordial gravitational
  waves and the H0-tension problem}},  {\em Phys. Rev.} {\bf D99} (2019), no.~4
  043519, [\href{http://arxiv.org/abs/1809.04501}{{\tt arXiv:1809.04501}}].

\bibitem{Banihashemi:2018has}
A.~Banihashemi, N.~Khosravi, and A.~H. Shirazi, {\it {Ginzburg-Landau Theory of
  Dark Energy: A Framework to Study Both Temporal and Spatial Cosmological
  Tensions Simultaneously}},  {\em Phys. Rev.} {\bf D99} (2019), no.~8 083509,
  [\href{http://arxiv.org/abs/1810.11007}{{\tt arXiv:1810.11007}}].

\bibitem{Poulin:2018cxd}
V.~Poulin, T.~L. Smith, T.~Karwal, and M.~Kamionkowski, {\it {Early Dark Energy
  Can Resolve The Hubble Tension}},  {\em Phys. Rev. Lett.} {\bf 122} (2019),
  no.~22 221301, [\href{http://arxiv.org/abs/1811.04083}{{\tt
  arXiv:1811.04083}}].

\bibitem{Kreisch:2019yzn}
C.~D. Kreisch, F.-Y. Cyr-Racine, and O.~Doré, {\it {The Neutrino Puzzle:
  Anomalies, Interactions, and Cosmological Tensions}},
  \href{http://arxiv.org/abs/1902.00534}{{\tt arXiv:1902.00534}}.

\bibitem{Raveri:2019mxg}
M.~Raveri, {\it {Reconstructing Gravity on Cosmological Scales}},  {\em Phys.
  Rev.} {\bf D101} (2020), no.~8 083524,
  [\href{http://arxiv.org/abs/1902.01366}{{\tt arXiv:1902.01366}}].

\bibitem{Agrawal:2019lmo}
P.~Agrawal, F.-Y. Cyr-Racine, D.~Pinner, and L.~Randall, {\it {Rock 'n' Roll
  Solutions to the Hubble Tension}},
  \href{http://arxiv.org/abs/1904.01016}{{\tt arXiv:1904.01016}}.

\bibitem{Li:2019san}
X.~Li, A.~Shafieloo, V.~Sahni, and A.~A. Starobinsky, {\it {Revisiting
  Metastable Dark Energy and Tensions in the Estimation of Cosmological
  Parameters}},  {\em Astrophys. J.} {\bf 887} (2019) 153,
  [\href{http://arxiv.org/abs/1904.03790}{{\tt arXiv:1904.03790}}].

\bibitem{Yang:2019jwn}
W.~Yang, S.~Pan, A.~Paliathanasis, S.~Ghosh, and Y.~Wu, {\it {Observational
  constraints of a new unified dark fluid and the $H_0$ tension}},  {\em Mon.
  Not. Roy. Astron. Soc.} {\bf 490} (2019), no.~2 2071--2085,
  [\href{http://arxiv.org/abs/1904.10436}{{\tt arXiv:1904.10436}}].

\bibitem{Keeley:2019esp}
R.~E. Keeley, S.~Joudaki, M.~Kaplinghat, and D.~Kirkby, {\it {Implications of a
  transition in the dark energy equation of state for the $H_0$ and $\sigma_8$
  tensions}},  {\em JCAP} {\bf 1912} (2019) 035,
  [\href{http://arxiv.org/abs/1905.10198}{{\tt arXiv:1905.10198}}].

\bibitem{Lin:2019qug}
M.-X. Lin, G.~Benevento, W.~Hu, and M.~Raveri, {\it {Acoustic Dark Energy:
  Potential Conversion of the Hubble Tension}},  {\em Phys. Rev.} {\bf D100}
  (2019), no.~6 063542, [\href{http://arxiv.org/abs/1905.12618}{{\tt
  arXiv:1905.12618}}].

\bibitem{Li:2019ypi}
X.~Li and A.~Shafieloo, {\it {A Simple Phenomenological Emergent Dark Energy
  Model can Resolve the Hubble Tension}},  {\em Astrophys. J.} {\bf 883}
  (2019), no.~1 L3, [\href{http://arxiv.org/abs/1906.08275}{{\tt
  arXiv:1906.08275}}]. [Astrophys. J. Lett.883,L3(2019)].

\bibitem{Rossi:2019lgt}
M.~Rossi, M.~Ballardini, M.~Braglia, F.~Finelli, D.~Paoletti, A.~A.
  Starobinsky, and C.~Umilta, {\it {Cosmological constraints on post-Newtonian
  parameters in effectively massless scalar-tensor theories of gravity}},  {\em
  Phys. Rev.} {\bf D100} (2019), no.~10 103524,
  [\href{http://arxiv.org/abs/1906.10218}{{\tt arXiv:1906.10218}}].

\bibitem{DiValentino:2019exe}
E.~Di~Valentino, R.~Z. Ferreira, L.~Visinelli, and U.~Danielsson, {\it {Late
  time transitions in the quintessence field and the $H_0$ tension}},  {\em
  Phys. Dark Univ.} {\bf 26} (2019) 100385,
  [\href{http://arxiv.org/abs/1906.11255}{{\tt arXiv:1906.11255}}].

\bibitem{Archidiacono:2019wdp}
M.~Archidiacono, D.~C. Hooper, R.~Murgia, S.~Bohr, J.~Lesgourgues, and M.~Viel,
  {\it {Constraining Dark Matter-Dark Radiation interactions with CMB, BAO, and
  Lyman-$\alpha$}},  {\em JCAP} {\bf 1910} (2019) 055,
  [\href{http://arxiv.org/abs/1907.01496}{{\tt arXiv:1907.01496}}].

\bibitem{Kazantzidis:2019dvk}
L.~Kazantzidis and L.~Perivolaropoulos, {\it {Is gravity getting weaker at low
  z? Observational evidence and theoretical implications}},
  \href{http://arxiv.org/abs/1907.03176}{{\tt arXiv:1907.03176}}.

\bibitem{Desmond:2019ygn}
H.~Desmond, B.~Jain, and J.~Sakstein, {\it {Local resolution of the Hubble
  tension: The impact of screened fifth forces on the cosmic distance ladder}},
   {\em Phys. Rev.} {\bf D100} (2019), no.~4 043537,
  [\href{http://arxiv.org/abs/1907.03778}{{\tt arXiv:1907.03778}}]. [Erratum:
  Phys. Rev.D101,no.6,069904(2020)].

\bibitem{Nesseris:2019fwr}
S.~Nesseris, D.~Sapone, and S.~Sypsas, {\it {Evaporating primordial black holes
  as varying dark energy}},  {\em Phys. Dark Univ.} {\bf 27} (2020) 100413,
  [\href{http://arxiv.org/abs/1907.05608}{{\tt arXiv:1907.05608}}].

\bibitem{Vagnozzi:2019ezj}
S.~Vagnozzi, {\it {New physics in light of the $H_0$ tension: an alternative
  view}},  \href{http://arxiv.org/abs/1907.07569}{{\tt arXiv:1907.07569}}.

\bibitem{Pan:2019hac}
S.~Pan, W.~Yang, E.~Di~Valentino, A.~Shafieloo, and S.~Chakraborty, {\it
  {Reconciling $H_0$ tension in a six parameter space?}},
  \href{http://arxiv.org/abs/1907.12551}{{\tt arXiv:1907.12551}}.

\bibitem{Visinelli:2019qqu}
L.~Visinelli, S.~Vagnozzi, and U.~Danielsson, {\it {Revisiting a negative
  cosmological constant from low-redshift data}},  {\em Symmetry} {\bf 11}
  (2019), no.~8 1035, [\href{http://arxiv.org/abs/1907.07953}{{\tt
  arXiv:1907.07953}}].

\bibitem{Cai:2019bdh}
Y.-F. Cai, M.~Khurshudyan, and E.~N. Saridakis, {\it {Model-independent
  reconstruction of $f(T)$ gravity from Gaussian Processes}},  {\em Astrophys.
  J.} {\bf 888} (2020) 62, [\href{http://arxiv.org/abs/1907.10813}{{\tt
  arXiv:1907.10813}}].

\bibitem{Xiao:2019ccl}
L.~Xiao, L.~Zhang, R.~An, C.~Feng, and B.~Wang, {\it {Fractional Dark Matter
  decay: cosmological imprints and observational constraints}},  {\em JCAP}
  {\bf 2001} (2020) 045, [\href{http://arxiv.org/abs/1908.02668}{{\tt
  arXiv:1908.02668}}].

\bibitem{Smith:2019ihp}
T.~L. Smith, V.~Poulin, and M.~A. Amin, {\it {Oscillating scalar fields and the
  Hubble tension: a resolution with novel signatures}},  {\em Phys. Rev.} {\bf
  D101} (2020), no.~6 063523, [\href{http://arxiv.org/abs/1908.06995}{{\tt
  arXiv:1908.06995}}].

\bibitem{Ghosh:2019tab}
S.~Ghosh, R.~Khatri, and T.~S. Roy, {\it {Dark Neutrino interactions phase out
  the Hubble tension}},  \href{http://arxiv.org/abs/1908.09843}{{\tt
  arXiv:1908.09843}}.

\bibitem{Sola:2019jek}
J.~Solà~Peracaula, A.~Gomez-Valent, J.~de~Cruz~Pérez, and C.~Moreno-Pulido,
  {\it {Brans–Dicke Gravity with a Cosmological Constant Smoothes Out
  $\Lambda$CDM Tensions}},  {\em Astrophys. J.} {\bf 886} (2019), no.~1 L6,
  [\href{http://arxiv.org/abs/1909.02554}{{\tt arXiv:1909.02554}}].

\bibitem{Escudero:2019gvw}
M.~Escudero and S.~J. Witte, {\it {A CMB Search for the Neutrino Mass Mechanism
  and its Relation to the $H_0$ Tension}},  {\em Eur. Phys. J.} {\bf C80}
  (2020), no.~4 294, [\href{http://arxiv.org/abs/1909.04044}{{\tt
  arXiv:1909.04044}}].

\bibitem{Yan:2019gbw}
S.-F. Yan, P.~Zhang, J.-W. Chen, X.-Z. Zhang, Y.-F. Cai, and E.~N. Saridakis,
  {\it {Interpreting cosmological tensions from the effective field theory of
  torsional gravity}},  \href{http://arxiv.org/abs/1909.06388}{{\tt
  arXiv:1909.06388}}.

\bibitem{Sakstein:2019fmf}
J.~Sakstein and M.~Trodden, {\it {Early dark energy from massive neutrinos -- a
  natural resolution of the Hubble tension}},  {\em Phys. Rev. Lett.} {\bf 124}
  (2020), no.~16 161301, [\href{http://arxiv.org/abs/1911.11760}{{\tt
  arXiv:1911.11760}}].

\bibitem{Anchordoqui:2019amx}
L.~A. Anchordoqui, I.~Antoniadis, D.~Lüst, J.~F. Soriano, and T.~R. Taylor,
  {\it {$H_0$ tension and the String Swampland}},  {\em Phys. Rev.} {\bf D101}
  (2020) 083532, [\href{http://arxiv.org/abs/1912.00242}{{\tt
  arXiv:1912.00242}}].

\bibitem{Frusciante:2019puu}
N.~Frusciante, S.~Peirone, L.~Atayde, and A.~De~Felice, {\it {Phenomenology of
  the generalized cubic covariant Galileon model and cosmological bounds}},
  {\em Phys. Rev.} {\bf D101} (2020), no.~6 064001,
  [\href{http://arxiv.org/abs/1912.07586}{{\tt arXiv:1912.07586}}].

\bibitem{Akarsu:2019hmw}
O.~Akarsu, J.~D. Barrow, L.~A. Escamilla, and J.~A. Vazquez, {\it {Graduated
  dark energy: Observational hints of a spontaneous sign switch in the
  cosmological constant}},  {\em Phys. Rev. D} {\bf 101} (2020), no.~6 063528,
  [\href{http://arxiv.org/abs/1912.08751}{{\tt arXiv:1912.08751}}].

\bibitem{Ye:2020btb}
G.~Ye and Y.-S. Piao, {\it {Is the Hubble tension a hint of AdS phase around
  recombination?}},  {\em Phys. Rev.} {\bf D101} (2020), no.~8 083507,
  [\href{http://arxiv.org/abs/2001.02451}{{\tt arXiv:2001.02451}}].

\bibitem{Krishnan:2020obg}
C.~Krishnan, E.~O. Colgáin, Ruchika, A.~A. Sen, M.~M. Sheikh-Jabbari, and
  T.~Yang, {\it {Is there an early Universe solution to Hubble tension?}},
  \href{http://arxiv.org/abs/2002.06044}{{\tt arXiv:2002.06044}}.

\bibitem{DAgostino:2020dhv}
R.~D'Agostino and R.~C. Nunes, {\it {Measurements of $H_0$ in modified gravity
  theories: The role of lensed quasars in the late-time Universe}},  {\em Phys.
  Rev.} {\bf D101} (2020) 103505, [\href{http://arxiv.org/abs/2002.06381}{{\tt
  arXiv:2002.06381}}].

\bibitem{Benevento:2020fev}
G.~Benevento, W.~Hu, and M.~Raveri, {\it {Can Late Dark Energy Transitions
  Raise the Hubble constant?}},  \href{http://arxiv.org/abs/2002.11707}{{\tt
  arXiv:2002.11707}}.

\bibitem{Desmond:2020wep}
H.~Desmond and J.~Sakstein, {\it {Screened fifth forces lower the
  TRGB-calibrated Hubble constant too}},
  \href{http://arxiv.org/abs/2003.12876}{{\tt arXiv:2003.12876}}.

\bibitem{Akarsu:2020yqa}
O.~Akarsu, N.~Katirci, S.~Kumar, R.~C. Nunes, B.~Ozturk, and S.~Sharma, {\it
  {Rastall gravity extension of the standard $\Lambda$CDM model: Theoretical
  features and observational constraints}},
  \href{http://arxiv.org/abs/2004.04074}{{\tt arXiv:2004.04074}}.

\bibitem{Haridasu:2020xaa}
B.~S. Haridasu and M.~Viel, {\it {Late-time decaying dark matter: constraints
  and implications for the $H_0$-tension}},
  \href{http://arxiv.org/abs/2004.07709}{{\tt arXiv:2004.07709}}.

\bibitem{Alestas:2020mvb}
G.~Alestas, L.~Kazantzidis, and L.~Perivolaropoulos, {\it {$H_0$ Tension,
  Phantom Dark Energy and Cosmological Parameter Degeneracies}},
  \href{http://arxiv.org/abs/2004.08363}{{\tt arXiv:2004.08363}}.

\bibitem{Braglia:2020iik}
M.~Braglia, M.~Ballardini, W.~T. Emond, F.~Finelli, A.~E. Gumrukcuoglu,
  K.~Koyama, and D.~Paoletti, {\it {A larger value for $H_0$ by an evolving
  gravitational constant}},  \href{http://arxiv.org/abs/2004.11161}{{\tt
  arXiv:2004.11161}}.

\bibitem{Ballardini:2020iws}
M.~Ballardini, M.~Braglia, F.~Finelli, D.~Paoletti, A.~A. Starobinsky, and
  C.~Umiltà, {\it {Scalar-tensor theories of gravity, neutrino physics, and
  the $H_0$ tension}},  \href{http://arxiv.org/abs/2004.14349}{{\tt
  arXiv:2004.14349}}.

\bibitem{Elizalde:2020mfs}
E.~Elizalde, M.~Khurshudyan, S.~D. Odintsov, and R.~Myrzakulov, {\it {An
  analysis of the $H_{0}$ tension problem in a universe with a viscous dark
  fluid}},  \href{http://arxiv.org/abs/2006.01879}{{\tt arXiv:2006.01879}}.

\bibitem{Lemos:2018smw}
P.~Lemos, E.~Lee, G.~Efstathiou, and S.~Gratton, {\it {Model independent $H(z)$
  reconstruction using the cosmic inverse distance ladder}},  {\em Mon. Not.
  Roy. Astron. Soc.} {\bf 483} (2019), no.~4 4803--4810,
  [\href{http://arxiv.org/abs/1806.06781}{{\tt arXiv:1806.06781}}].

\bibitem{Aylor:2018drw}
K.~Aylor, M.~Joy, L.~Knox, M.~Millea, S.~Raghunathan, and W.~L.~K. Wu, {\it
  {Sounds Discordant: Classical Distance Ladder \& $\Lambda$CDM -based
  Determinations of the Cosmological Sound Horizon}},  {\em Astrophys. J.} {\bf
  874} (2019), no.~1 4, [\href{http://arxiv.org/abs/1811.00537}{{\tt
  arXiv:1811.00537}}].

\bibitem{Knox:2019rjx}
L.~Knox and M.~Millea, {\it {Hubble constant hunter’s guide}},  {\em Phys.
  Rev.} {\bf D101} (2020), no.~4 043533,
  [\href{http://arxiv.org/abs/1908.03663}{{\tt arXiv:1908.03663}}].

\bibitem{Jones:2017udy}
D.~Jones et~al., {\it {Measuring Dark Energy Properties with Photometrically
  Classified Pan-STARRS Supernovae. II. Cosmological Parameters}},  {\em
  Astrophys.\ J.} {\bf 857} (2018), no.~1 51,
  [\href{http://arxiv.org/abs/1710.00846}{{\tt arXiv:1710.00846}}].

\bibitem{Efstathiou:1999tm}
G.~Efstathiou, {\it {Constraining the equation of state of the universe from
  distant type Ia supernovae and cosmic microwave background anisotropies}},
  {\em Mon. Not. Roy. Astron. Soc.} {\bf 310} (1999) 842--850,
  [\href{http://arxiv.org/abs/astro-ph/9904356}{{\tt astro-ph/9904356}}].

\bibitem{Jassal:2004ej}
H.~Jassal, J.~Bagla, and T.~Padmanabhan, {\it {WMAP constraints on low redshift
  evolution of dark energy}},  {\em Mon. Not. Roy. Astron. Soc.} {\bf 356}
  (2005) L11--L16, [\href{http://arxiv.org/abs/astro-ph/0404378}{{\tt
  astro-ph/0404378}}].

\bibitem{Barboza:2008rh}
E.~M. Barboza, Jr. and J.~S. Alcaniz, {\it {A parametric model for dark
  energy}},  {\em Phys. Lett.} {\bf B666} (2008) 415--419,
  [\href{http://arxiv.org/abs/0805.1713}{{\tt arXiv:0805.1713}}].

\bibitem{Ma:2011nc}
J.-Z. Ma and X.~Zhang, {\it {Probing the dynamics of dark energy with novel
  parametrizations}},  {\em Phys. Lett.} {\bf B699} (2011) 233--238,
  [\href{http://arxiv.org/abs/1102.2671}{{\tt arXiv:1102.2671}}].

\bibitem{Pantazis:2016nky}
G.~Pantazis, S.~Nesseris, and L.~Perivolaropoulos, {\it {Comparison of thawing
  and freezing dark energy parametrizations}},  {\em Phys. Rev.} {\bf D93}
  (2016), no.~10 103503, [\href{http://arxiv.org/abs/1603.02164}{{\tt
  arXiv:1603.02164}}].

\bibitem{Escamilla-Rivera:2016qwv}
C.~Escamilla-Rivera, {\it {Status on bidimensional dark energy
  parameterizations using SNe Ia JLA and BAO datasets}},  {\em Galaxies} {\bf
  4} (2016), no.~3 8, [\href{http://arxiv.org/abs/1605.02702}{{\tt
  arXiv:1605.02702}}].

\bibitem{Yang:2017alx}
W.~Yang, S.~Pan, and A.~Paliathanasis, {\it {Latest astronomical constraints on
  some non-linear parametric dark energy models}},  {\em Mon. Not. Roy. Astron.
  Soc.} {\bf 475} (2018), no.~2 2605--2613,
  [\href{http://arxiv.org/abs/1708.01717}{{\tt arXiv:1708.01717}}].

\bibitem{Pan:2017zoh}
S.~Pan, E.~N. Saridakis, and W.~Yang, {\it {Observational Constraints on
  Oscillating Dark-Energy Parametrizations}},  {\em Phys. Rev.} {\bf D98}
  (2018), no.~6 063510, [\href{http://arxiv.org/abs/1712.05746}{{\tt
  arXiv:1712.05746}}].

\bibitem{Vagnozzi:2018jhn}
S.~Vagnozzi, S.~Dhawan, M.~Gerbino, K.~Freese, A.~Goobar, and O.~Mena, {\it
  {Constraints on the sum of the neutrino masses in dynamical dark energy
  models with $w(z) \geq -1$ are tighter than those obtained in $\Lambda$CDM}},
   {\em Phys. Rev.} {\bf D98} (2018), no.~8 083501,
  [\href{http://arxiv.org/abs/1801.08553}{{\tt arXiv:1801.08553}}].

\bibitem{Yang:2018qmz}
W.~Yang, S.~Pan, E.~Di~Valentino, E.~N. Saridakis, and S.~Chakraborty, {\it
  {Observational constraints on one-parameter dynamical dark-energy
  parametrizations and the $H_0$ tension}},  {\em Phys. Rev.} {\bf D99} (2019),
  no.~4 043543, [\href{http://arxiv.org/abs/1810.05141}{{\tt
  arXiv:1810.05141}}].

\bibitem{Perkovic:2020mph}
D.~Perkovic and H.~Stefancic, {\it {Barotropic fluid compatible
  parametrizations of dark energy}},
  \href{http://arxiv.org/abs/2004.05342}{{\tt arXiv:2004.05342}}.

\bibitem{Hee:2016nho}
S.~Hee, J.~A. Vázquez, W.~J. Handley, M.~P. Hobson, and A.~N. Lasenby, {\it
  {Constraining the dark energy equation of state using Bayes theorem and the
  Kullback-Leibler divergence}},  {\em Mon. Not. Roy. Astron. Soc.} {\bf 466}
  (2017), no.~1 369--377, [\href{http://arxiv.org/abs/1607.00270}{{\tt
  arXiv:1607.00270}}].

\bibitem{Shafieloo:2018gin}
A.~Shafieloo, B.~L'Huillier, and A.~A. Starobinsky, {\it {Falsifying
  $\Lambda$CDM: Model-independent tests of the concordance model with eBOSS
  DR14Q and Pantheon}},  {\em Phys. Rev.} {\bf D98} (2018), no.~8 083526,
  [\href{http://arxiv.org/abs/1804.04320}{{\tt arXiv:1804.04320}}].

\bibitem{Wang:2018fng}
Y.~Wang, L.~Pogosian, G.-B. Zhao, and A.~Zucca, {\it {Evolution of dark energy
  reconstructed from the latest observations}},  {\em Astrophys. J.} {\bf 869}
  (2018) L8, [\href{http://arxiv.org/abs/1807.03772}{{\tt arXiv:1807.03772}}].

\bibitem{Gerardi:2019obr}
F.~Gerardi, M.~Martinelli, and A.~Silvestri, {\it {Reconstruction of the Dark
  Energy equation of state from latest data: the impact of theoretical
  priors}},  {\em JCAP} {\bf 1907} (2019) 042,
  [\href{http://arxiv.org/abs/1902.09423}{{\tt arXiv:1902.09423}}].

\bibitem{DiazRivero:2019ukx}
A.~Diaz~Rivero, V.~Miranda, and C.~Dvorkin, {\it {Observable Predictions for
  Massive-Neutrino Cosmologies with Model-Independent Dark Energy}},  {\em
  Phys. Rev.} {\bf D100} (2019), no.~6 063504,
  [\href{http://arxiv.org/abs/1903.03125}{{\tt arXiv:1903.03125}}].

\bibitem{Amendola:1999er}
L.~Amendola, {\it {Coupled quintessence}},  {\em Phys. Rev.} {\bf D62} (2000)
  043511, [\href{http://arxiv.org/abs/astro-ph/9908023}{{\tt
  astro-ph/9908023}}].

\bibitem{Mangano:2002gg}
G.~Mangano, G.~Miele, and V.~Pettorino, {\it {Coupled quintessence and the
  coincidence problem}},  {\em Mod. Phys. Lett.} {\bf A18} (2003) 831--842,
  [\href{http://arxiv.org/abs/astro-ph/0212518}{{\tt astro-ph/0212518}}].

\bibitem{Farrar:2003uw}
G.~R. Farrar and P.~J.~E. Peebles, {\it {Interacting dark matter and dark
  energy}},  {\em Astrophys. J.} {\bf 604} (2004) 1--11,
  [\href{http://arxiv.org/abs/astro-ph/0307316}{{\tt astro-ph/0307316}}].

\bibitem{Pettorino:2004zt}
V.~Pettorino, C.~Baccigalupi, and G.~Mangano, {\it {Extended quintessence with
  an exponential coupling}},  {\em JCAP} {\bf 0501} (2005) 014,
  [\href{http://arxiv.org/abs/astro-ph/0412334}{{\tt astro-ph/0412334}}].

\bibitem{Barrow:2006hia}
J.~D. Barrow and T.~Clifton, {\it {Cosmologies with energy exchange}},  {\em
  Phys. Rev.} {\bf D73} (2006) 103520,
  [\href{http://arxiv.org/abs/gr-qc/0604063}{{\tt gr-qc/0604063}}].

\bibitem{Amendola:2006dg}
L.~Amendola, G.~Camargo~Campos, and R.~Rosenfeld, {\it {Consequences of dark
  matter-dark energy interaction on cosmological parameters derived from SNIa
  data}},  {\em Phys. Rev.} {\bf D75} (2007) 083506,
  [\href{http://arxiv.org/abs/astro-ph/0610806}{{\tt astro-ph/0610806}}].

\bibitem{He:2008tn}
J.-H. He and B.~Wang, {\it {Effects of the interaction between dark energy and
  dark matter on cosmological parameters}},  {\em JCAP} {\bf 0806} (2008) 010,
  [\href{http://arxiv.org/abs/0801.4233}{{\tt arXiv:0801.4233}}].

\bibitem{Pettorino:2008ez}
V.~Pettorino and C.~Baccigalupi, {\it {Coupled and Extended Quintessence:
  theoretical differences and structure formation}},  {\em Phys. Rev.} {\bf
  D77} (2008) 103003, [\href{http://arxiv.org/abs/0802.1086}{{\tt
  arXiv:0802.1086}}].

\bibitem{Valiviita:2008iv}
J.~Valiviita, E.~Majerotto, and R.~Maartens, {\it {Instability in interacting
  dark energy and dark matter fluids}},  {\em JCAP} {\bf 0807} (2008) 020,
  [\href{http://arxiv.org/abs/0804.0232}{{\tt arXiv:0804.0232}}].

\bibitem{Baldi:2008ay}
M.~Baldi, V.~Pettorino, G.~Robbers, and V.~Springel, {\it {Hydrodynamical
  N-body simulations of coupled dark energy cosmologies}},  {\em Mon. Not. Roy.
  Astron. Soc.} {\bf 403} (2010) 1684--1702,
  [\href{http://arxiv.org/abs/0812.3901}{{\tt arXiv:0812.3901}}].

\bibitem{Gavela:2009cy}
M.~B. Gavela, D.~Hernandez, L.~Lopez~Honorez, O.~Mena, and S.~Rigolin, {\it
  {Dark coupling}},  {\em JCAP} {\bf 0907} (2009) 034,
  [\href{http://arxiv.org/abs/0901.1611}{{\tt arXiv:0901.1611}}]. [Erratum:
  JCAP1005,E01(2010)].

\bibitem{Majerotto:2009np}
E.~Majerotto, J.~Valiviita, and R.~Maartens, {\it {Adiabatic initial conditions
  for perturbations in interacting dark energy models}},  {\em Mon. Not. Roy.
  Astron. Soc.} {\bf 402} (2010) 2344--2354,
  [\href{http://arxiv.org/abs/0907.4981}{{\tt arXiv:0907.4981}}].

\bibitem{Gavela:2010tm}
M.~B. Gavela, L.~Lopez~Honorez, O.~Mena, and S.~Rigolin, {\it {Dark Coupling
  and Gauge Invariance}},  {\em JCAP} {\bf 1011} (2010) 044,
  [\href{http://arxiv.org/abs/1005.0295}{{\tt arXiv:1005.0295}}].

\bibitem{Baldi:2010td}
M.~Baldi and V.~Pettorino, {\it {High-z massive clusters as a test for
  dynamical coupled dark energy}},  {\em Mon. Not. Roy. Astron. Soc.} {\bf 412}
  (2011) L1, [\href{http://arxiv.org/abs/1006.3761}{{\tt arXiv:1006.3761}}].

\bibitem{Chimento:2011pk}
L.~P. Chimento and M.~G. Richarte, {\it {Interacting dark matter and modified
  holographic Ricci dark energy induce a relaxed Chaplygin gas}},  {\em Phys.
  Rev.} {\bf D84} (2011) 123507, [\href{http://arxiv.org/abs/1107.4816}{{\tt
  arXiv:1107.4816}}].

\bibitem{Clemson:2011an}
T.~Clemson, K.~Koyama, G.-B. Zhao, R.~Maartens, and J.~Valiviita, {\it
  {Interacting Dark Energy -- constraints and degeneracies}},  {\em Phys. Rev.}
  {\bf D85} (2012) 043007, [\href{http://arxiv.org/abs/1109.6234}{{\tt
  arXiv:1109.6234}}].

\bibitem{Amendola:2011ie}
L.~Amendola, V.~Pettorino, C.~Quercellini, and A.~Vollmer, {\it {Testing
  coupled dark energy with next-generation large-scale observations}},  {\em
  Phys. Rev.} {\bf D85} (2012) 103008,
  [\href{http://arxiv.org/abs/1111.1404}{{\tt arXiv:1111.1404}}].

\bibitem{Bettoni:2012xv}
D.~Bettoni, V.~Pettorino, S.~Liberati, and C.~Baccigalupi, {\it {Non-minimally
  coupled dark matter: effective pressure and structure formation}},  {\em
  JCAP} {\bf 1207} (2012) 027, [\href{http://arxiv.org/abs/1203.5735}{{\tt
  arXiv:1203.5735}}].

\bibitem{Chimento:2012zz}
L.~P. Chimento and M.~G. Richarte, {\it {Interacting dark matter and modified
  holographic Ricci dark energy plus a noninteracting cosmic component}},  {\em
  Phys. Rev.} {\bf D85} (2012) 127301,
  [\href{http://arxiv.org/abs/1207.1492}{{\tt arXiv:1207.1492}}].

\bibitem{Pettorino:2012ts}
V.~Pettorino, L.~Amendola, C.~Baccigalupi, and C.~Quercellini, {\it
  {Constraints on coupled dark energy using CMB data from WMAP and SPT}},  {\em
  Phys. Rev.} {\bf D86} (2012) 103507,
  [\href{http://arxiv.org/abs/1207.3293}{{\tt arXiv:1207.3293}}].

\bibitem{Chimento:2012aea}
L.~P. Chimento and M.~G. Richarte, {\it {Dark matter, dark energy, and dark
  radiation coupled with a transversal interaction}},  {\em Phys. Rev.} {\bf
  D86} (2012) 103501, [\href{http://arxiv.org/abs/1210.5505}{{\tt
  arXiv:1210.5505}}].

\bibitem{Chimento:2013se}
L.~P. Chimento, M.~Forte, and M.~G. Richarte, {\it {Modified holographic Ricci
  dark energy coupled to interacting dark matter and a non interacting baryonic
  component}},  {\em Eur. Phys. J.} {\bf C73} (2013), no.~1 2285,
  [\href{http://arxiv.org/abs/1301.2737}{{\tt arXiv:1301.2737}}].

\bibitem{Chimento:2013qja}
L.~P. Chimento and M.~G. Richarte, {\it {Dark radiation and dark matter coupled
  to holographic Ricci dark energy}},  {\em Eur. Phys. J.} {\bf C73} (2013),
  no.~4 2352, [\href{http://arxiv.org/abs/1303.3356}{{\tt arXiv:1303.3356}}].

\bibitem{Pettorino:2013oxa}
V.~Pettorino, {\it {Testing modified gravity with Planck: the case of coupled
  dark energy}},  {\em Phys. Rev.} {\bf D88} (2013) 063519,
  [\href{http://arxiv.org/abs/1305.7457}{{\tt arXiv:1305.7457}}].

\bibitem{Chimento:2013ira}
L.~P. Chimento and M.~G. Richarte, {\it {Nonbaryonic dark matter and scalar
  field coupled with a transversal interaction plus decoupled radiation}},
  {\em Eur. Phys. J.} {\bf C73} (2013) 2497,
  [\href{http://arxiv.org/abs/1308.0860}{{\tt arXiv:1308.0860}}].

\bibitem{Chimento:2013rya}
L.~P. Chimento, M.~G. Richarte, and I.~E. Sánchez~García, {\it {Interacting
  dark sector with variable vacuum energy}},  {\em Phys. Rev.} {\bf D88} (2013)
  087301, [\href{http://arxiv.org/abs/1310.5335}{{\tt arXiv:1310.5335}}].

\bibitem{Nunes:2014qoa}
R.~C. Nunes and E.~M. Barboza, {\it {Dark matter-dark energy interaction for a
  time-dependent EoS parameter}},  {\em Gen. Rel. Grav.} {\bf 46} (2014) 1820,
  [\href{http://arxiv.org/abs/1404.1620}{{\tt arXiv:1404.1620}}].

\bibitem{Richarte:2014yva}
M.~G. Richarte and L.~Xu, {\it {Interacting parametrized post-Friedmann
  method}},  {\em Gen. Rel. Grav.} {\bf 48} (2016), no.~4 39,
  [\href{http://arxiv.org/abs/1407.4348}{{\tt arXiv:1407.4348}}].

\bibitem{Casas:2015qpa}
S.~Casas, L.~Amendola, M.~Baldi, V.~Pettorino, and A.~Vollmer, {\it {Fitting
  and forecasting coupled dark energy in the non-linear regime}},  {\em JCAP}
  {\bf 1601} (2016) 045, [\href{http://arxiv.org/abs/1508.07208}{{\tt
  arXiv:1508.07208}}].

\bibitem{Murgia:2016ccp}
R.~Murgia, S.~Gariazzo, and N.~Fornengo, {\it {Constraints on the Coupling
  between Dark Energy and Dark Matter from CMB data}},  {\em JCAP} {\bf 04}
  (2016), no.~04 014, [\href{http://arxiv.org/abs/1602.01765}{{\tt
  arXiv:1602.01765}}].

\bibitem{Nunes:2016dlj}
R.~C. Nunes, S.~Pan, and E.~N. Saridakis, {\it {New constraints on interacting
  dark energy from cosmic chronometers}},  {\em Phys. Rev.} {\bf D94} (2016),
  no.~2 023508, [\href{http://arxiv.org/abs/1605.01712}{{\tt
  arXiv:1605.01712}}].

\bibitem{Kumar:2016zpg}
S.~Kumar and R.~C. Nunes, {\it {Probing the interaction between dark matter and
  dark energy in the presence of massive neutrinos}},  {\em Phys. Rev.} {\bf
  D94} (2016), no.~12 123511, [\href{http://arxiv.org/abs/1608.02454}{{\tt
  arXiv:1608.02454}}].

\bibitem{Pan:2016ngu}
S.~Pan and G.~S. Sharov, {\it {A model with interaction of dark components and
  recent observational data}},  {\em Mon. Not. Roy. Astron. Soc.} {\bf 472}
  (2017), no.~4 4736--4749, [\href{http://arxiv.org/abs/1609.02287}{{\tt
  arXiv:1609.02287}}].

\bibitem{Sharov:2017iue}
G.~S. Sharov, S.~Bhattacharya, S.~Pan, R.~C. Nunes, and S.~Chakraborty, {\it {A
  new interacting two fluid model and its consequences}},  {\em Mon. Not. Roy.
  Astron. Soc.} {\bf 466} (2017), no.~3 3497--3506,
  [\href{http://arxiv.org/abs/1701.00780}{{\tt arXiv:1701.00780}}].

\bibitem{Benisty:2017eqh}
D.~Benisty and E.~I. Guendelman, {\it {Interacting Diffusive Unified Dark
  Energy and Dark Matter from Scalar Fields}},  {\em Eur. Phys. J.} {\bf C77}
  (2017), no.~6 396, [\href{http://arxiv.org/abs/1701.08667}{{\tt
  arXiv:1701.08667}}].

\bibitem{Kumar:2017dnp}
S.~Kumar and R.~C. Nunes, {\it {Echo of interactions in the dark sector}},
  {\em Phys. Rev.} {\bf D96} (2017), no.~10 103511,
  [\href{http://arxiv.org/abs/1702.02143}{{\tt arXiv:1702.02143}}].

\bibitem{Guo:2017hea}
R.-Y. Guo, Y.-H. Li, J.-F. Zhang, and X.~Zhang, {\it {Weighing neutrinos in the
  scenario of vacuum energy interacting with cold dark matter: application of
  the parameterized post-Friedmann approach}},  {\em JCAP} {\bf 1705} (2017)
  040, [\href{http://arxiv.org/abs/1702.04189}{{\tt arXiv:1702.04189}}].

\bibitem{DiValentino:2017iww}
E.~Di~Valentino, A.~Melchiorri, and O.~Mena, {\it {Can interacting dark energy
  solve the $H_0$ tension?}},  {\em Phys. Rev. D} {\bf 96} (2017), no.~4
  043503, [\href{http://arxiv.org/abs/1704.08342}{{\tt arXiv:1704.08342}}].

\bibitem{Yang:2017zjs}
W.~Yang, S.~Pan, and J.~D. Barrow, {\it {Large-scale Stability and Astronomical
  Constraints for Coupled Dark-Energy Models}},  {\em Phys. Rev.} {\bf D97}
  (2018), no.~4 043529, [\href{http://arxiv.org/abs/1706.04953}{{\tt
  arXiv:1706.04953}}].

\bibitem{Yang:2017ccc}
W.~Yang, S.~Pan, and D.~F. Mota, {\it {Novel approach toward the large-scale
  stable interacting dark-energy models and their astronomical bounds}},  {\em
  Phys. Rev.} {\bf D96} (2017), no.~12 123508,
  [\href{http://arxiv.org/abs/1709.00006}{{\tt arXiv:1709.00006}}].

\bibitem{Costa:2018aoy}
A.~A. Costa, R.~C.~G. Landim, B.~Wang, and E.~Abdalla, {\it {Interacting Dark
  Energy: Possible Explanation for 21-cm Absorption at Cosmic Dawn}},  {\em
  Eur. Phys. J.} {\bf C78} (2018), no.~9 746,
  [\href{http://arxiv.org/abs/1803.06944}{{\tt arXiv:1803.06944}}].

\bibitem{Yang:2018pej}
W.~Yang, S.~Pan, and A.~Paliathanasis, {\it {Cosmological constraints on an
  exponential interaction in the dark sector}},  {\em Mon. Not. Roy. Astron.
  Soc.} {\bf 482} (2019), no.~1 1007--1016,
  [\href{http://arxiv.org/abs/1804.08558}{{\tt arXiv:1804.08558}}].

\bibitem{Yang:2018euj}
W.~Yang, S.~Pan, E.~Di~Valentino, R.~C. Nunes, S.~Vagnozzi, and D.~F. Mota,
  {\it {Tale of stable interacting dark energy, observational signatures, and
  the $H_0$ tension}},  {\em JCAP} {\bf 1809} (2018) 019,
  [\href{http://arxiv.org/abs/1805.08252}{{\tt arXiv:1805.08252}}].

\bibitem{Yang:2018uae}
W.~Yang, A.~Mukherjee, E.~Di~Valentino, and S.~Pan, {\it {Interacting dark
  energy with time varying equation of state and the $H_0$ tension}},  {\em
  Phys. Rev. D} {\bf 98} (2018), no.~12 123527,
  [\href{http://arxiv.org/abs/1809.06883}{{\tt arXiv:1809.06883}}].

\bibitem{Li:2018ydj}
H.-L. Li, L.~Feng, J.-F. Zhang, and X.~Zhang, {\it {Models of vacuum energy
  interacting with cold dark matter: Constraints and comparison}},  {\em Sci.
  China Phys. Mech. Astron.} {\bf 62} (2019), no.~12 120411,
  [\href{http://arxiv.org/abs/1812.00319}{{\tt arXiv:1812.00319}}].

\bibitem{Martinelli:2019dau}
M.~Martinelli, N.~B. Hogg, S.~Peirone, M.~Bruni, and D.~Wands, {\it
  {Constraints on the interacting vacuum-geodesic CDM scenario}},  {\em Mon.
  Not. Roy. Astron. Soc.} {\bf 488} (2019), no.~3 3423--3438,
  [\href{http://arxiv.org/abs/1902.10694}{{\tt arXiv:1902.10694}}].

\bibitem{Paliathanasis:2019hbi}
A.~Paliathanasis, S.~Pan, and W.~Yang, {\it {Dynamics of nonlinear interacting
  dark energy models}},  {\em Int. J. Mod. Phys.} {\bf D28} (2019), no.~12
  1950161, [\href{http://arxiv.org/abs/1903.02370}{{\tt arXiv:1903.02370}}].

\bibitem{Kumar:2019wfs}
S.~Kumar, R.~C. Nunes, and S.~K. Yadav, {\it {Dark sector interaction: a remedy
  of the tensions between CMB and LSS data}},  {\em Eur. Phys. J.} {\bf C79}
  (2019), no.~7 576, [\href{http://arxiv.org/abs/1903.04865}{{\tt
  arXiv:1903.04865}}].

\bibitem{Pan:2019jqh}
S.~Pan, W.~Yang, C.~Singha, and E.~N. Saridakis, {\it {Observational
  constraints on sign-changeable interaction models and alleviation of the
  $H_0$ tension}},  {\em Phys. Rev.} {\bf D100} (2019), no.~8 083539,
  [\href{http://arxiv.org/abs/1903.10969}{{\tt arXiv:1903.10969}}].

\bibitem{Li:2019loh}
C.~Li, X.~Ren, M.~Khurshudyan, and Y.-F. Cai, {\it {Implications of the
  possible 21-cm line excess at cosmic dawn on dynamics of interacting dark
  energy}},  {\em Phys. Lett.} {\bf B801} (2020) 135141,
  [\href{http://arxiv.org/abs/1904.02458}{{\tt arXiv:1904.02458}}].

\bibitem{Anagnostopoulos:2019myt}
F.~K. Anagnostopoulos, D.~Benisty, S.~Basilakos, and E.~I. Guendelman, {\it
  {Dark energy and dark matter unification from dynamical space time:
  observational constraints and cosmological implications}},  {\em JCAP} {\bf
  1906} (2019) 003, [\href{http://arxiv.org/abs/1904.05762}{{\tt
  arXiv:1904.05762}}].

\bibitem{Yang:2019vni}
W.~Yang, S.~Vagnozzi, E.~Di~Valentino, R.~C. Nunes, S.~Pan, and D.~F. Mota,
  {\it {Listening to the sound of dark sector interactions with gravitational
  wave standard sirens}},  {\em JCAP} {\bf 1907} (2019), no.~07 037,
  [\href{http://arxiv.org/abs/1905.08286}{{\tt arXiv:1905.08286}}].

\bibitem{Yang:2019uzo}
W.~Yang, O.~Mena, S.~Pan, and E.~Di~Valentino, {\it {Dark sectors with
  dynamical coupling}},  {\em Phys. Rev. D} {\bf 100} (2019), no.~8 083509,
  [\href{http://arxiv.org/abs/1906.11697}{{\tt arXiv:1906.11697}}].

\bibitem{Pan:2019gop}
S.~Pan, W.~Yang, E.~Di~Valentino, E.~N. Saridakis, and S.~Chakraborty, {\it
  {Interacting scenarios with dynamical dark energy: Observational constraints
  and alleviation of the $H_0$ tension}},  {\em Phys. Rev.} {\bf D100} (2019),
  no.~10 103520, [\href{http://arxiv.org/abs/1907.07540}{{\tt
  arXiv:1907.07540}}].

\bibitem{Nakamura:2019phn}
S.~Nakamura, R.~Kase, and S.~Tsujikawa, {\it {Coupled vector dark energy}},
  {\em JCAP} {\bf 1912} (2019) 032,
  [\href{http://arxiv.org/abs/1907.12216}{{\tt arXiv:1907.12216}}].

\bibitem{DiValentino:2019ffd}
E.~Di~Valentino, A.~Melchiorri, O.~Mena, and S.~Vagnozzi, {\it {Interacting
  dark energy after the latest Planck, DES, and $H_0$ measurements: an
  excellent solution to the $H_0$ and cosmic shear tensions}},
  \href{http://arxiv.org/abs/1908.04281}{{\tt arXiv:1908.04281}}.

\bibitem{Benetti:2019lxu}
M.~Benetti, W.~Miranda, H.~A. Borges, C.~Pigozzo, S.~Carneiro, and J.~S.
  Alcaniz, {\it {Looking for interactions in the cosmological dark sector}},
  {\em JCAP} {\bf 1912} (2019) 023,
  [\href{http://arxiv.org/abs/1908.07213}{{\tt arXiv:1908.07213}}].

\bibitem{Kase:2019veo}
R.~Kase and S.~Tsujikawa, {\it {Scalar-Field Dark Energy Nonminimally and
  Kinetically Coupled to Dark Matter}},  {\em Phys. Rev.} {\bf D101} (2020),
  no.~6 063511, [\href{http://arxiv.org/abs/1910.02699}{{\tt
  arXiv:1910.02699}}].

\bibitem{Yang:2019uog}
W.~Yang, S.~Pan, R.~C. Nunes, and D.~F. Mota, {\it {Dark calling Dark:
  Interaction in the dark sector in presence of neutrino properties after
  Planck CMB final release}},  {\em JCAP} {\bf 2004} (2020) 008,
  [\href{http://arxiv.org/abs/1910.08821}{{\tt arXiv:1910.08821}}].

\bibitem{Aljaf:2019ilr}
M.~Aljaf, D.~Gregoris, and M.~Khurshudyan, {\it {Phase space analysis and
  singularity classification for linearly interacting dark energy models}},
  {\em Eur. Phys. J.} {\bf C80} (2020), no.~2 112,
  [\href{http://arxiv.org/abs/1911.00747}{{\tt arXiv:1911.00747}}].

\bibitem{Cheng:2019bkh}
G.~Cheng, Y.~Ma, F.~Wu, J.~Zhang, and X.~Chen, {\it {Testing interacting dark
  matter and dark energy model with cosmological data}},
  \href{http://arxiv.org/abs/1911.04520}{{\tt arXiv:1911.04520}}.

\bibitem{Pan:2020zza}
S.~Pan, G.~S. Sharov, and W.~Yang, {\it {Field theoretic interpretations of
  interacting dark energy scenarios and recent observations}},
  \href{http://arxiv.org/abs/2001.03120}{{\tt arXiv:2001.03120}}.

\bibitem{Yang:2020uga}
W.~Yang, E.~Di~Valentino, O.~Mena, S.~Pan, and R.~C. Nunes, {\it {All-inclusive
  interacting dark sector cosmologies}},  {\em Phys. Rev. D} {\bf 101} (2020),
  no.~8 083509, [\href{http://arxiv.org/abs/2001.10852}{{\tt
  arXiv:2001.10852}}].

\bibitem{Lucca:2020zjb}
M.~Lucca and D.~C. Hooper, {\it {Tensions in the dark: shedding light on Dark
  Matter-Dark Energy interactions}},
  \href{http://arxiv.org/abs/2002.06127}{{\tt arXiv:2002.06127}}.

\bibitem{Hogg:2020rdp}
N.~B. Hogg, M.~Bruni, R.~Crittenden, M.~Martinelli, and S.~Peirone, {\it
  {Latest evidence for a late time vacuum -- geodesic CDM interaction}},
  \href{http://arxiv.org/abs/2002.10449}{{\tt arXiv:2002.10449}}.

\bibitem{Amendola:2020ldb}
L.~Amendola and S.~Tsujikawa, {\it {Scaling solutions and weak gravity in dark
  energy with energy and momentum couplings}},
  \href{http://arxiv.org/abs/2003.02686}{{\tt arXiv:2003.02686}}.

\bibitem{Benisty:2020nql}
D.~Benisty, E.~I. Guendelman, E.~Nissimov, and S.~Pacheva, {\it {$\Lambda$CDM
  as a Noether Symmetry in Cosmology}},
  \href{http://arxiv.org/abs/2003.13146}{{\tt arXiv:2003.13146}}.

\bibitem{Gomez-Valent:2020mqn}
A.~Gomez-Valent, V.~Pettorino, and L.~Amendola, {\it {Update on Coupled Dark
  Energy and the $H_0$ tension}},  \href{http://arxiv.org/abs/2004.00610}{{\tt
  arXiv:2004.00610}}.

\bibitem{Wang:2016lxa}
B.~Wang, E.~Abdalla, F.~Atrio-Barandela, and D.~Pavon, {\it {Dark Matter and
  Dark Energy Interactions: Theoretical Challenges, Cosmological Implications
  and Observational Signatures}},  {\em Rept. Prog. Phys.} {\bf 79} (2016),
  no.~9 096901, [\href{http://arxiv.org/abs/1603.08299}{{\tt
  arXiv:1603.08299}}].

\bibitem{DiValentino:2019jae}
E.~Di~Valentino, A.~Melchiorri, O.~Mena, and S.~Vagnozzi, {\it {Nonminimal dark
  sector physics and cosmological tensions}},  {\em Phys. Rev. D} {\bf 101}
  (2020), no.~6 063502, [\href{http://arxiv.org/abs/1910.09853}{{\tt
  arXiv:1910.09853}}].

\bibitem{Nojiri:2006ri}
S.~Nojiri and S.~D. Odintsov, {\it {Introduction to modified gravity and
  gravitational alternative for dark energy}},  {\em eConf} {\bf C0602061}
  (2006) 06, [\href{http://arxiv.org/abs/hep-th/0601213}{{\tt
  hep-th/0601213}}]. [Int. J. Geom. Meth. Mod. Phys.4,115(2007)].

\bibitem{Amendola:2006we}
L.~Amendola, R.~Gannouji, D.~Polarski, and S.~Tsujikawa, {\it {Conditions for
  the cosmological viability of f(R) dark energy models}},  {\em Phys. Rev.}
  {\bf D75} (2007) 083504, [\href{http://arxiv.org/abs/gr-qc/0612180}{{\tt
  gr-qc/0612180}}].

\bibitem{Hu:2007nk}
W.~Hu and I.~Sawicki, {\it {Models of f(R) Cosmic Acceleration that Evade
  Solar-System Tests}},  {\em Phys. Rev.} {\bf D76} (2007) 064004,
  [\href{http://arxiv.org/abs/0705.1158}{{\tt arXiv:0705.1158}}].

\bibitem{Cognola:2007zu}
G.~Cognola, E.~Elizalde, S.~Nojiri, S.~D. Odintsov, L.~Sebastiani, and
  S.~Zerbini, {\it {A Class of viable modified f(R) gravities describing
  inflation and the onset of accelerated expansion}},  {\em Phys. Rev.} {\bf
  D77} (2008) 046009, [\href{http://arxiv.org/abs/0712.4017}{{\tt
  arXiv:0712.4017}}].

\bibitem{Saridakis:2009bv}
E.~N. Saridakis, {\it {Horava-Lifshitz Dark Energy}},  {\em Eur. Phys. J.} {\bf
  C67} (2010) 229--235, [\href{http://arxiv.org/abs/0905.3532}{{\tt
  arXiv:0905.3532}}].

\bibitem{Lim:2010yk}
E.~A. Lim, I.~Sawicki, and A.~Vikman, {\it {Dust of Dark Energy}},  {\em JCAP}
  {\bf 1005} (2010) 012, [\href{http://arxiv.org/abs/1003.5751}{{\tt
  arXiv:1003.5751}}].

\bibitem{Deffayet:2010qz}
C.~Deffayet, O.~Pujolas, I.~Sawicki, and A.~Vikman, {\it {Imperfect Dark Energy
  from Kinetic Gravity Braiding}},  {\em JCAP} {\bf 1010} (2010) 026,
  [\href{http://arxiv.org/abs/1008.0048}{{\tt arXiv:1008.0048}}].

\bibitem{Clifton:2011jh}
T.~Clifton, P.~G. Ferreira, A.~Padilla, and C.~Skordis, {\it {Modified Gravity
  and Cosmology}},  {\em Phys. Rept.} {\bf 513} (2012) 1--189,
  [\href{http://arxiv.org/abs/1106.2476}{{\tt arXiv:1106.2476}}].

\bibitem{Chamseddine:2013kea}
A.~H. Chamseddine and V.~Mukhanov, {\it {Mimetic Dark Matter}},  {\em JHEP}
  {\bf 11} (2013) 135, [\href{http://arxiv.org/abs/1308.5410}{{\tt
  arXiv:1308.5410}}].

\bibitem{Addazi:2016oob}
A.~Addazi, S.~Capozziello, and S.~Odintsov, {\it {Born–Infeld condensate as a
  possible origin of neutrino masses and dark energy}},  {\em Phys. Lett.} {\bf
  B760} (2016) 611--616, [\href{http://arxiv.org/abs/1607.05706}{{\tt
  arXiv:1607.05706}}].

\bibitem{Sebastiani:2016ras}
L.~Sebastiani, S.~Vagnozzi, and R.~Myrzakulov, {\it {Mimetic gravity: a review
  of recent developments and applications to cosmology and astrophysics}},
  {\em Adv. High Energy Phys.} {\bf 2017} (2017) 3156915,
  [\href{http://arxiv.org/abs/1612.08661}{{\tt arXiv:1612.08661}}].

\bibitem{Barvinsky:2017pmm}
A.~O. Barvinsky and A.~{\relax Yu}. Kamenshchik, {\it {Darkness without dark
  matter and energy -- generalized unimodular gravity}},  {\em Phys. Lett.}
  {\bf B774} (2017) 59--63, [\href{http://arxiv.org/abs/1705.09470}{{\tt
  arXiv:1705.09470}}].

\bibitem{Renk:2017rzu}
J.~Renk, M.~Zumalacarregui, F.~Montanari, and A.~Barreira, {\it {Galileon
  gravity in light of ISW, CMB, BAO and H$_0$ data}},  {\em JCAP} {\bf 1710}
  (2017) 020, [\href{http://arxiv.org/abs/1707.02263}{{\tt arXiv:1707.02263}}].

\bibitem{Dutta:2017fjw}
J.~Dutta, W.~Khyllep, E.~N. Saridakis, N.~Tamanini, and S.~Vagnozzi, {\it
  {Cosmological dynamics of mimetic gravity}},  {\em JCAP} {\bf 1802} (2018)
  041, [\href{http://arxiv.org/abs/1711.07290}{{\tt arXiv:1711.07290}}].

\bibitem{Calza:2018ohl}
M.~Calza, M.~Rinaldi, and L.~Sebastiani, {\it {A special class of solutions in
  $F(R)$-gravity}},  {\em Eur. Phys. J.} {\bf C78} (2018), no.~3 178,
  [\href{http://arxiv.org/abs/1802.00329}{{\tt arXiv:1802.00329}}].

\bibitem{Casalino:2018tcd}
A.~Casalino, M.~Rinaldi, L.~Sebastiani, and S.~Vagnozzi, {\it {Mimicking dark
  matter and dark energy in a mimetic model compatible with GW170817}},  {\em
  Phys. Dark Univ.} {\bf 22} (2018) 108,
  [\href{http://arxiv.org/abs/1803.02620}{{\tt arXiv:1803.02620}}].

\bibitem{Saridakis:2018unr}
E.~N. Saridakis, K.~Bamba, R.~Myrzakulov, and F.~K. Anagnostopoulos, {\it
  {Holographic dark energy through Tsallis entropy}},  {\em JCAP} {\bf 1812}
  (2018) 012, [\href{http://arxiv.org/abs/1806.01301}{{\tt arXiv:1806.01301}}].

\bibitem{Kase:2018aps}
R.~Kase and S.~Tsujikawa, {\it {Dark energy in Horndeski theories after
  GW170817: A review}},  {\em Int. J. Mod. Phys. D} {\bf 28} (2019), no.~05
  1942005, [\href{http://arxiv.org/abs/1809.08735}{{\tt arXiv:1809.08735}}].

\bibitem{Odintsov:2019evb}
S.~D. Odintsov and V.~K. Oikonomou, {\it {Unification of Inflation with Dark
  Energy in $f(R)$ Gravity and Axion Dark Matter}},  {\em Phys. Rev.} {\bf D99}
  (2019), no.~10 104070, [\href{http://arxiv.org/abs/1905.03496}{{\tt
  arXiv:1905.03496}}].

\bibitem{Hogas:2019ywm}
M.~Hogas, F.~Torsello, and E.~Mörtsell, {\it {On the Stability of Bimetric
  Structure Formation}},  {\em JCAP} {\bf 2004} (2020) 046,
  [\href{http://arxiv.org/abs/1910.01651}{{\tt arXiv:1910.01651}}].

\bibitem{Ade:2015rim}
{\bf Planck} Collaboration, P.~Ade et~al., {\it {Planck 2015 results. XIV. Dark
  energy and modified gravity}},  {\em Astron.\ Astrophys.} {\bf 594} (2016)
  A14, [\href{http://arxiv.org/abs/1502.01590}{{\tt arXiv:1502.01590}}].

\bibitem{Bertschinger:2008zb}
E.~Bertschinger and P.~Zukin, {\it {Distinguishing Modified Gravity from Dark
  Energy}},  {\em Phys. Rev. D} {\bf 78} (2008) 024015,
  [\href{http://arxiv.org/abs/0801.2431}{{\tt arXiv:0801.2431}}].

\bibitem{Martinelli:2010wn}
M.~Martinelli, E.~Calabrese, F.~De~Bernardis, A.~Melchiorri, L.~Pagano, and
  R.~Scaramella, {\it {Constraining Modified Gravity with Euclid}},  {\em Phys.
  Rev.} {\bf D83} (2011) 023012, [\href{http://arxiv.org/abs/1010.5755}{{\tt
  arXiv:1010.5755}}].

\bibitem{Baker:2014zva}
T.~Baker, P.~G. Ferreira, C.~D. Leonard, and M.~Motta, {\it {New Gravitational
  Scales in Cosmological Surveys}},  {\em Phys. Rev. D} {\bf 90} (2014), no.~12
  124030, [\href{http://arxiv.org/abs/1409.8284}{{\tt arXiv:1409.8284}}].

\bibitem{DiValentino:2015bja}
E.~Di~Valentino, A.~Melchiorri, and J.~Silk, {\it {Cosmological hints of
  modified gravity?}},  {\em Phys.\ Rev.\ D} {\bf 93} (2016), no.~2 023513,
  [\href{http://arxiv.org/abs/1509.07501}{{\tt arXiv:1509.07501}}].

\bibitem{Calabrese:2008rt}
E.~Calabrese, A.~Slosar, A.~Melchiorri, G.~F. Smoot, and O.~Zahn, {\it {Cosmic
  Microwave Weak lensing data as a test for the dark universe}},  {\em Phys.
  Rev. D} {\bf 77} (2008) 123531, [\href{http://arxiv.org/abs/0803.2309}{{\tt
  arXiv:0803.2309}}].

\bibitem{Handley:2019tkm}
W.~Handley, {\it {Curvature tension: evidence for a closed universe}},
  \href{http://arxiv.org/abs/1908.09139}{{\tt arXiv:1908.09139}}.

\bibitem{DiValentino:2019qzk}
E.~Di~Valentino, A.~Melchiorri, and J.~Silk, {\it {Planck evidence for a closed
  Universe and a possible crisis for cosmology}},  {\em Nat. Astron.} {\bf 4}
  (2019), no.~2 196--203, [\href{http://arxiv.org/abs/1911.02087}{{\tt
  arXiv:1911.02087}}].

\bibitem{Efstathiou:2020wem}
G.~Efstathiou and S.~Gratton, {\it {The evidence for a spatially flat
  Universe}},  \href{http://arxiv.org/abs/2002.06892}{{\tt arXiv:2002.06892}}.

\bibitem{Efstathiou:2019mdh}
G.~Efstathiou and S.~Gratton, {\it {A Detailed Description of the CamSpec
  Likelihood Pipeline and a Reanalysis of the Planck High Frequency Maps}},
  \href{http://arxiv.org/abs/1910.00483}{{\tt arXiv:1910.00483}}.

\bibitem{Ivezic:2008fe}
{\bf LSST} Collaboration, v.~Z. Ivezi\'c et~al., {\it {LSST: from Science
  Drivers to Reference Design and Anticipated Data Products}},  {\em Astrophys.
  J.} {\bf 873} (2019), no.~2 111, [\href{http://arxiv.org/abs/0805.2366}{{\tt
  arXiv:0805.2366}}].

\end{thebibliography}\endgroup
\bibliographystyle{JHEP}
\end{document}